\documentclass[fleqn,usenatbib]{mnras}

\usepackage{newtxtext,newtxmath}

\usepackage[T1]{fontenc}
\usepackage{ae,aecompl}

\DeclareRobustCommand{\VAN}[3]{#2}
\let\VANthebibliography\thebibliography
\def\thebibliography{\DeclareRobustCommand{\VAN}[3]{##3}\VANthebibliography}

\usepackage{graphicx}	

\title[A panoramic view of NGC 6822]{A panoramic view of the Local Group dwarf galaxy NGC 6822}

\author[S. Zhang et al.]{
Shumeng Zhang$^{1}$\thanks{E-mail: u5692595@anu.edu.au},
Dougal Mackey$^{1}$,
Gary S. Da Costa$^{1}$
\\
$^{1}$Research School of Astronomy and Astrophysics, Australian National University, Canberra, ACT 2611, Australia
}

\date{Accepted XXX. Received YYY; in original form ZZZ}

\pubyear{2021}

\begin{document}
\label{firstpage}
\pagerange{\pageref{firstpage}--\pageref{lastpage}}
\maketitle

\begin{abstract}
We present a panoramic survey of the isolated Local Group dwarf irregular galaxy NGC 6822. Our photometry reaches $\sim2-3$ magnitudes deeper than most previous studies and spans the widest area around the dwarf compared to any prior work. We observe no stellar over-densities in the outskirts of NGC 6822 to $V\sim 30$\ mag$\,$arcsec$^{-2}$ and a projected radius of $16.5$\ kpc. This indicates that NGC 6822 has not experienced any recent interaction with a companion galaxy, despite previous suggestions to the contrary. Similarly, we find no evidence for any dwarf satellites of NGC 6822 to a limiting luminosity $M_V\approx -5$. NGC 6822 contains a disk of H{\sc i} gas and young stars, oriented at $\sim 60\degr$ to an extended spheroid composed of old stellar populations. We observe no correlation between the distribution of young stars and spheroid members. Our imaging allows us to trace the spheroid to nearly $11$\ kpc along its major axis, commensurate with the extent of the NGC 6822 globular cluster system. We find that the spheroid becomes increasingly flattened at larger radii, and its position angle twists by up to $40\degr$. We use {\it Gaia} EDR3 astrometry to measure a proper motion for NGC 6822, and then sample its orbital parameter space. While this galaxy has spent the majority of its life in isolation, we find that it likely passed within the virial radius of the Milky Way $\sim3-4$\ Gyr ago. This may explain the apparent flattening and twisting observed in the outskirts of its spheroid.
\end{abstract}




\section{Introduction}
Dwarf galaxies are the most common type of galaxy in the Universe, and the building blocks of larger systems. As such, they play a crucial role as probes of the astrophysics of galaxy formation and evolution, especially in the context of the prevailing $\Lambda$ plus Cold Dark Matter ($\Lambda$CDM) framework \citep[see e.g.,][]{bullock:17}. Dwarf galaxies in the Local Group are particularly important to this endeavour, as they are sufficiently close that they may be studied, both photometrically and spectroscopically, on a star-by-star basis.

Because the halo mass function in $\Lambda$CDM is scale free, all dark matter halos that are large enough to host galaxies should be surrounded by substructures \citep[e.g.,][]{moore:99,diemand:07,springel:08}. This is true both for giant galaxies like the Milky Way, as well as much smaller systems \citep[e.g.,][]{diemand:08,li:08}. The practical consequence is that we should expect to see dwarf galaxies hosting systems of even smaller satellites \citep[e.g.,][]{wheeler:15}. Moreover, we might expect to see low-mass stellar halos surrounding dwarf galaxies, since the build-up of such structures is at least partially due to the accumulation of tidally disrupted satellites over cosmic time \citep[e.g.][]{bullock:05,cooper:10}. 

Experimental verification of these predictions is difficult. Even around large galaxies such as the Milky Way and Andromeda, stellar halos are extremely faint \citep[e.g.,][]{cooper:10,ibata:14}, and any such structure around a dwarf galaxy might reasonably be expected to have a commensurately lower luminosity. Unsurprisingly, evidence for their existence is therefore scarce \citep[e.g.,][]{minniti:96,minniti:99,md:12,amorisco:14,mcmonigal:16}. Similarly, while a few Magellanic-type dwarfs in the Local Volume are known to host companions \citep[e.g.,][]{carlin:16,carlin:21}, the mass scale for "satellites of satellites" is concentrated in the regime of ultra-faint systems \citep{wheeler:15}, which have so far proven largely impossible to detect outside the Milky Way \citep[for example, all known dwarfs around M31 have $M_V\leq -6$;][]{martin:16}.

The Large and Small Magellanic Clouds (LMC/SMC), as the most massive dwarf satellites of the Milky Way, and among the closest, are a natural location to try and test the above ideas. However, even in these systems the observations are challenging -- both galaxies span large angles on the sky such that very deep and wide photometric surveys are required. Numerous studies have searched without success for a stellar halo surrounding the LMC, although a multitude of perturbations to its outer disk are evident \citep[e.g.,][]{saha:10,balbinot:2015,mackey:16,mackey:2018,belokurov:19,mag:21,cullinane:21}. Nonetheless, the LMC may possess a kinematically hot component \citep{minniti:03}; moreover, \citet{belokurov:16} photometrically detected a number of streamlike features in the extreme outskirts of the Clouds (at radii $25-50$\ kpc, well beyond the structures seen in the LMC outer disk). Subsequent spectroscopic observations confirmed that at least two of these are plausibly associated with the Magellanic system \citep{navarrete:19}, suggesting that it may indeed be surrounded by an exceptionally low-luminosity stellar envelope. The difficulty in detecting, characterising, and interpreting such a structure is compounded by the fact that the LMC and SMC are interacting both with each other and with the Milky Way; at a minimum, contamination by the Galactic stellar halo is a significant problem \citep{navarrete:19}. 

Despite the difficulty in determining the presence of a halo-like component around the Magellanic Clouds, the existence of Magellanic satellites is well established. Many of the ultra-faint dwarf galaxies uncovered in recent years by deep wide-field surveys of the southern hemisphere \citep[e.g.,][]{drlica:15,koposov:15,bechtol:15} have been convincingly demonstrated to be members, or ex-members, of the Magellanic system \citep[e.g.,][]{jethwa:16,dooley:17,sales:17,kallivayalil:18,erkal:20}. However, again the Magellanic-Milky Way interaction complicates interpretation -- association of satellites with the Clouds, or otherwise, must be done probabilistically, and depends on various assumptions such as the mass of the LMC \citep[e.g., ][]{erkal:20}.

One way of mitigating the difficulties introduced by the fact that the LMC and SMC are interacting with the Milky Way is to move to more isolated targets. NGC 6822, sometimes known as Barnard's Galaxy \citep{barnard:84}, is the most massive of the handful of systems that sit just outside the Milky Way's virial radius. It is a gas-rich dwarf irregular galaxy with a stellar mass $M_{*}\approx 10^8 M_\odot$ \citep{mcconnachie:12}, perhaps a quarter that of the SMC, plus in excess of $\sim 10^8 M_\odot$ of H{\sc i} gas \citep[e.g.,][]{deblok:00}. As such, it is an excellent location to search for low-mass satellites \citep[e.g.,][]{wheeler:15,dooley:17b}, and/or a stellar halo component.

Many previous studies have revealed NGC 6822 to be a fascinating object in its own right. It possesses both an extended elliptical spheroid composed of old stars, and an irregular disk-like H{\sc i} distribution hosting active star formation \citep[e.g.,][]{deblok:00,deblok:03,deblok:06,komiyama:03,demers:06,battinelli:06,fusco:14,higgs:21}. The main axis of the H{\sc i} disk is angled at approximately $60\degr$ to the major axis of the spheroid, leading to speculation that NGC 6822 might have experienced a recent merger event or tidal interaction, or even be a polar ring galaxy \citep[e.g.,][]{deblok:00,deblok:06,demers:06,battinelli:06}. 

However, {\it Hubble Space Telescope} imaging at various locations in the system by \citet{cannon:12}, including the putative companion galaxy, revealed no obvious variation in stellar populations, leading these authors to suggest that the H{\sc i} distribution most likely reflects a warped disk inclined to the line of sight. This idea is supported by the fact that the spheroid is observed to exhibit {\it prolate} rotation \citep[i.e., about its major axis,][]{thompson:16,belland:20}, which is around approximately the same axis as the rotation of the H{\sc i} disk as measured by, e.g., \citet{weldrake:03}. NGC 6822 also hosts a globular cluster system numbering $8$ objects \citep[e.g.,][]{hubble:25,hwang:11,huxor:13}. This system is unusual in that the clusters are largely distributed along the major axis of the spheroid (see Figure \ref{f:pointings}); it may also exhibit weak net rotation, albeit with poorly constrained orientation \citep{veljanoski:15}. 

Several recent studies have attempted to measure a proper motion for NGC 6822 using astrometry from the {\it Gaia} satellite \citep[e.g.,][]{mcconnachie:21,battaglia:21}. The implied orbital parameter space indicates that NGC 6822 has likely been an isolated system for the majority of its life; however, there is a non-negligible chance that it has passed within the virial radius of th Milky Way in the past few Gyr. In this case, NGC 6822 may constitute a "backsplash" galaxy -- such systems are expected to be relatively common around giant galaxies like the Milky Way \citep[e.g.,][]{gill:05,teyssier:12,buck:19}.

In this paper, we present deep panoramic imaging of NGC 6822, extending to a maximum radius of $\sim 2\degr$ from the centre of the dwarf. This is a much wider area than covered by previous surveys. Our main aim is to investigate the extreme outskirts of this system for the first time, searching for (i) stellar streams or overdensities, or a halo-like component that may indicate recent merger activity; (ii) tidal disturbances that may indicate a previous interaction with the Milky Way; and (iii) low-luminosity satellites. Throughout this work we assume that the centre of NGC 6822 is at $\alpha = 19^{\rm h}\,44^{\rm m}\,56.6^{\rm s}, \delta = -14\degr\,47\arcmin\,21\arcsec$ \citep{mcconnachie:12}, and that it sits at a distance of $472$\ kpc \citep[derived as a weighted average of the Cepheid, RR Lyrae, and tip of the red giant branch (TRGB) distances compiled in][]{gorski:11}. At this distance, one arcsecond subtends $\approx 2.29$\ pc.

\section{Observations and Data reduction}
\subsection{Observations}  
We observed NGC 6822 and the surrounding area with the Dark Energy Camera (DECam) on the 4m Blanco Telescope at Cerro Tololo Inter-American Observatory (CTIO) in Chile, and with Megacam on the 6.5m Magellan Clay Telescope at Las Campanas Observatory in Chile. Table \ref{t:log} presents our observing log.

\begin{figure*}
\begin{minipage}{0.49\textwidth}
\centering
\includegraphics[width=75mm]{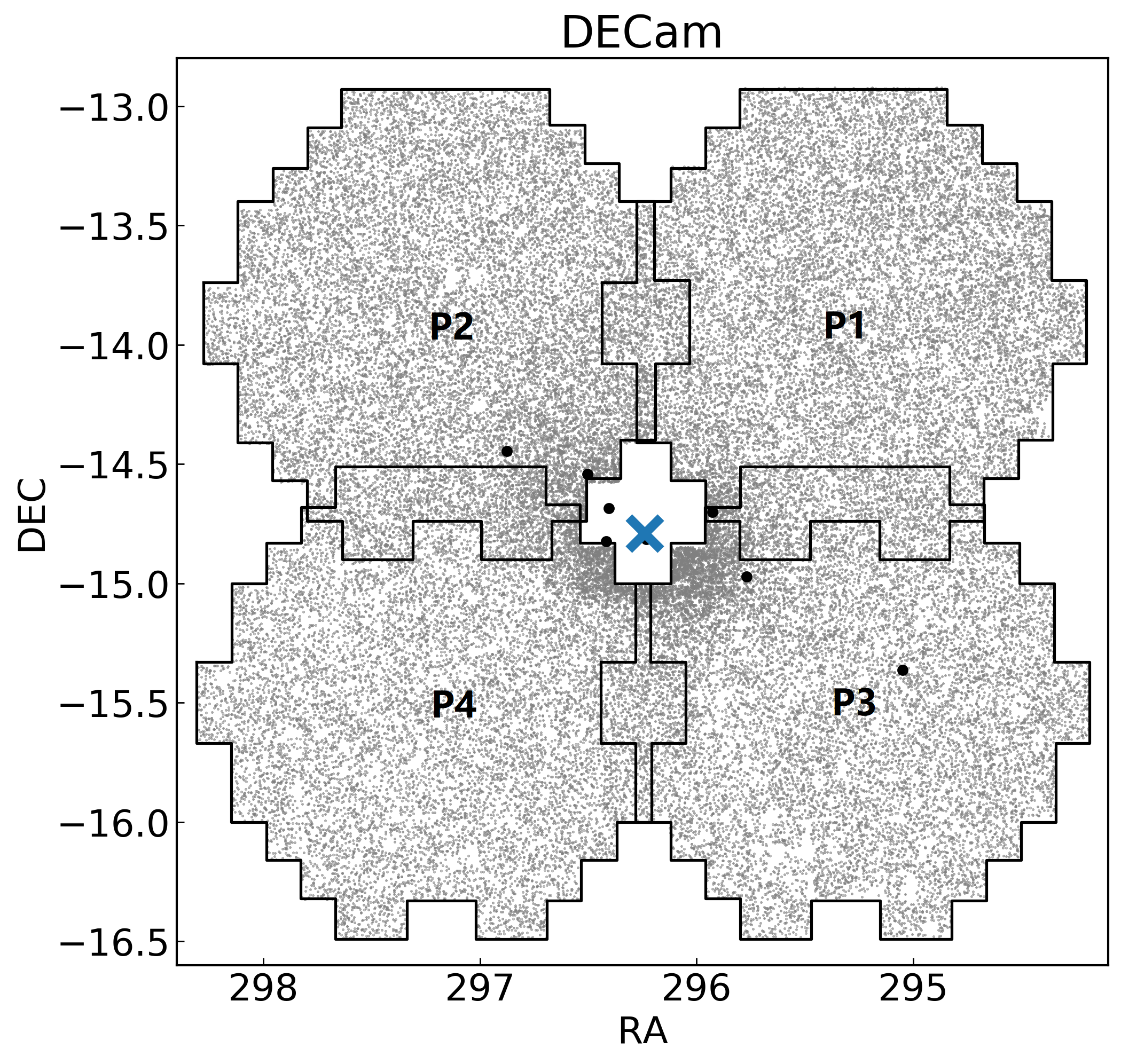}
\end{minipage}\hfill
\begin{minipage}{0.49\textwidth}
\centering
\includegraphics[width=85mm]{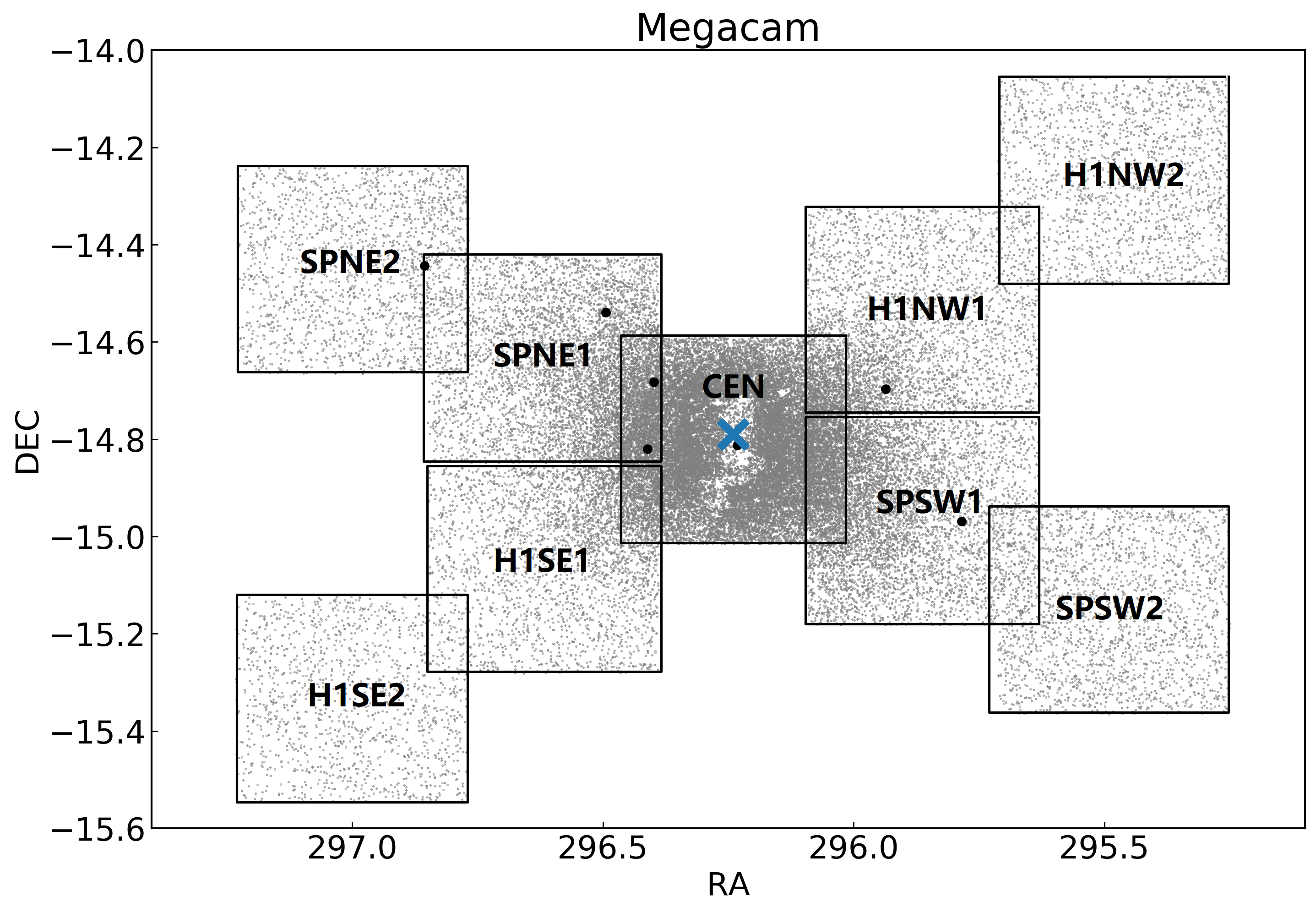}
\end{minipage}
\caption{Maps showing the pointings for our DECam (left panel) and Megacam (right panel) mosaics. North is up and east to the left. Each pointing is labelled, and the position of the centre of NGC 6822 is marked with a blue cross. Note that the geometric centre of the DECam mosaic is mildly offset from the centre of NGC 6822. Also shown are red giant branch stars passing the membership criteria outlined in Section \ref{ss:cmds} (light grey points); the spheroid of NGC 6822 is clearly visible in both data sets. The eight known globular clusters belonging to NGC 6822 are indicated with black dots.}
\label{f:pointings}
\end{figure*}

DECam is a mosaic imager comprising 62 CCDs spanning a $3$\ deg$^2$ field with a pixel scale of $0.27\arcsec$ pixel$^{-1}$ \citep{flaugher:15}. Its very wide field-of-view allowed us to probe the outskirts of NGC 6822 to radii well beyond the limit of $\sim50$\ arcmin explored by \citet{battinelli:06}. In total we observed four DECam pointings, named P1-P4, arranged in a square pattern around the centre of NGC 6822 as shown in Figure \ref{f:pointings}\footnote{Note that due to a small observing error, the geometric centre of the DECam mosaic is mildly offset from the centre of NGC 6822.}. We used the $g$- and $i$-band filters. Observations for P2-P4 were carried out on the nights of 2013 July 11-13 with typical seeing of $1.0-1.4\arcsec$, under program 2013A-0617 (PI: D. Mackey). Changing conditions precluded observations for P1 at that time, and P1 was observed during a subsequent run (program 2013B-0611, PI: D. Mackey) on 2013 September 24 with seeing $0.9-1.1\arcsec$. At each pointing we obtained $4\times900$s images in $g$ and $11\times600$s images in $i$, with the aim of obtaining photometry reaching at least to the depth of the NGC 6822 red clump at $g\approx 25$ and $i\approx 24.5$.

\begin{table}
	\centering
	\caption{DECam and Megacam observing logs.}
	\label{t:log}
	\begin{tabular}{lclll}
		\hline
		Field & Filter & Exposures & Date & Seeing \\
		\hline
		DECam\vspace{0.5mm} & & & & \\
		P1 & $g$ & $4\times900$s & 2013 Sep 24 & $1.0-1.1\arcsec$ \\
		   & $i$ & $11\times600$s & 2013 Sep 24 & $0.9-1.0\arcsec$ \\
		P2 & $g$ & $4\times900$s & 2013 Jul 11 & $1.3-1.5\arcsec$ \\
		   & $i$ & $11\times600$s & 2013 Jul 11 & $1.1-1.3\arcsec$ \\
		P3 & $g$ & $4\times900$s & 2013 Jul 12 & $1.1-1.3\arcsec$ \\
		   & $i$ & $11\times600$s & 2013 Jul 12 & $0.9-1.2\arcsec$ \\
		P4 & $g$ & $4\times900$s & 2013 Jul 12 & $1.1-1.4\arcsec$ \\
		   & $i$ & $11\times600$s & 2013 Jul 13 & $1.0-1.2\arcsec$ \\
        \hline
		Megacam\vspace{0.5mm} & & & & \\
	    CEN   & $g$ & $3\times200$s & 2013 Jun 13 & $0.7-0.8\arcsec$ \\ 
	          & $i$ & $3\times420$s & 2013 Jun 13 & $0.6-0.7\arcsec$ \\ 
	    SPNE1 & $g$ & $3\times200$s & 2013 Jun 13 & $0.7-0.9\arcsec$ \\ 
	          & $i$ & $3\times420$s & 2013 Jun 13 & $0.6-0.7\arcsec$ \\ 
	    SPNE2 & $g$ & $3\times200$s & 2013 Jun 13 & $0.7-0.8\arcsec$ \\ 
	          & $i$ & $3\times420$s & 2013 Jun 13 & $0.6\arcsec$ \\ 
	    SPSW1 & $g$ & $3\times200$s & 2013 Jun 13 & $0.7-0.9\arcsec$ \\ 
	          & $i$ & $3\times420$s & 2013 Jun 13 & $0.6-0.7\arcsec$ \\ 
	    SPSW2 & $g$ & $3\times200$s & 2013 Jun 13 & $0.7-0.8\arcsec$ \\ 
	          & $i$ & $3\times420$s & 2013 Jun 13 & $0.6-0.7\arcsec$ \\
	    H1NW1 & $g$ & $3\times200$s & 2013 Jun 14 & $0.6-0.8\arcsec$ \\ 
	          & $i$ & $3\times420$s & 2013 Jun 14 & $0.6\arcsec$ \\ 
	    H1NW2 & $g$ & $3\times200$s & 2013 Jun 14 & $0.8-0.9\arcsec$ \\ 
	          & $i$ & $3\times420$s & 2013 Jun 14 & $0.7\arcsec$ \\ 
	    H1SE1 & $g$ & $3\times200$s & 2013 Jun 14 & $0.8-1.0\arcsec$ \\ 
	          & $i$ & $3\times420$s & 2013 Jun 14 & $0.6-0.8\arcsec$ \\ 
	    H1SE2 & $g$ & $3\times200$s & 2013 Jun 14 & $0.9-1.0\arcsec$ \\ 
	          & $i$ & $3\times420$s & 2013 Jun 14 & $0.7-0.8\arcsec$ \\ 
        \hline
	\end{tabular}
\end{table}

Megacam utilises $36$ CCDs arranged in a $9\times4$ array, allowing for a $25\arcmin \times 25\arcmin$ field-of-view with a pixel scale of $0.16\arcsec$ when binned $2\times2$ \citep{mcleod:15}. The smaller area covered by Megacam means that it is not as well suited to contiguously tile as large a region as spanned by our DECam data. Instead we obtained a mosaic of $9$ pointings arranged in an `X' shape as shown in Figure \ref{f:pointings}, with one arm tracing the major axis of the spheroid mapped by \citet{battinelli:06} and the other following the H{\sc i} disk of \citet{deblok:00}. These data include the central parts of NGC 6822, which were not covered by our DECam observations (although even with the higher image quality of our Megacam data this region remains largely unresolved due to signficant crowding). We again used the $g$- and $i$-band filters. All data were taken on the nights of 2013 June 13-14, with typical seeing of $0.6-1.0\arcsec$. At each pointing we obtained $3\times200$s images in $g$ and $3\times420$s in $i$ with the aim of reaching at least a magnitude below the NGC 6822 red clump in uncrowded regions. As indicated in Table \ref{t:log} and Figure \ref{f:pointings}, our Megacam fields are labelled according to their location (`SP' for spheroid pointings and `H1' for H{\sc i} pointings; `CEN' for the central pointing) and direction relative to the centre of NGC 6822 (NE for north-east, NW for north-west, and so on).

\subsection{Data Reduction \& Photometry}
Our raw DECam data were processed by the DECam community pipeline \citep{valdes:14}, and we downloaded the reduced images from the NOAO data archive. Processed Megacam images were provided by the OIR Telescope Data Centre at the Smithsonian Astrophysical Observatory via the dedicated instrument pipeline \citep[see][]{mcleod:15}. Both pipelines incorporate standard reduction tasks including debiasing and dark correction, flat-fielding, defringing (Megacam $i$-band images, in particular, exhibit noticeable fringing at the $\approx 1\%$ level), illumination correction, cosmic ray and bad pixel masking, astrometric calibration, distortion correction, and remapping. The pipelines create stacked images for all pointings; however as explained below, we preferred to photometer the individual corrected and resampled exposures.

We followed a procedure very similar to that outlined in \citet{koposov:15, koposov:18}. For a given pointing and filter we first created image stacks on a per-CCD basis using the {\sc swarp} software \citep{bertin:10}. We then used {\sc sextractor} \citep{bertin:96} to measure the location of all sources on a given CCD stack. Next, we ran {\sc sextractor} on each of the aligned individual CCD images, and with this list of detections used {\sc psfex} \citep{bertin:11} to create a point spread function (PSF) model for each individual image. Finally, we used {\sc sextractor} in double image mode to conduct forced photometry on each of the individual CCD images using the list of sources measured from the stacked frame as the input list, the stacked frame itself as the reference image, and the appropriate PSF model per individual frame.

This resulted in a list of between $1$ and $N$ measurements of brightness, shape, etc, for each detected source, where $N$ is the number of exposures for a given position and filter. We first calibrated the individual photometric measurements to the PanSTARRS1 (PS1) second data release (DR2) \citep[e.g.,][]{flewelling:20}, and then combined the data for each star in a weighted average. We chose the PS1 catalogue since this provides well-calibrated photometry (to the $\sim 1$ per cent level) in both the $g$ and $i$ bands across the entire region of interest.

To perform the calibration, we took all stars in the pointing under consideration (i.e., across all $\sim 60$ CCDs for DECam, or 36 CCDs for Megacam) and matched those with PS1 magnitudes in the range $16.0-20.5$ in $g$, or $17.0-21.0$ in $i$. The bright end of this range is set by the saturation level of our photometry, and the faint end by the precision of PS1 photometry. We also chose only stars with $0.5 < (g-i)_{\rm PS1} < 1.3$ to minimize the effect of non-zero colour terms between the DECam and Megacam filters, and PS1. Across the colour interval of interest (matching the approximate extent of the NGC 6822 red giant branch) these non-zero terms induce systematics at a level comparable to the size of our individual uncertainties (a few $\times 0.01$ mag) and we therefore elected not to correct them. In any case, our analysis (presented in Section \ref{s:analysis}) is largely differential in nature such that correction for small colour terms is not necessary.

The above procedure determines the mean offset between instrumental magnitudes and calibrated magnitudes across a single pointing. We then calculated corrections on a per-CCD basis using the observed residuals after the mean offset had been applied. This let us account for e.g., small differences in gain between CCDs. Once all the individual measurements for a given star were on the same (calibrated) scale we calculated an error-weighted average to determine the final magnitude and uncertainty. Relevant shape parameters provided by {\sc sextractor} (such as {\tt fwhm}, {\tt ellipticity}, {\tt spread\_model}, and {\tt spreaderr\_model}) were also averaged.

Next, we performed an internal cross-match on the source list for each pointing, to identify detections of the same source on different CCDs. This can occur for objects falling near the edge of a CCD due to the dither pattern used to cover the inter-chip gaps on both DECam and Megacam. Data for any multiply-detected sources were combined in another weighted average.

Finally, we cross-matched the $g$- and $i$-band catalogues for each pointing to form a final catalogue containing two-filter photometry for each source. We performed star-galaxy separation using the method described in \citet{koposov:15}, which adopts $|${\tt spread\_model}$| <$ {\tt spreaderr\_model} $+\,\,0.003$ as the threshold for stars \citep[see also][]{desai:12,myeong:17}. We applied this criterion across both filters. 

To create final catalogues for our DECam observations, and for our Megacam observations, we concatenated the stellar catalogues for all pointings, being sure to identify and merge stars observed more than once in the overlap regions between pointings. We kept the DECam and Megacam catalogues separate for our subsequent analysis, since they have different photometric characteristics (e.g., the Megacam data are substantially deeper than the DECam data). However, we note that the median offset between the DECam and Megacam photometry, derived using stars brighter than $23$rd magnitude in overlapping regions of the two mosaics, is smaller than $0.03$\ mag for both the $g$- and $i$-band filters. We accounted for the effects of interstellar dust along the line-of-sight to NGC 6822 using the \citet{schlegel:98} maps and the appropriate reddening coefficients from \citet{schlafly:11}.

\section{Results and Analysis}
\label{s:analysis}
\subsection{Colour-magnitude diagrams}
\label{ss:cmds}

\begin{figure*}
\begin{center}
\includegraphics[trim=0mm 0mm 24mm 0mm, clip, height=87mm]{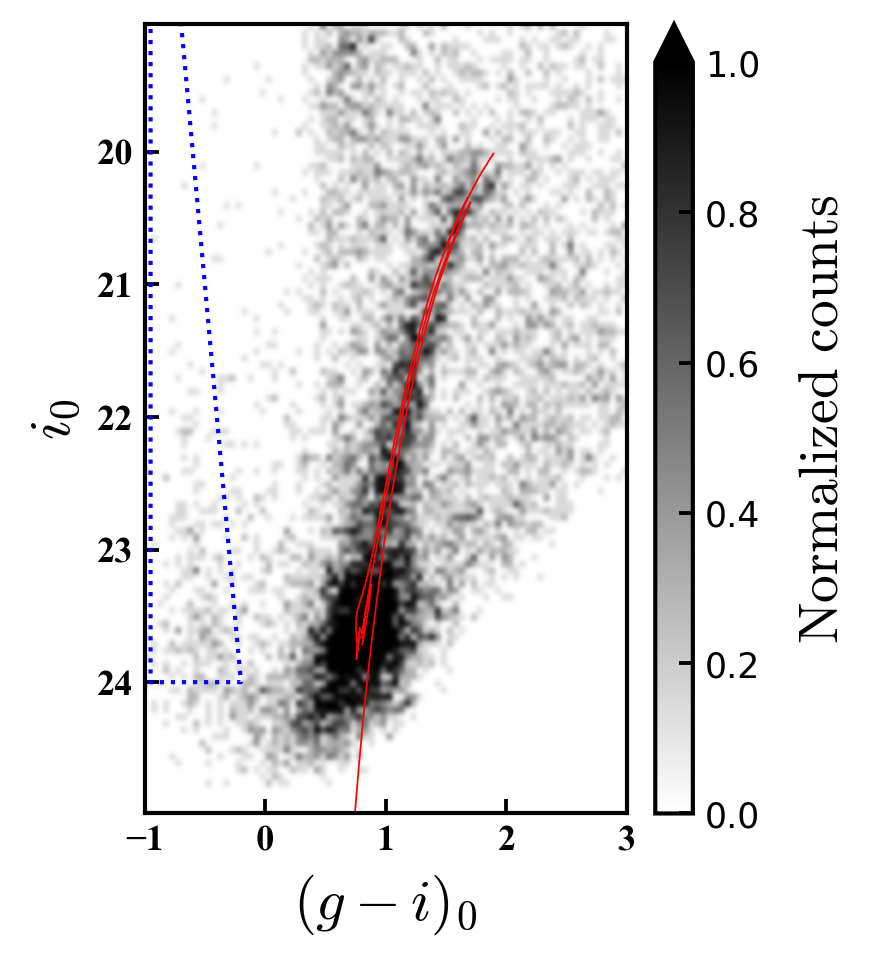}
\hspace{1mm}
\includegraphics[trim=0mm 0mm 24mm 0mm, clip, height=87mm]{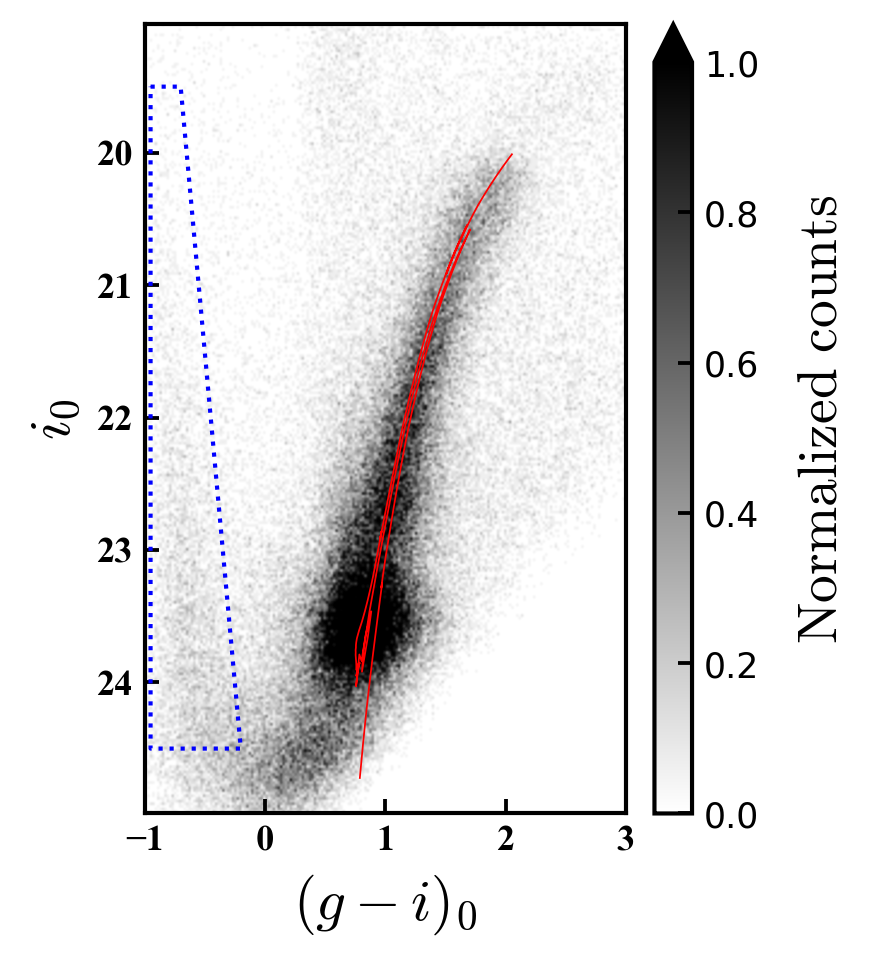}
\hspace{1mm}
\includegraphics[height=86.5mm]{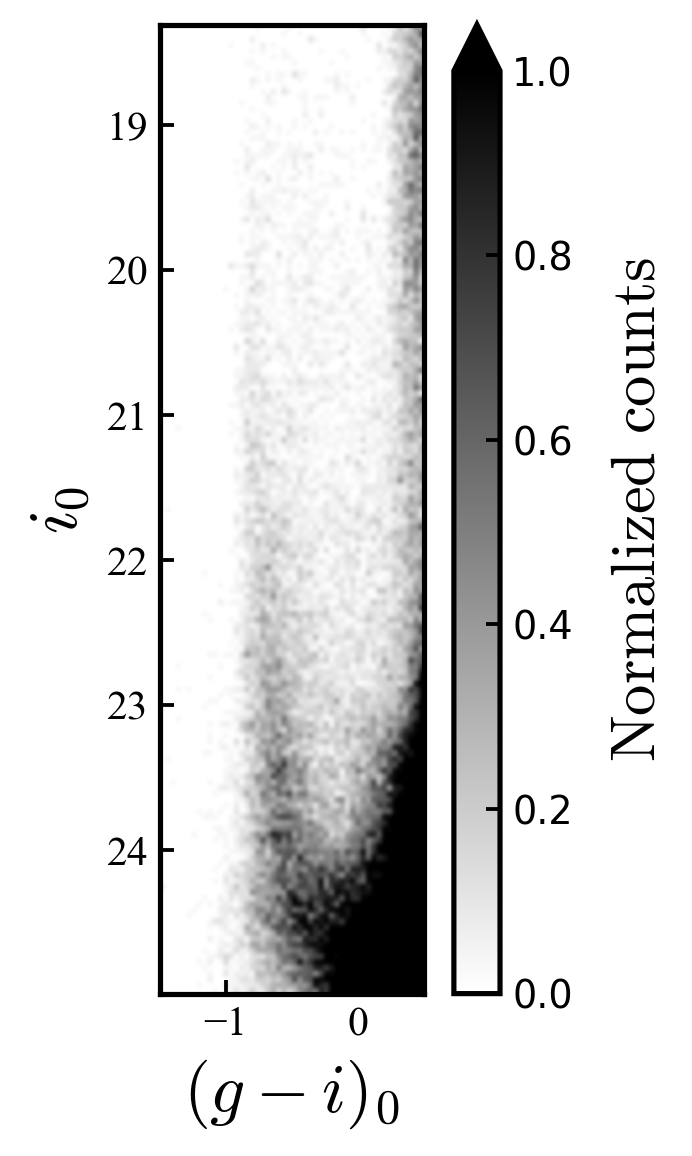}
\end{center}
\caption{Example dereddened Hess diagrams for DECam pointing P4 (left panel) and Megacam pointing CEN (centre and right panels). Only stars within $0.5\degr$ of the centre of NGC 6822 are plotted. In the left and centre panels, the red giant branch and red clump belonging to NGC 6822 are prominent, while young main sequence populations are also visible. The red lines show our fiducial MIST isochrone, which has age $3$\ Gyr and $[$Fe$/$H$] = -1.2$. Our selection boxes for young stars are also marked (blue dotted regions). The right panel shows only the young populations in our Megacam data, with enhanced contrast.}
\label{f:cmds}
\end{figure*}

Figure \ref{f:cmds} shows Hess diagrams for DECam pointing P4, and the central Megacam pointing CEN. Only stars within $0.5\degr$ of the centre of NGC 6822 are plotted. The red giant branch (RGB) and red clump for intermediate-age and old populations in NGC 6822 are prominent, as are much younger populations sitting to the blue. 

The photometry for P4 reaches to the bottom of the red clump ($g_0\sim 24.7$ and $i_0 \sim 24$) and is typical of most of the DECam fields, with only P2 shallower by $\sim 0.4$\ mag due to the somewhat poorer conditions under which the images for that field were taken (see Table \ref{t:log}). The photometry for CEN reaches about $0.5$ mag below the red clump ($g_0\sim 25.2$ and $i_0 \sim 24.5$) and is typical of the majority of the Megacam data. Indeed, most of the Megacam fields have CMDs reaching a few tenths of a magnitude fainter than this, due to the effects of crowding in the central regions of NGC 6822. Overall, our measurements are $\sim 2-3$ mag deeper than the CFHT/MegaCam data presented by \citet{battinelli:06}, and comparable in depth (although spanning a much wider area) to the Subaru/Suprime-Cam photometry described by \citet{deblok:06} \citep[see also][]{komiyama:03}.

Even at the central locations plotted in Figure \ref{f:cmds}, which are dominated by NGC 6822 members, it is clear that the level of contamination by foreground stars is non-negligible. This is due to the fact that NGC 6822 is projected at low Galactic latitude and towards the Galactic centre ($l\sim 25\degr$, $b\sim -18\degr$); these foreground populations make it difficult to trace the structure of the dwarf to low surface brightness. To help minimise the effects of contamination we introduce a simple membership criterion based on the proximity of a star to the red giant branch and/or red clump of NGC 6822. This is similar to the scheme used by \citet{roderick:15,roderick:16a,roderick:16b} to study the outer structure of various Milky Way satellite dwarfs from DECam imaging \citep[see also the globular cluster studies of][]{kuzma:16,kuzma:18}.

We first find an isochrone from the MIST set of models \citep{choi:16} that closely describes the RGB locus.  This is shown in Figure \ref{f:cmds} and has an age of $3$\ Gyr and $[$Fe$/$H$] = -1.2$. While we use this stellar track purely as an empirical description of the RGB, we note that these parameters are similar to measurements of the mean field age and metallicity obtained by many previous studies -- e.g., \citet{davidge:03,battinelli:11,kirby:13,fusco:14,swan:16}.
	
\begin{figure}
\centering
\includegraphics[width=0.7\columnwidth]{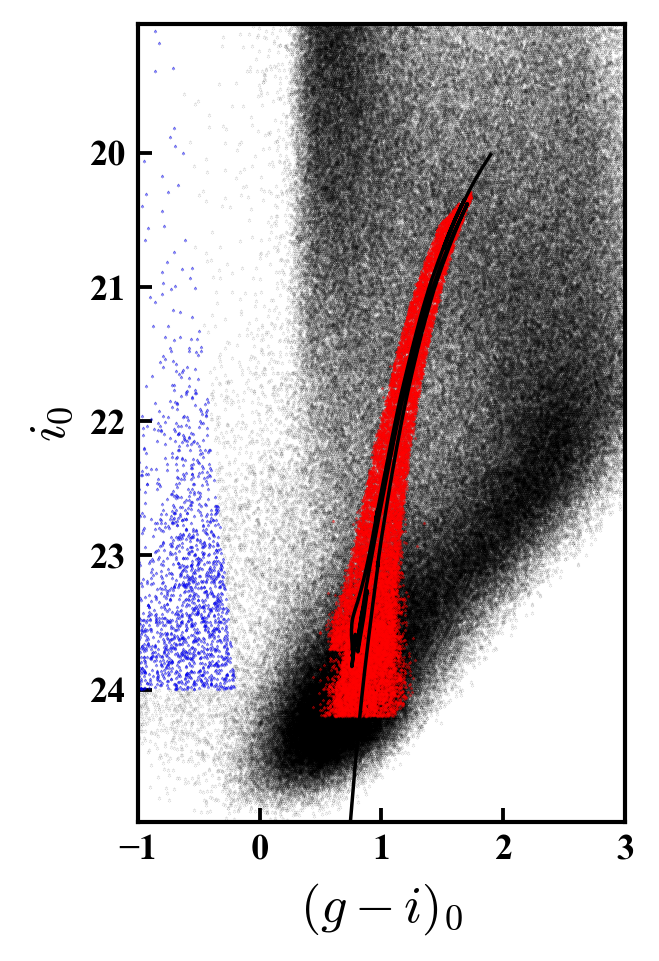}
\caption{Colour-magnitude diagram for the entire DECam field P4, showing red giant and red clump stars classified as NGC 6822 members (red points) according to their proximity to the fiducial isochrone (black line). Specifically, these stars have weight $w\geq 0.5$ as described in Section \ref{ss:cmds}. The faint limit for RGB members, where the histogram of stellar brightnesses begins to turn over, sits at $i_0 \approx 24.2$ for this particular field. Black points show non-members (i.e., with $w<0.5$), while stars falling in the selection box for young populations are marked with blue. This box is dominated by contaminants due to the large area of the field, and the fact it mostly avoids the regions where young populations are found.}
\label{f:weight}
\end{figure}
 
We used this isochrone to calculate a weight value, $w$, for each stellar source detected across the field. The weight quantifies the likelihood that a star is a member of the galaxy, based on its horizontal distance away from the fiducial line relative to the size of the photometric uncertainty. Specifically, we calculated the weight value per star using the Gaussian distribution $N(x; \mu, \sigma)$, where $\mu(i_0)$ represents the colour of the fiducial line; $\sigma(i_0)$ is the photometric uncertainty in colour as a function of magnitude, derived from the distribution of individual photometric errors produced by {\sc sextractor}; and $x=\Delta(g-i)_0$ is the horizontal distance between the star and the isochrone. The calculation is normalized to have $w=1.0$ on the fiducial line. It is then possible to tune the selection of NGC 6822 stars by varying the threshold weight above which stars are considered members. 

An example of this weighting scheme for field P4 is shown in Figure \ref{f:weight} where we have classified as members all stars with $w\geq 0.5$ (red points), and non-members all stars with $w<0.5$ (black points). Placing the threshold at $w=0.5$ is somewhat stricter than used by e.g., \citet{roderick:16a,roderick:16b}, but is mandated by the heavy foreground contamination around the dwarf. At this stage we set the faint limit for each field by determining the approximate magnitude at which the histogram of stellar brightnesses begins to turn over, at colours matching the RGB and red clump of NGC 6822. Around the faint limit, a sizeable fraction of our "stellar" sources are in fact unresolved background galaxies which we need to account for later in the analysis.

We also compiled a sample of young main-sequence stars using the selection regions shown in Figure \ref{f:cmds}. At these blue colours the density of foreground contamination is greatly reduced such that it is sufficient simply to draw a box around the region occupied by the young stellar populations in NGC 6822. An example of the young selection in P4 is also plotted in Figure \ref{f:weight} (blue points). When selecting young stars we set a slightly brighter faint limit than for the older populations, to avoid sampling the red clump and the main region of contamination by background galaxies.
  
\subsection{Spatial distribution of stellar populations}
\label{ss:maps}
We use our colour-magnitude selections to explore the distribution of different stellar populations within NGC 6822. 

\subsubsection{Young populations}
Figure \ref{f:hi} shows the spatial distribution of young stars from our Megacam catalogue, relative to the integrated H{\sc i} column density map constructed by \citet{deblok:00} \citep[see also][]{weldrake:03,deblok:06}. While a number of previous studies have convincingly demonstrated that the youngest populations in NGC 6822 (with ages $<100$\ Myr) closely trace the H{\sc i}, especially the regions where the gas is densest \citep[e.g.,][]{komiyama:03,deblok:03,deblok:06}, our Megacam photometry is up to $\sim 1$\ mag deeper on the blue side of the CMD than previous data, providing more robust statistics.

The results shown in Figure \ref{f:hi} strongly support the idea that the young populations in NGC 6822 are closely confined to regions of high H{\sc i} density. These include the central disk (inside $\sim 0.25\degr$), especially to the south-east near the edge of the giant H{\sc i} hole identified by \citet{deblok:00}, and in the far north-west near the H{\sc i} "cloud" \citep[also seen by][]{deblok:00}. The apparent dearth of young stars near the very centre of NGC 6822 is due to crowding in our Megacam images. However, there is also a very clear under-density of young stars coincident with the H{\sc i} hole which is not due to crowding. This has previously been noted by other authors \citep[e.g.,][]{komiyama:03,deblok:06}, but shows up strongly in our data; \citet{cannon:12} used a Hubble Space Telescope pointing to show that there has been only low-level star formation at this location during the past $\approx 500$\ Myr.

Away from the gas, very few blue sources are seen. Spatially, they are distributed approximately uniformly and do not follow the luminosity function of the young stars towards the centre of NGC 6822 (i.e., they are not increasingly distributed towards fainter magnitudes). We conclude that these sources constitute a low-density contaminating population, most likely blue horizontal branch stars in the halo or thick disk of the Milky Way (although background quasars may also contribute). Similarly, our DECam photometry reveals no unexpected concentrations of blue stars at larger radii, although we note that the H{\sc i} data from \citet{deblok:00} do not extend much beyond the region shown in Figure \ref{f:hi}. 

\subsubsection{Intermediate-age and old populations}
We now consider the spatial distribution of intermediate-age and old populations in NGC 6822, as traced by the RGB and red clump on the CMD. For these stars, contamination is still an important issue -- despite our weighting scheme, there are clearly still underlying contaminants, especially unresolved background galaxies, that happen to sit coincident with the region of interest (see Figure \ref{f:weight}).

\begin{figure}
\centering
\includegraphics[width=0.99\columnwidth]{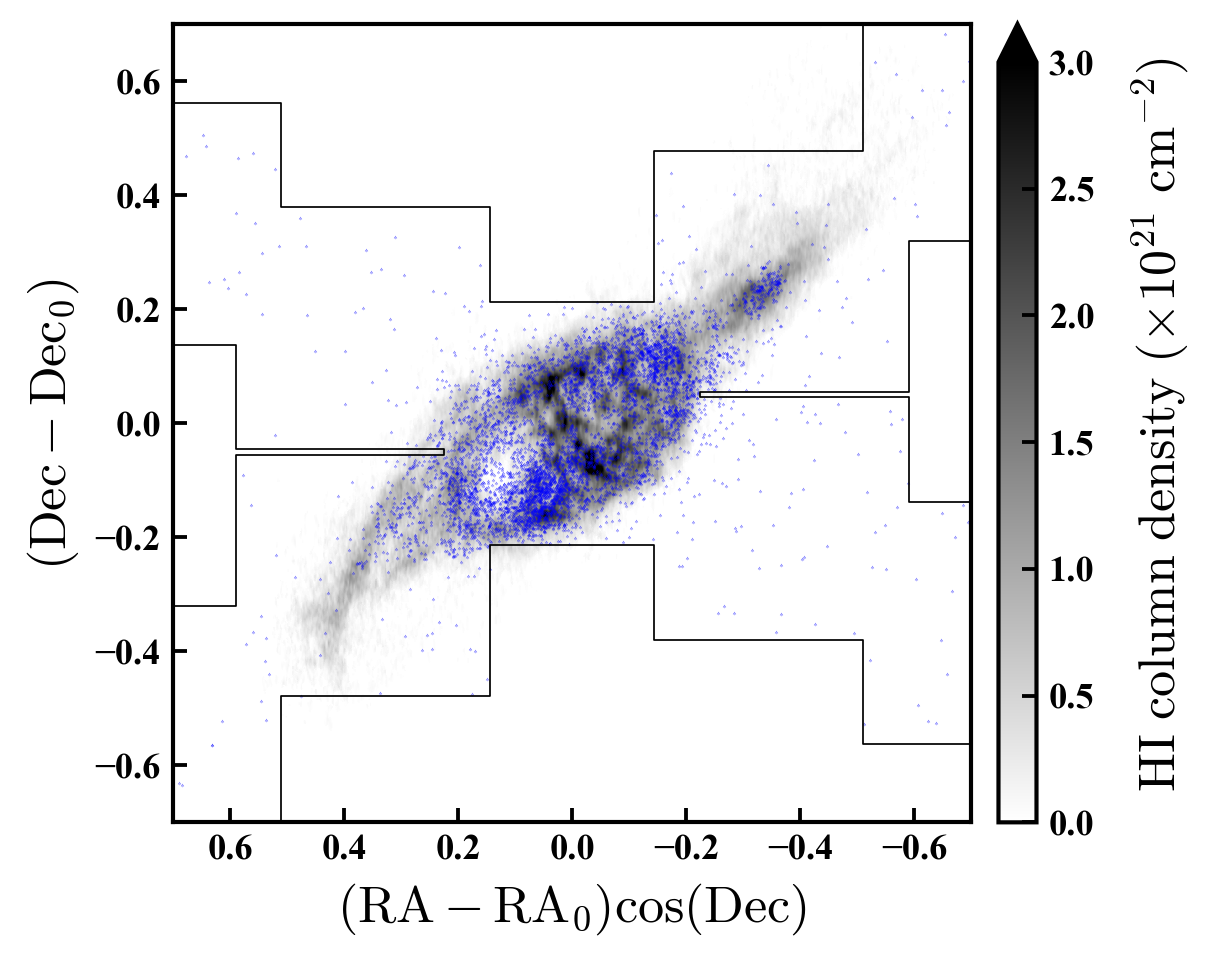}
\caption{Spatial distribution of young stars (blue points) selected from our Megacam CMDs using the box shown in the left-hand panel of Figure \ref{f:cmds}, relative to the H{\sc i} column density map from \citet{deblok:00} \citep[see also][]{weldrake:03,deblok:06}. North is up and east to the left. This map, and all subsequent maps, use a galaxy-centric coordinate system as indicated. The black outline shows the edges of our Megacam mosaic. As seen by a number of previous studies, the young stars closely trace the distribution of H{\sc i} in NGC 6822.}
\label{f:hi}
\end{figure}

We account for this residual contamination by directly correcting the spatial density map of RGB and red clump stars. We first divide the area around NGC 6822 into a grid of bins. Some experimentation revealed an optimal bin size of $12 \times 12$ arcsec, which strikes a balance between too-fine resolution (which is noisy) and too-coarse resolution (which loses detail). We then divided all RGB and red clump stars classified as "members" into these bins to create a 2D density map. At this stage, we adjust our faint limit to be that of the worst field for the DECam and Megacam maps, respectively, to avoid artifical changes in density across each mosaic.  While the Megacam maps all have a similar depth such that the overall faint limit is very similar to that in any given field, the DECam data are limited to the depth of P2 (which is $\sim 0.4$\ mag shallower than the other three pointings, as described in Section \ref{ss:cmds}). 

Next, we created a list of sources with weight $0.1 < w < 0.35$, and used this to make a second 2D density map, this time of the contaminating populations.  In defining the appropriate weight limits for this procedure we were careful to try and select, as far as possible, a sample with a similar colour-magnitude distribution to the "member" sample.  The contaminating sources selected using these limits define a region surrounding the member sample on the CMD. The upper limit of $w=0.35$ maintains a small but clear separation on the CMD between the two groups, to minimise the number of genuine members scattered into the contamination region by photometric uncertainty. The lower limit excludes populations that sit far from the member region on the CMD, which clearly have a different colour-magnitude distribution. In defining our contamination sample, we also applied the same faint limits as defined above for the two mosaics.

Following the procedure outlined below \citep[as introduced in e.g.,][]{roderick:15,roderick:16a,roderick:16b,kuzma:16,kuzma:18} we used the contamination map to correct the map of NGC 6822 members:

\begin{enumerate}
\item{We first make a version of the contamination map that is scaled so that the mean bin density is $1.0$.}
\item{Then we divide the density map for members by this normalized contamination map. This "flat-fielding" process accounts for large-scale spatial variations across the field of view.}
\item{Next we create another version of the contamination map that is scaled so that the mean bin density in the outskirts (at large radii where there is no evidence of NGC 6822 populations) is equal to the mean bin density in the outskirts of the flat-fielded member map from the previous step. Although \citet{battinelli:06} traced the NGC 6822 spheroid to a radius of $\sim 40\arcmin$, our radial surface density profile (presented in Section \ref{ss:profile} below) shows that we must choose stars outside $\sim 60-80\arcmin$ for this step (depending on whether we are using our Megacam or DECam catalogue).}
\item{Finally, we subtract this scaled contamination map from the flat-fielded member map, to create a new density map where the mean bin density at large radii is close to zero, and the contamination background has been removed. Note that this step implicitly assumes that the contamination is distributed approximately uniformly across the field of view. This is a good approximation for the Megacam mosiac, and acceptable for the much larger DECam mosaic.}
\end{enumerate}

\begin{figure*}
\centering              
\includegraphics[width=\columnwidth]{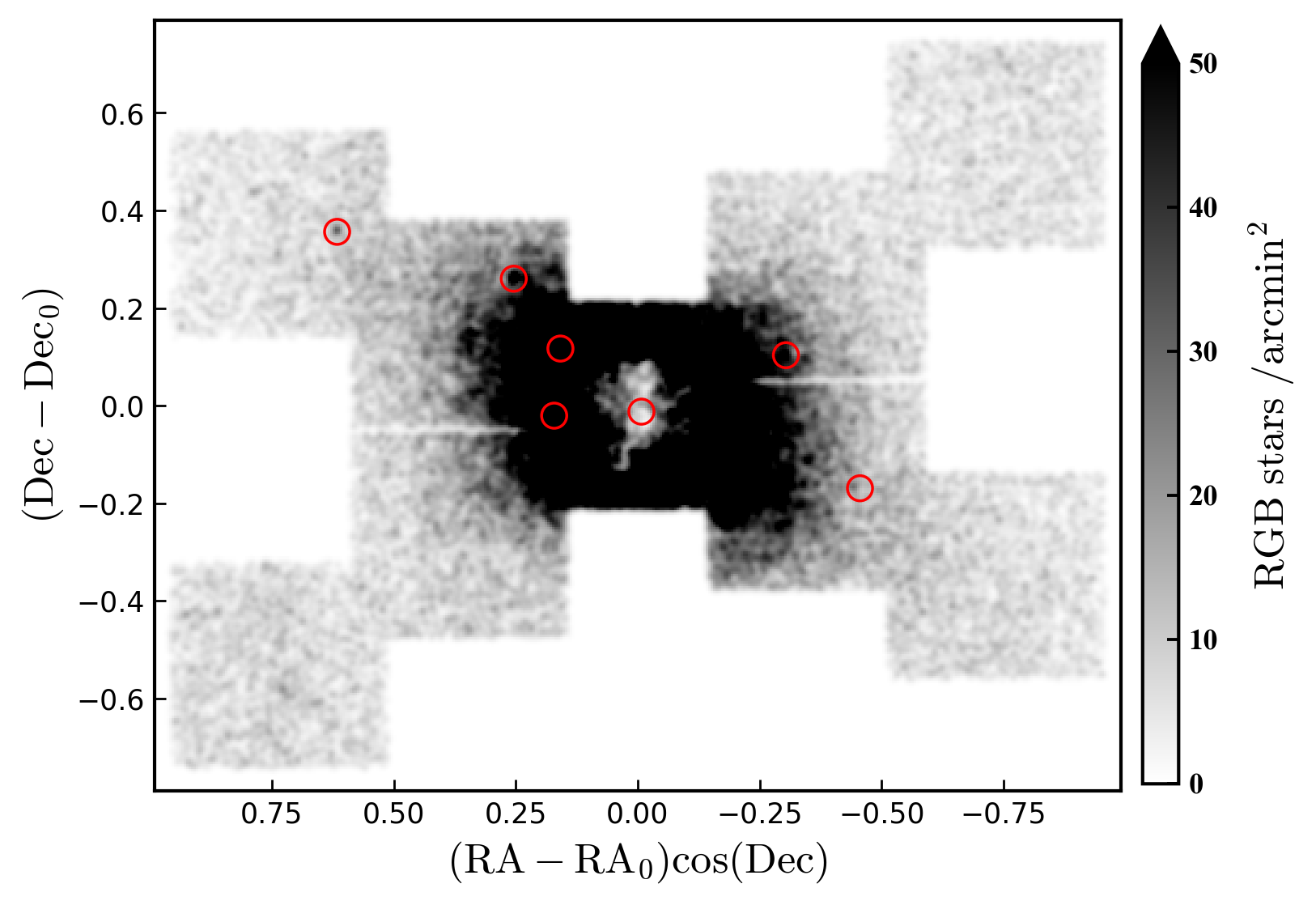}
\hspace{1mm}
\includegraphics[width=1.05\columnwidth]{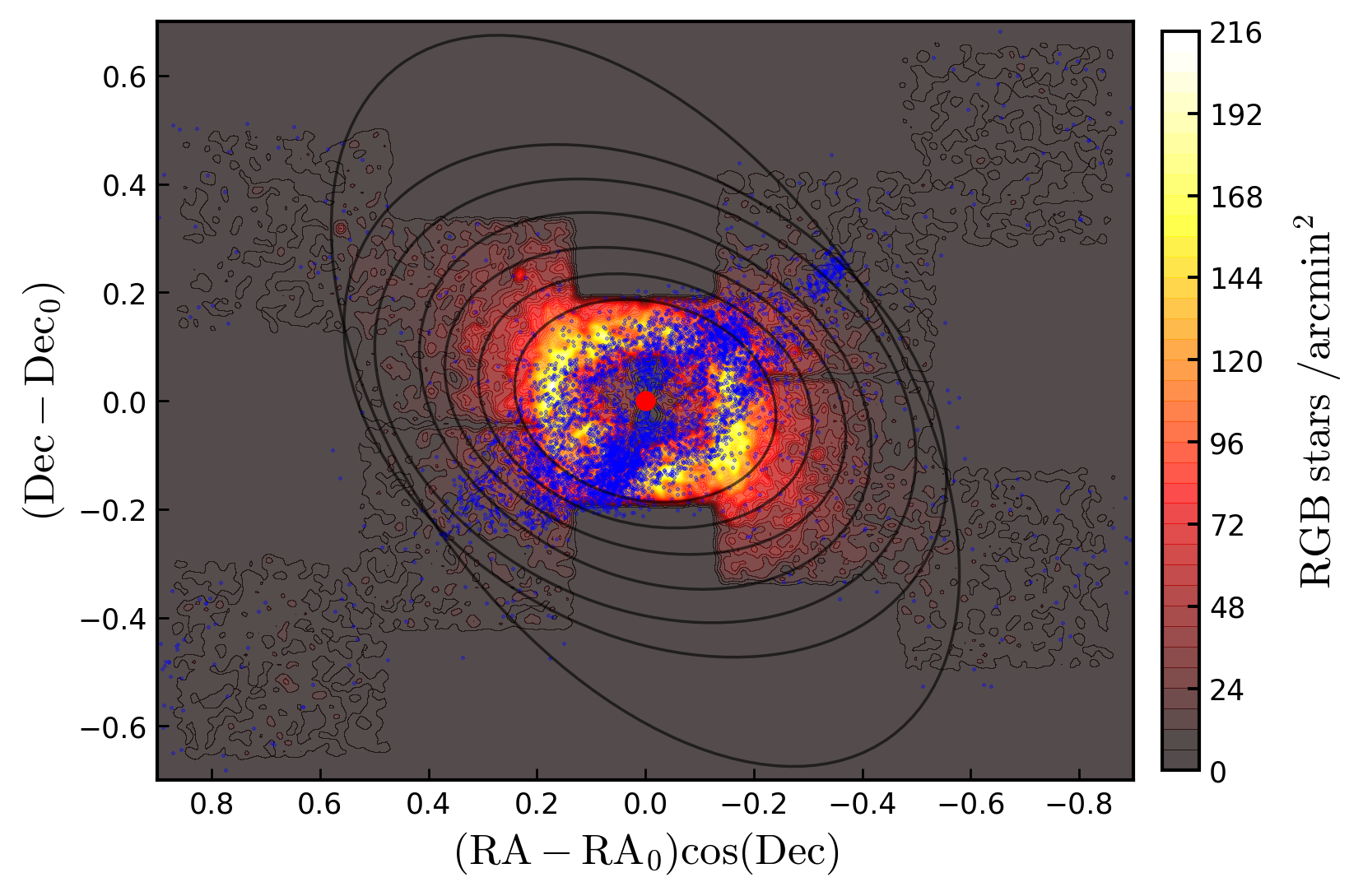}
\caption{A binned density map (left panel) and contoured density map (right panel) for intermediate-age and old stars in NGC 6822 from our Megacam data. North is up and east to the left. The red circles in the left panel show the location of seven of the eight known globular clusters in the NGC 6822 system (one lies beyond the edge of the map), at least three of which are clearly visible in the data. The blue points in the right panel show the previously-defined young star selection. The ellipses in the right panel show results from the isodensity contour-fitting process described in Section \ref{ss:spheroid}.}
\label{f:megamap}
\end{figure*}

\begin{figure*}
\centering
\includegraphics[width=\columnwidth]{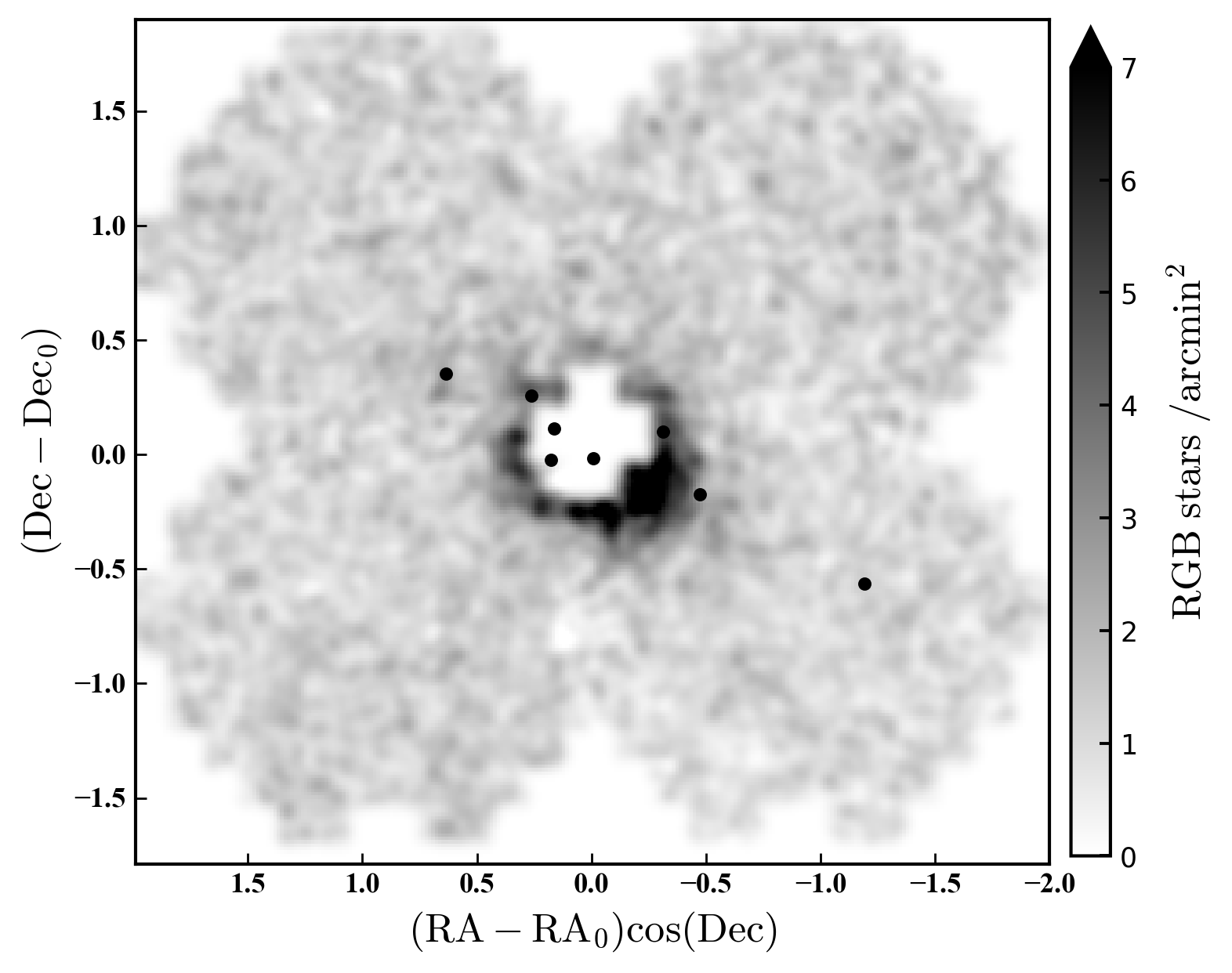}
\hspace{1mm}
\includegraphics[width=1.05\columnwidth]{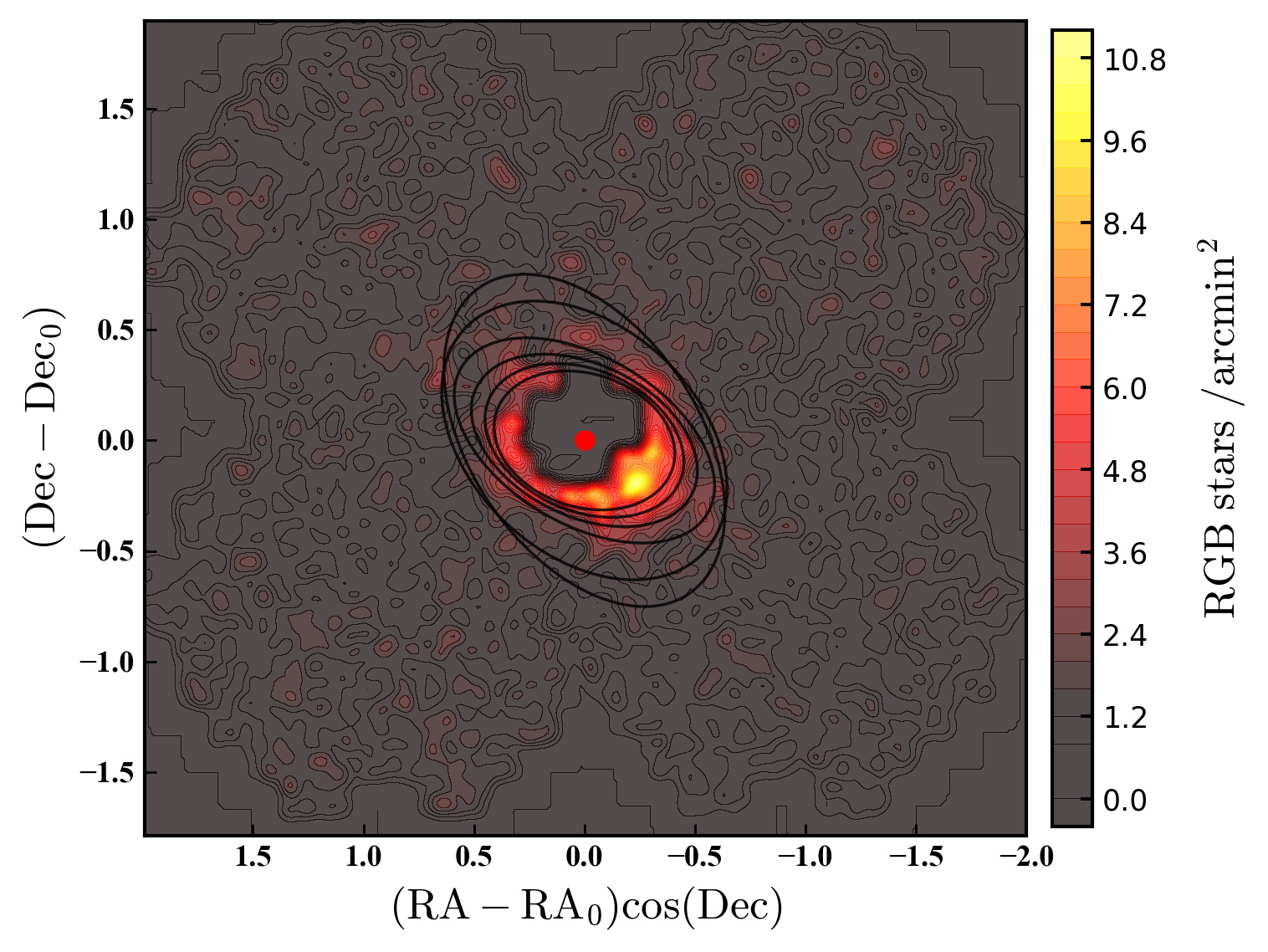}
\caption{A binned density map (left panel) and contoured density map (right panel) for intermediate-age and old stars in NGC 6822 from our DECam data. North is up and east to the left. The black circles in the left panel show the location of the eight known globular clusters in the NGC 6822 system, but these do not show up in the map. The ellipses in the right panel show results from the isodensity contour fitting described in Section \ref{ss:spheroid}.}
\label{f:decmap}
\end{figure*}

The resulting contamination-subtracted density maps for intermediate-age and old stars in NGC 6822, as traced by the red giant branch and the red clump, are shown in Figures \ref{f:megamap} and \ref{f:decmap} for the Megacam data and DECam data, respectively. These maps have had a small degree of spatial smoothing applied, using a Gaussian kernel of width $2.5$\ bins (i.e., $30\arcsec$). In each set of plots we show the "raw" smoothed density map and a contoured version.

These maps clearly show the extended elliptical spheroid of NGC 6822. This structure can easily be traced by eye to at least $\sim 0.6$ degrees ($\sim 5$\ kpc) along its major axis, consistent with the results of \citet{battinelli:06} who were able to follow the spheroid to $\approx 40\arcmin = 0.67$\ deg. Because our Megacam data were taken under better conditions than our DECam data, the faint detection limit of the Megacam photometry is fainter and the Megacam map has overall higher signal-to-noise than the DECam map (although is limited by the extent of the mosaic). Nonetheless, the two independently-created maps appear consistent, and this is confirmed quantitatively in the following sections.

The DECam map appears to show an overdensity of spheroid stars to the south-west near coordinates $(-0.2,\,-0.2)$, but with no counterpart present to the north-east. Rather than representing a significant asymmetry in the spheroid we believe this is partly due to the slight mis-centering of the DECam mosiac relative to the true centre of NGC 6822. This mis-centering means that the map probes to smaller radii (higher stellar density) in the south-west than in the north-east, giving the appearance of an asymmetry. The somewhat lower image quality for pointing P2 (to the north-east) also plays a significant role, as these inner regions constitute the most crowded parts of the DECam mosaic. We note that no obvious asymmetries are present in the Megacam map at comparable radii.

NGC 6822 has eight known globular clusters \citep[e.g.,][]{hwang:11,huxor:13,veljanoski:15}, which we plot on our maps. In the Megacam map (Figure \ref{f:megamap}), where the seeing was very good, some of the richer clusters are partially resolved into stars and show up as clear over-densities. However this is not the case for the slightly lower resolution DECam images. A by-eye examination of all our stacked images revealed no obvious undiscovered clusters.

On the contoured Megacam map we also plot the young star selection described above (see Figure \ref{f:hi}). The axis of young stars (and the H{\sc i} that they follow) has a position angle of $\approx 125\degr$, oriented at $\sim 60\degr$ compared to the major axis of the spheroid traced by RGB and red clump stars (see Section \ref{ss:spheroid}). This confirms the results of numerous earlier studies. However, our data very clearly show that over-densities of young stars are completely uncorrelated with the older populations. For example, the strong clump of young stars sitting on the H{\sc i} "cloud" to the north-west shows no coincident over-density or clump of older populations. This is consistent with the {\it Hubble Space Telescope} observations of \citet{cannon:12}.

As described above, the depth of the DECam map is limited by pointing P2 which has lower image quality than the other three pointings. In applying the faint limit of P2 to the other portions of the mosaic, we inevitably lose some of the information available at these other locations. We therefore constructed similar spatial density maps for each of the four pointings individually, but applying their appropriate respective faint limits, to check that we were not missing any low surface brightness substructures. We found that indeed we did not miss any such structures by applying the faint limit of P2 to our overall map; the primary difference between the individual DECam maps and the mosaic DECam map is the possibility of a very low significance extension of the spheroid along the major axis in pointing P3, which approaches the most distant known globular cluster in the system \citep[SC1,][]{hwang:11}.

We conclude this section by emphasising that all the maps presented above clearly show that apart from the spheroid, which appears mostly very smooth, there are no significant stellar streams or over-densities visible in the region around NGC 6822 out to a maximum projected radius of $\approx 2$\ degrees ($\sim 16.5$\ kpc). In Section \ref{ss:profile} we show that our limiting surface-density for detecting the spheroid is a  few ($\sim 1-3$) red giant stars per square arcminute. This corresponds, very approximately, to a faint surface brightness limit of $V\sim 30$\ mag$\,$arcsec$^{-2}$. 

\subsection{Structure of the NGC 6822 spheroid}
\label{ss:spheroid}
Having mapped the spheroid of NGC 6822 we would like to quantify its properties. In this section we consider the ellipticity and orientation of the spheroid as a function of radius, and in the next section we consider its radial surface density profile.

To measure the shape and orientation of the spheroid as a function of radius, we developed a method of fitting ellipses to isodensity contours. We first smooth the mosaic density maps presented above with a much wider Gaussian kernel to suppress fluctuations, then determine the $(x,y)$ loci of points that define isodensity contours at varying density levels, and fit ellipses to these sets to quantify their shape (ellipticity) and orientation (major-axis position angle). The lower the density level of a contour, the larger the projected radius of the measurement. 

\begin{figure}
\centering
\includegraphics[width=0.95\columnwidth]{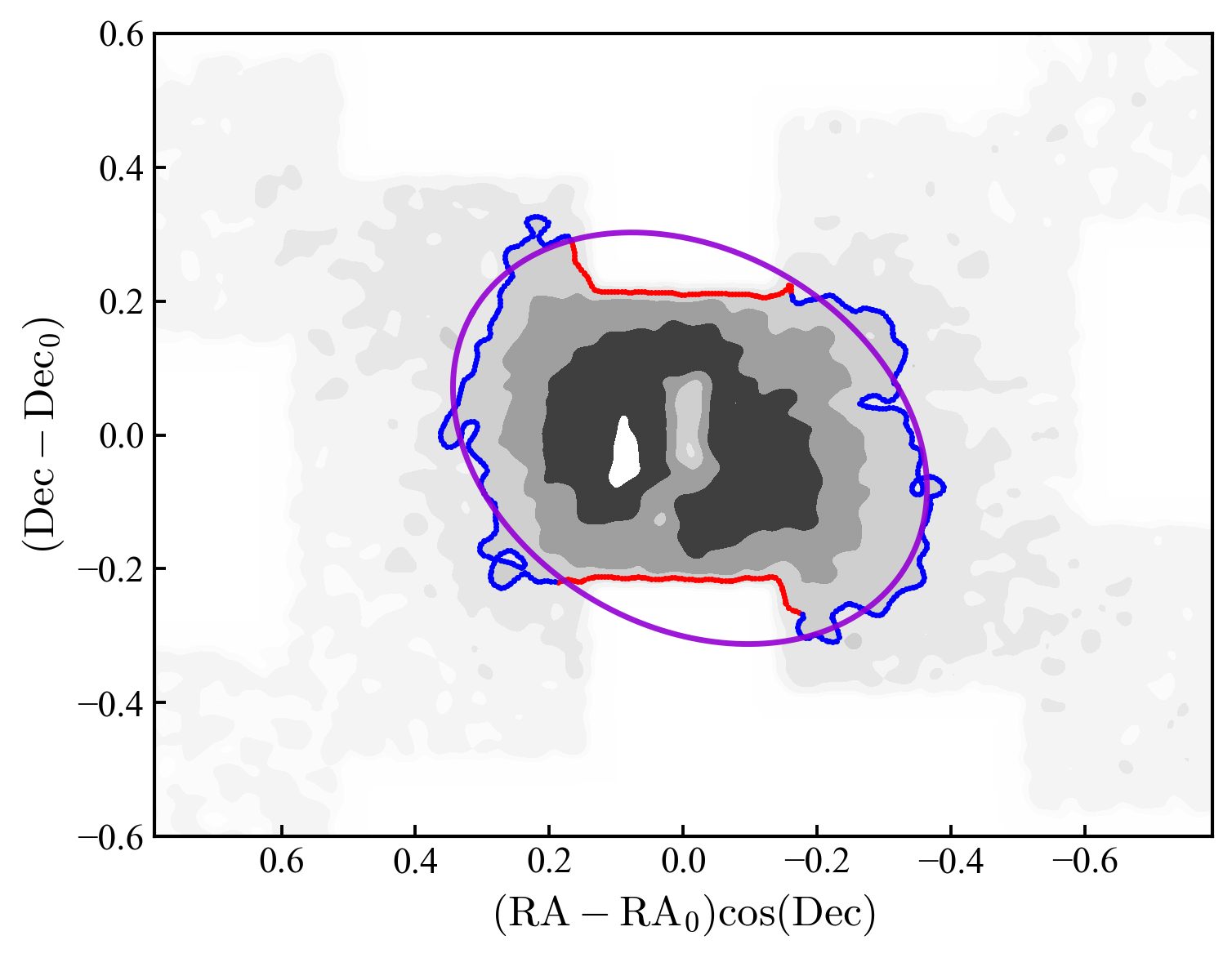}
\caption{Example of fitting an ellipse to an isodensity contour on our Megacam density map. From darkest to lightest, the grey contours trace density levels of $40,\,20,\,10,\,5,$ and $2$ stars per arcmin$^{2}$. The blue line traces the contour with a density of $10$ stars per arcmin$^{2}$. Directly above and below the centre of NGC 6822, this contour intersects with the edge of the Megacam footprint, and we have masked these regions (red) before fitting an ellipse to the contour. The purple line shows the best-fit ellipse.}
\label{f:ellipse}
\end{figure}

We show an example of this process in Figure \ref{f:ellipse} where we trace the locus of the contour corresponding to $10$ stars arcmin$^{-2}$ in our Megacam map. This contour is well defined but intersects with the edge of the mosaic above and below the centre of the galaxy. We mask these segments and then fit an ellipse to the remaining points, of the form $\alpha x^2 + \beta xy + \gamma y^2 = 1$, using a least-squares function in Python. The following equations convert the best-fit coefficients into the effective major- and minor-axis radii ($a$ and $b$, respectively), position angle ($\theta$), and ellipticity ($e$):
\begin{eqnarray}
a & = & -\frac{\left(2(4\alpha\gamma - \beta^2)(\alpha+\gamma) + \sqrt{((\alpha-\gamma)^2 + \beta^2)}\right)^{0.5}}{(\beta^2-4\alpha\gamma)}\\
b & = & -\frac{\left(2(4\alpha\gamma - \beta^2)(\alpha+\gamma) - \sqrt{((\alpha-\gamma)^2 + \beta^2)}\right)^{0.5}}{(\beta^2-4\alpha\gamma)}\\
\theta & = & 90 - \arctan\left(\frac{\gamma-\alpha-\sqrt{((\alpha-\gamma)^2 + \beta^2)}}{\beta}\right)\\
e & = & 1-\frac{b}{a}   
\end{eqnarray}

This method assumes that the centre of NGC 6822 is fixed. It is possible to allow the central position to vary freely by increasing the number of fit parameters to five; however (i) in general we found such fits to be ill-constrained, partially due to degeneracies in the parameters and partially due to the lack of central data (due to the DECam mosaic shape, and Megacam crowding), and (ii) we did not observe any compelling evidence to justify this additional complexity (e.g., by observing mismatching fits on either side of the galaxy for a given isodensity contour). 

To determine uncertainties for a given least-squares fit we computed the covariance matrix, and then used this to randomly generate $10^4$ normally-distributed (and appropriately-correlated) samples for the parameters $(\alpha,\,\beta,\,\gamma)$. We propagated these through the above equations to obtain the corresponding distributions on $(\theta,\,e)$, which we used to calculate the uncertainties on these quantities.

\begin{figure}
\centering
\includegraphics[width=0.95\columnwidth]{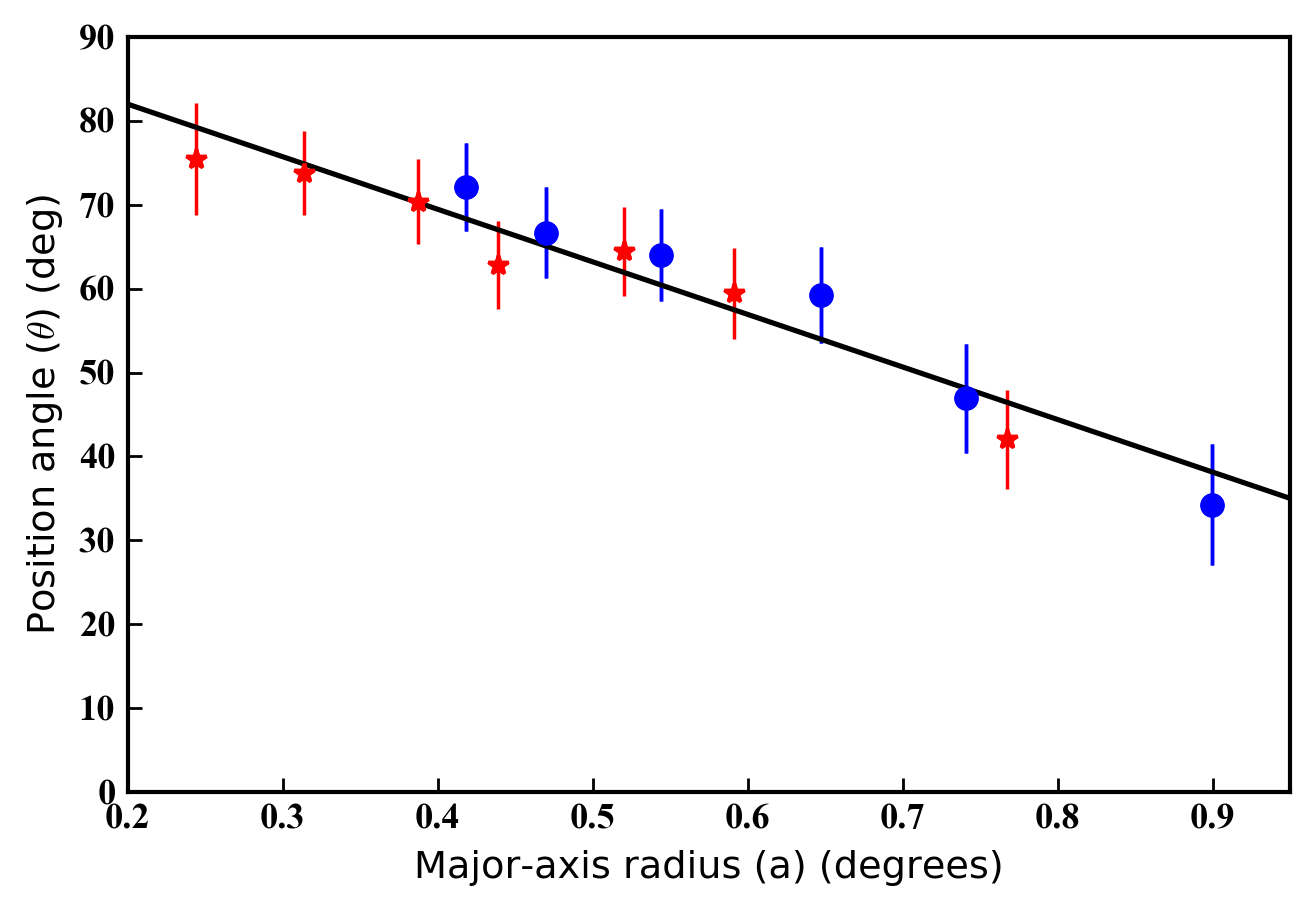}\\
\vspace{0.5mm}
\includegraphics[width=0.95\columnwidth]{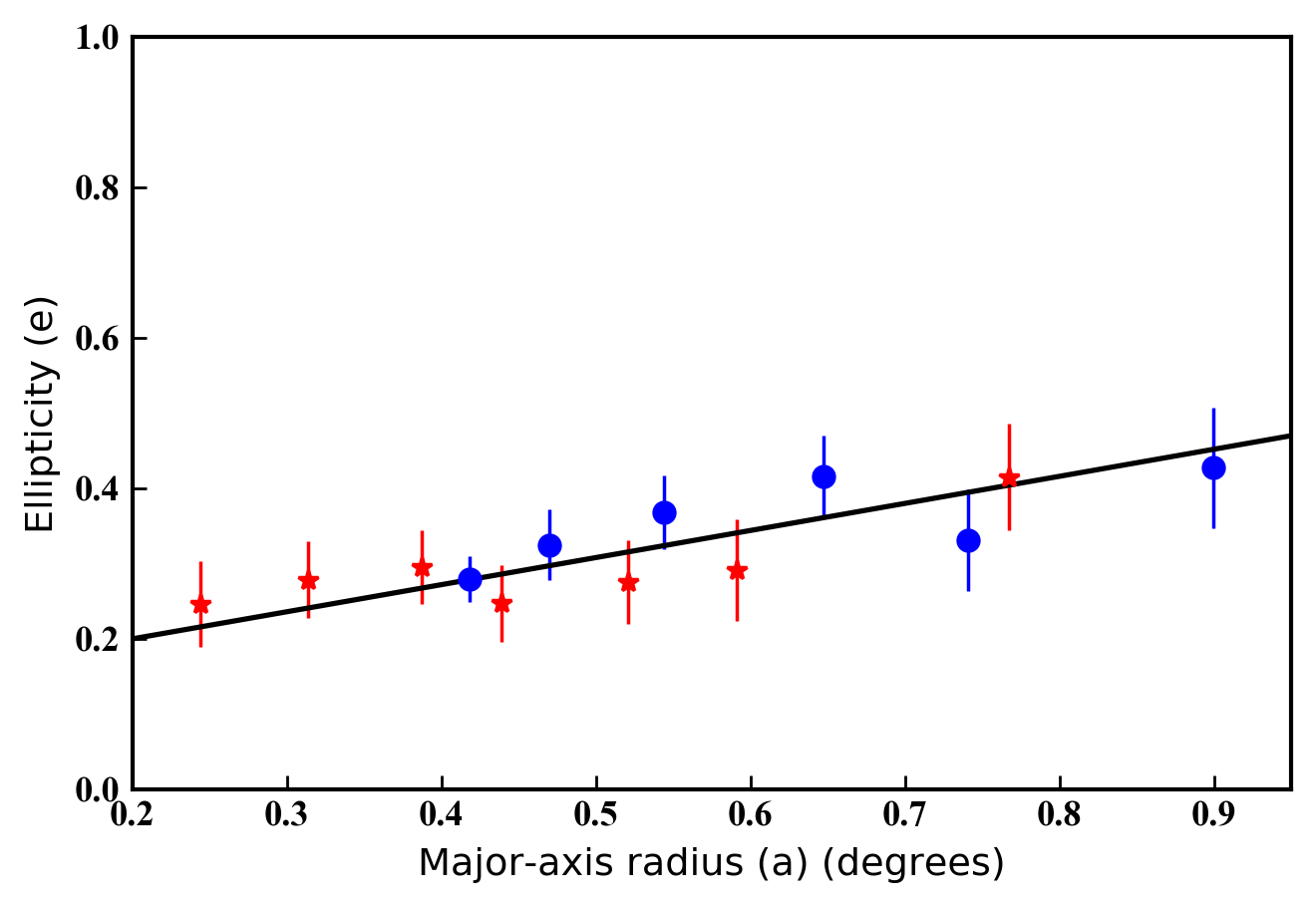}
\caption{Variation of the position angle of the major-axis (upper panel), and the ellipticity (lower panel) for the spheroid of NGC 6822, from our method of fitting ellipses to different isodensity contours. In both panels the $x$-axis shows the major-axis radius ($a$) in units of degrees. The position angle is reported in degrees east of north. In both panels the red stars stand for Megacam data and the blue dots represent DECam data. The black lines show linear fits to the observed trends, as discussed in the text.}
\label{f:ellip}
\end{figure}

We show the measured variation in the ellipticity and major-axis position angle of the spheroid, as a function of major-axis radius, in Figure \ref{f:ellip}. There is generally close agreement between the independent Megacam and DECam measurements. For both data sets we are able to trace the contours to large radii approaching one degree ($8.2$\ kpc) from the centre of NGC 6822. Beyond this the contours become too noisy, and the azimuthal coverage of the Megacam mosaic too limited, to enable reliable fits. The uncertainties on both position angle and ellipticity gradually increase with radius, reflecting the commensurate decrease in the signal-to-noise of the star-count data.

Our results show that both the ellipticity and the orientation of the NGC 6822 spheroid change with radius. The ellipticity increases from around $0.20$ in the inner parts of the spheroid, to $\sim 0.45$ in the outer parts. The position angle of the major axis decreases from around $80$\ degrees east of north in the inner parts of the spheroid, to $\sim 40$ degrees in the outer parts. Both parameters appear to vary approximately linearly with radius; the best-fit lines shown in Figure \ref{f:ellip} have slopes of $\frac{d\theta}{da} \approx -62.6\degr$ per degree, and $\frac{de}{da} \approx 0.36$ per degree, respectively. At intermediate radii ($\sim 0.5-0.6$\ deg) our results match well those of \citet{battinelli:06} who found $e\approx 0.36$ and $\theta\approx 65\degr$ in this region.

Overall, our measurements show that the spheroid of NGC 6822 is becoming flatter, and its orientation is twisting, at increasing distance from the centre of the galaxy. We illustrate this by plotting the two sets of ellipse solutions on top of the Megacam and DECam contour maps in Figures \ref{f:megamap} and \ref{f:decmap}.

\subsection{Radial density profiles}
\label{ss:profile}
We next construct radial profiles to measure how the surface density of the NGC 6822 spheroid declines with galactocentric distance. First, we split our Megacam and DECam member catalogues (i.e., the stars with weight $w>0.5$) into concentric elliptical annuli about the galaxy centre, by calculating the effective major-axis radius for each star. In doing this, we assume the average position angle and ellipticity measured in the previous section ($\theta= 62\degr$ and $e= 0.35$). For each annulus we calculate the stellar density by computing the area of the annulus and accounting for regions outside the survey (this is particularly important for the irregular Megacam mosaic). 

We next determine the mean background level from a wide region at very large radius and subtract this from the relevant profile to account for the residual contamination in our member lists. The outermost points in the density profiles are most sensitive to this subtraction, and uncertainty in the computed background propagates into a systematic error in the measured density decline of the NGC 6822 spheroid at large radius (discussed below). At small radii our profiles are affected by severe crowding (Megacam), or missing imaging (DECam). The innermost annulus for our DECam profile sits at a radius of $\approx 27\arcmin$, because the DECam mosaic does not cover the central portion of NGC 6822. While we are able to sample to smaller radii in our Megacam data, where the innermost data point sits at $\approx 9\arcmin$, inspection of the density maps in Figure \ref{f:megamap} shows that the region inside $\sim 6-10\arcmin$ (depending on direction) is clearly affected by incompleteness due to crowding.

\begin{figure}
\centering
\includegraphics[width=0.95\columnwidth]{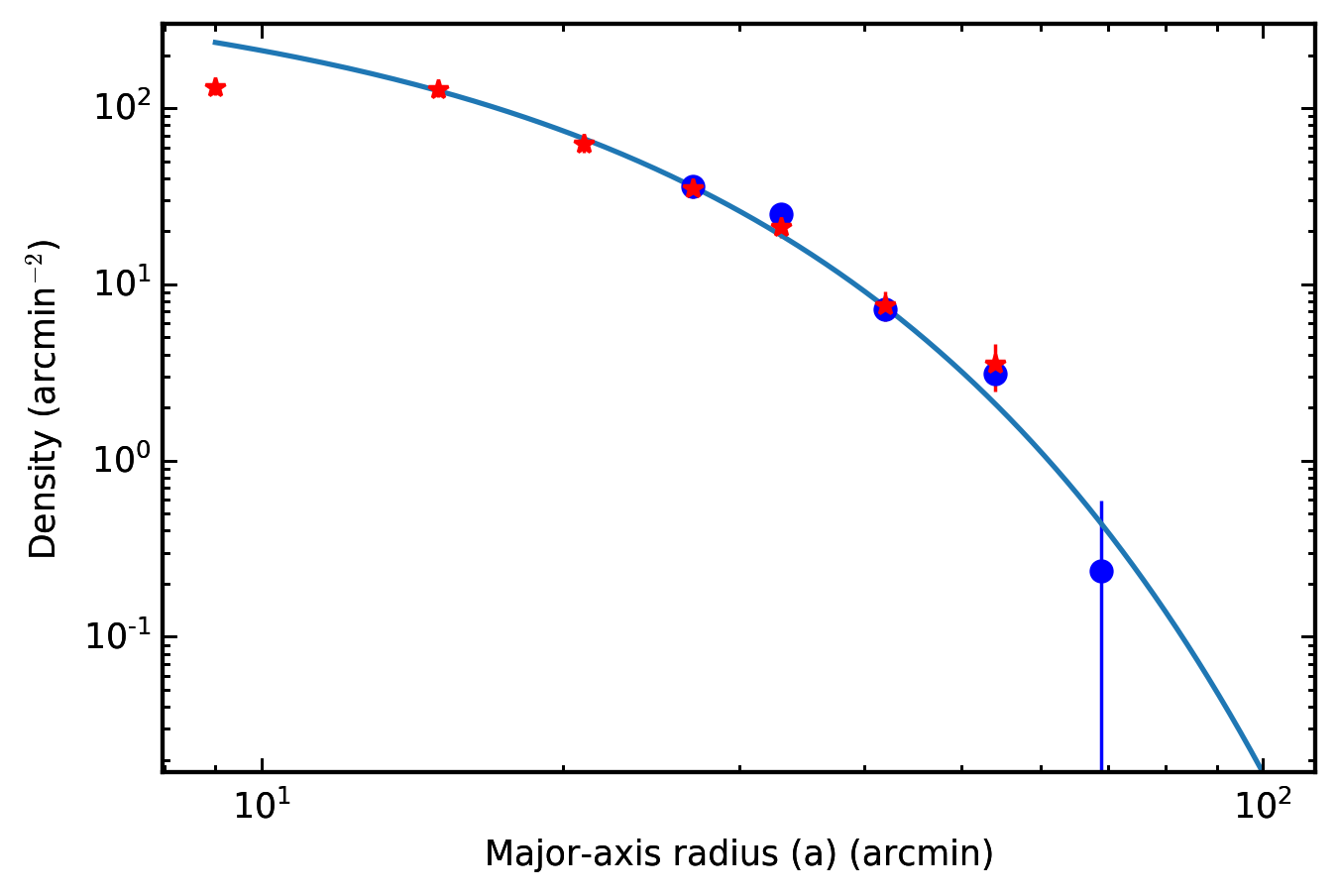}\\
\vspace{0.5mm}
\includegraphics[width=0.95\columnwidth]{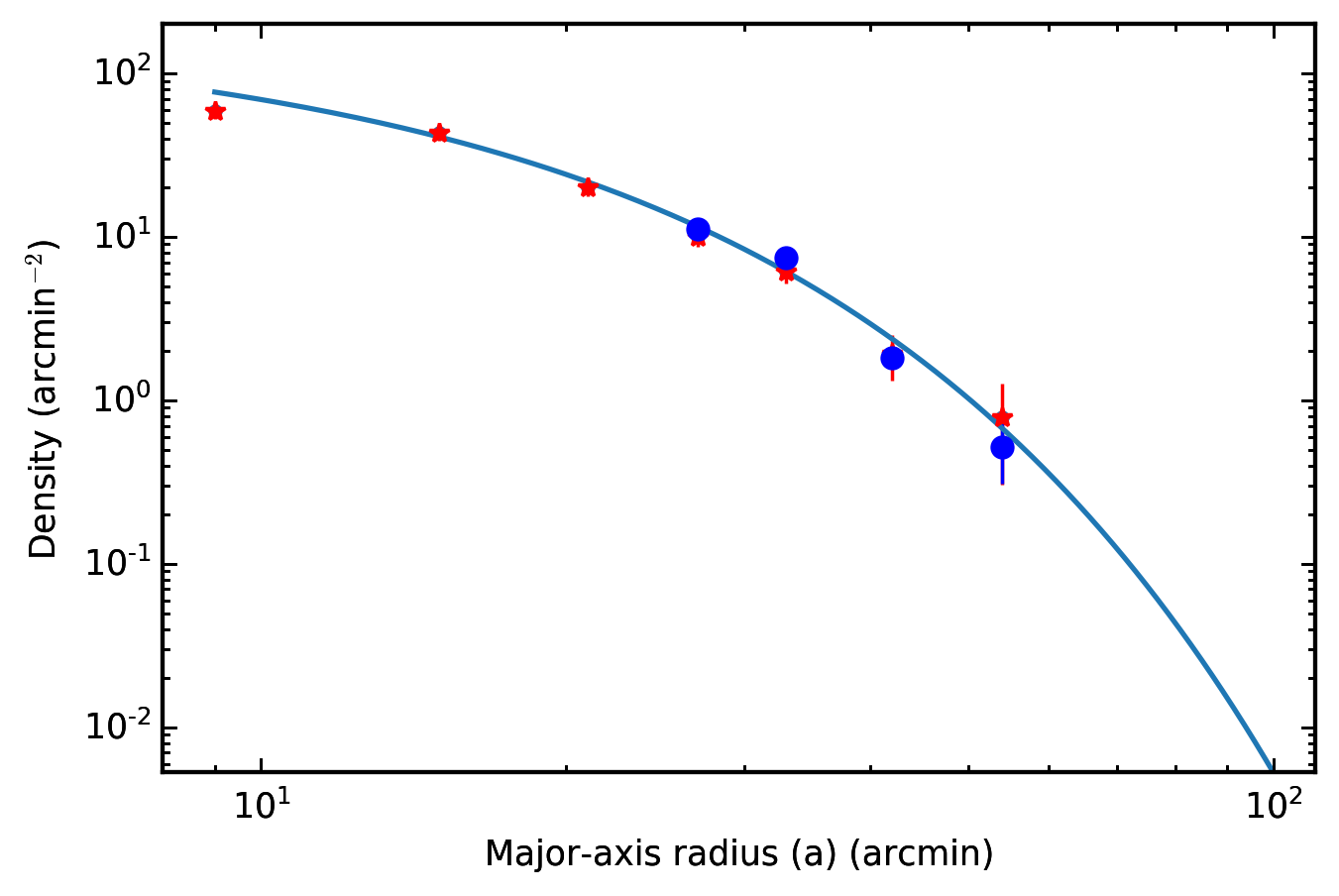}
\caption[Radial surface density profile for the NGC 6822 spheroid.]{This figure shows the radial surface density profile for the spheroid of NGC 6822. The upper panel is the profile when we count both RGB and red clump stars, while the lower panel is the profile counting only RGB stars. The y-axis is in units of number per square arcmin, and the x-axis is in units of arcmin. Blue dots show measurements from the DECam data, while red stars show measurements from the Megacam data. The blue lines show the best-fit exponential profiles as described in the text.}
\label{f:profile}
\end{figure}

Since the Megacam data and the DECam data have different faint limits, member selections, and degrees of contamination, the two sets of measurements are on different density scales. Because Megacam extends further inwards we decided to set this as the "base" scale, and then adjusted the DECam measurements vertically until the DECam profile smoothly joined the Megacam profile. In the overlap region the shapes of the two profiles show good agreement.

For each mosaic, we constructed two separate radial profiles utilising two different stellar selections. First, we adopted the pan-mosaic faint limits used to construct the density maps described in Section \ref{ss:maps} -- i.e., covering both the red giant branch and the red clump, pushing to the faintest possible limit at the potential cost of higher contamination. Second, we imposed a faint limit at $i_0=23.0$ -- i.e., selecting a much cleaner red-giant-only sample. 

The results are shown in Figure \ref{f:profile}. For the case where we included both RGB and red clump stars, the higher number of sources allowed us to push to a larger radius in the DECam data. The spheroid of NGC 6822 is still tentatively detected in the annulus spanning $\sim 60-80\arcmin$ ($\sim 8.2-11.0$\ kpc) from its centre. This is substantially further than in any previous work, and is comparable to the major-axis distance of the most remote known globular cluster in the system \citep[$10.7$\ kpc,][]{veljanoski:15}.

Following \citet{battinelli:06}, we attempted to fit the shape of each profile using an exponential function of the form:
\begin{equation}
    \mu(r) = A e^{-r/r_0}
\end{equation}
Here, the surface density is $\mu$, the (elliptical) radius is $r$, the (arbitrary) central density is $A$, and the scale length is $r_0$. As noted above, crowding is an issue at small radii in the Megacam data. The effect of this is evident in Figure \ref{f:profile}, with abrupt flattening in each profile suggested by the innermost data point. As a consequence, we ignored this datum when fitting the exponential function. For the profile including both RGB and red clump stars the best-fit parameters are $A = 606$\ arcmin$^{-2}$ and $r_0 = 9.6 \pm 0.3\arcmin$, while for the profile only including RGB stars the best-fit parameters are $A = 198$\ arcmin$^{-2}$ and $r_0 = 9.5 \pm 0.4\arcmin$. The inferred scale lengths agree closely, and correspond to a physical size $\approx 1.3 \pm 0.1$\ kpc. The good consistency suggests that residual contamination is not a major issue for either stellar selection; however, varying the background level within uncertainty suggests an additional systematic error of order $\pm 0.5\arcmin$ ($\pm 0.1$\ kpc) should be included when quoting the measured scale length. Our results are in good agreement with the outermost of the two-component exponential profile fit by \citet{battinelli:06}\footnote{Note that while our data are well fit by a single exponential function, \citet{battinelli:06} required two components as their profile extended inwards to smaller radii than ours; their outermost datum was at $\approx 45\arcmin$.}, who found a scale length $b = 10.0 \pm 1.1\arcmin$.

An important observation is that neither of our profiles shows any evidence for a break at a particular radius that would indicate a change in scale length, or a multiple component exponential profile. Such a break might be expected if NGC 6822 possessed a more extended stellar halo-like component in addition to the spheroid.

\section{Detection limit for faint satellites}
A by-eye analysis of our spatial density maps (Figures \ref{f:megamap} and \ref{f:decmap}) reveals no strong density peaks that might indicate the presence of one or more satellites of NGC 6822. Given that we can detect several of the globular clusters in the system via resolved star counts, at least with our Megacam data, we expect that if any dwarfs were present they should be visible in these maps down to some faint limit. In this section we aim to determine this faint limit in order to place quantitative constraints on the absence of satellite galaxies in the NGC 6822 system.

To do this, we insert artificial dwarfs into our data, spanning a grid of size and luminosity to identify the region of parameter space where a satellite is no longer visible in our maps. We consider luminosities in the range between $M_V = -10$ and $-4$. For each artificial dwarf we randomly generate stars according to a model from the Dartmouth Stellar Evolution Database \citep{dotter:08}. We assume an ancient metal-poor population with age $12$ Gyr and metallicity [Fe/H]$=-2.1$, which is appropriate for dwarfs in this luminosity range \citep[see the luminosity-metallicity relation described by e.g.,][]{kirby:13,simon:19}. We further assume [$\alpha$/Fe]$=+0.2$, which is appropriate for faint dwarfs in the Milky Way.

Given these specifications, the Dartmouth models provide the number of stars in different stellar luminosity bins according to a \citet{salpeter:55} initial mass function (IMF). This allows us to determine the number of observable stars ($N^*$) in a given artificial dwarf, by applying our pan-mosaic faint limits for either the Megacam or DECam data. This process is complicated by the fact that the Dartmouth luminosity functions do not include evolution past the helium flash. We therefore need to correct our calculations for missing sources on the horizontal branch and asymptotic giant branch, which typically number $\sim 25\%$ of the observed RGB population in an old stellar system.

We distribute the observable stars spatially by assuming that the radial density profile for each artificial dwarf follows the simple exponential form proposed by \citet{martin:16}:
\begin{equation}
    \rho(r) = \frac{1.68^2}{2\pi r_h^2 (1-e)} N^* e^{-1.68 r/r_h}
\end{equation}
where $e$ is the ellipticity of the dwarf, $r$ is the distance along its major axis, and $r_h$ is its half-light radius. The compilation by \citet[][their Figure 2]{simon:19} shows that for dwarf satellites of the Milky Way, the half-light radius ranges from $r_h \approx 500$\ pc for objects at the upper end of our luminosity range ($M_V\sim -10$) down to $r_h \approx 50$\ pc for objects at the lower end of the range ($M_V\sim -4$). We therefore consider three characteristic sizes of $r_h = 50$, $100$, and $500$\ pc. For a given artificial dwarf we further select an ellipticity randomly from a uniform distribution spanning the range $0 \leq e \leq 0.6$\footnote{\citet{simon:19} shows that the mean ellipticity for classical dwarf satellites of the Milky Way is $\approx 0.35$ and for the "ultra-faint" satellites is  $\approx 0.5$.}, and a random position angle ($\theta$) for the major axis. 

Given these parameters, and the above equation, for each of the $N^*$ observable stars in an artificial dwarf we randomly generate a radius $r$ along the major axis, and a random position angle relative to the major axis. We place the dwarf at some location relative to the centre of NGC 6822, which allows us to calculate $(x,y)$ positions for all stars in the galaxy-centric coordinate system used for our density maps. It is then straightforward to add these stars to our catalogue and remake the map. In total we selected four different sites for each of the Megacam and DECam mosaics, each of them outside several scale lengths of the NGC 6822 spheroid (i.e., outside $\sim30-40\arcmin$). 

 \begin{figure}
\centering
\includegraphics[width=0.9\columnwidth]{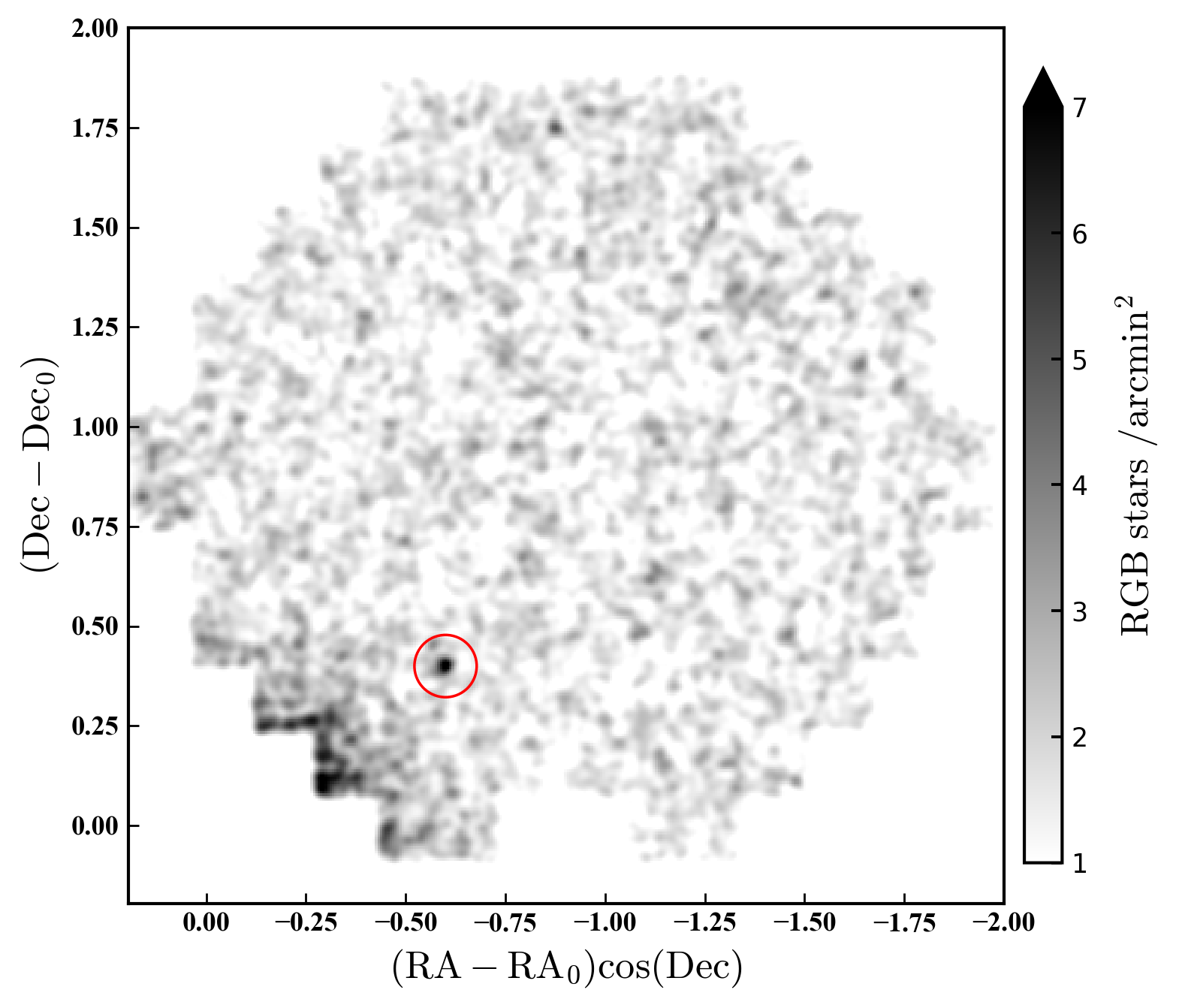}
\caption{An example artificial dwarf generated in the DECam density map for pointing P1. The red circle shows the location of the dwarf, which has a luminosity $M_V=-6$ and size $r_h = 100$\ pc. This object is prominent, with a detection significance of $\sim 10\sigma$.}
\label{f:exampledwarf}
\end{figure}

An example of the appearance of an artificial dwarf in the DECam map is shown in Figure \ref{f:exampledwarf}. With a luminosity $M_V=-6$ and half-light radius $r_h=100$\ pc, this object is prominent and would have been evident by eye almost anywhere in the DECam mosaic. We quantified the significance the density peak due to each dwarf in units of the standard deviation ($\sigma$) of map pixels surrounding the dwarf's location. For the example in Figure \ref{f:exampledwarf} the significance is $\sim 10\sigma$; in the literature, detection significance thresholds of $\sim 5-10\sigma$ are commonly utilised \citep[e.g.,][]{koposov:15,drlica:15,kim:15a,kim:15b}. 

\begin{figure}
\centering
\includegraphics[width=0.9\columnwidth]{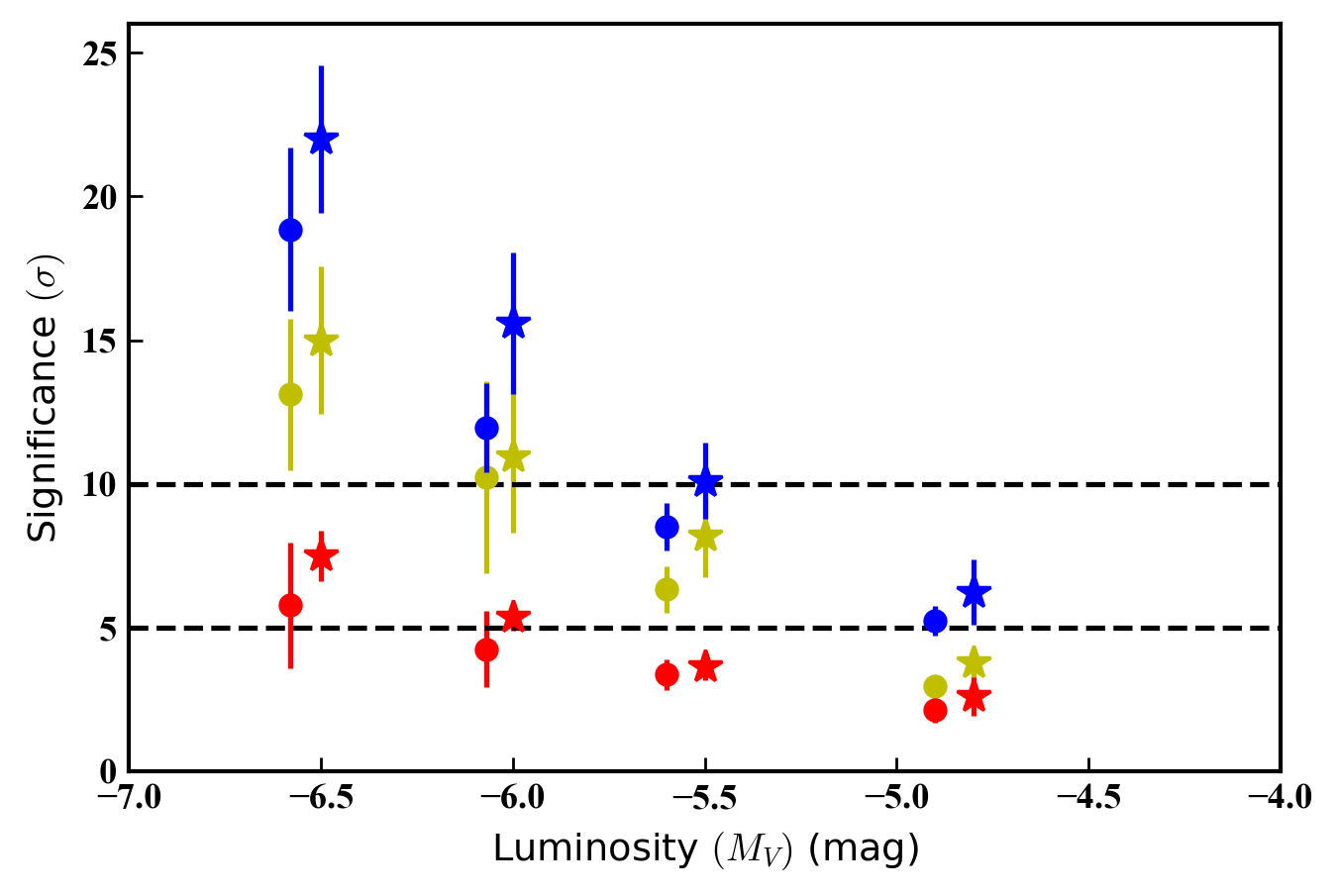}
\caption{Average detection significance (defined as described in the text) for our artificial dwarfs, as a function of luminosity and size. The points are coloured according to half-light radius -- those with $r_h=50$\ pc are blue, those with $r_h=100$\ pc are yellow, and those with $r_h=500$\ pc are red. Stars indicate results for dwarfs placed on the Megacam map, and circles for those on the DECam map. The latter have been systematically shifted slightly to the left to enhance visibility. Error bars indicate the scatter in detection significance across the different sampled locations. Significance thresholds of $5\sigma$ and $10\sigma$ (indicating the range commonly utilised in the literature) are marked with dashed horizontal lines.}
\label{f:dwarfresults}
\end{figure}

Figure \ref{f:dwarfresults} shows the detection significance for artificial dwarfs of different size and luminosity across our DECam and Megacam mosaics. In general, these objects have lower detection significance in the DECam map compared to the Megacam map, because the DECam photometry is shallower (i.e., for a given dwarf luminosity there are fewer observable stars). Moreover, there is a clear size dependence -- at fixed luminosity the more compact systems produce a stronger central peak resulting in a higher detection significance. Figure \ref{f:dwarfresults} demonstrates that the most diffuse dwarfs, with $r_h=500$\ pc, have become marginally detectable by $M_V\sim -6.5$, while more compact systems reach a $5\sigma$ threshold between $-4.8 \leq M_V \leq -5.5$. There is a small amount of variation from location to location, but in general the results have relatively small scatter.

In summary, the faintest detectable satellite dwarfs in our data would have $M_V\sim -5$ and half-light radii smaller than $\sim 100$\ pc. More diffuse objects would need to be brighter, $M_V\sim -6.5$. Another way of thinking about this is in terms of the limiting central surface brightness -- for a dwarf with $M_V= -5$ and $r_h = 100$\ pc at the distance of NGC 6822 the central surface brightness is $V\approx 28.6$ mag$\,$arcsec$^{-2}$, while for a system with $M_V= -6.5$ and $r_h = 500$\ pc it is fainter than $V\sim 30$ mag$\,$arcsec$^{-2}$. These values are in reasonable agreement with our previous estimate for the faint surface brightness limit of our survey (see Section \ref{ss:maps}). 

In physical terms our detection thresholds are relatively high -- many fainter satellites are known around the Milky Way, that we would not be sensitive to. This includes all the proposed ultra-faint companions of the Magellanic Clouds \citep[e.g.,][]{erkal:20}. The predictions of \citet{wheeler:15} suggest a $\sim 35\%$ chance of a galaxy like NGC 6822 possessing a dwarf satellite with stellar mass above $3000 M_\odot$ within its virial radius (which could be as large as $\sim 50$\ kpc). More encouragingly, the abundance-matching methods from \citet{dooley:17b} suggest that NGC 6822 could harbour up to $\sim 4$ satellites with stellar masses greater than $10^4 M_\odot$. Assuming a mass-to-light ratio of $\approx 1$ for an ancient metal-poor population, these two limiting stellar masses correspond to luminosities of $M_V \approx -3.9$ and $\approx -5.2$, respectively. The first of these is at least a magnitude fainter than our most optimistic detection limit, while the second sits near our faint limit. Importantly, our DECam data also only extend to less than half the expected virial radius of NGC 6822. Together, these results suggest that it would be beneficial for future deeper and even wider-field surveys -- such as the Legacy Survey of Space and Time (LSST) to be conducted by the Vera Rubin Observatory -- to revisit the question of satellites around this isolated system.

\section{Proper motion and orbit of NGC 6822}
One way of explaining the observed flattening and twisting of the outer spheroid in NGC 6822 might be if this galaxy had experienced a weak gravitational disturbance at some point in the past. The availability of {\it Gaia} Early Data Release 3 \citep[EDR3,][]{gaia:21} opens the possibility of measuring a proper motion for NGC 6822. This, combined with its position, and estimates of its distance and line-of-sight velocity, in turn allows us to calculate some simple orbits in order to test the extent to which NGC 6822 has been a truly isolated system.  

Previously, \citet{mcconnachie:21} used {\it Gaia} Data Release 2 \citep[DR2,][]{gaia:18} to address this problem, finding a systemic proper motion of $(\mu_\alpha,\,\mu_\delta)=(-0.02\pm0.02,\,0.00\pm0.02)$\ mas/yr. Here the reported uncertainties are random only, and in fact the measurement is dominated by systematics at the level of $\approx 0.035$\ mas/yr. Nonetheless, \citet{mcconnachie:21} concluded that NGC 6822 has likely fallen in towards the Milky Way from the edge of the Local Group, but has not had an interaction with the Milky Way (i.e., has not passed within its virial radius of $\sim 300$\ kpc) at $\approx 90\%$ confidence. Similarly, their orbit sampling suggested that NGC 6822 has not had an interaction with M31 at $\approx 99.99\%$ confidence.

On angular scales of a few degrees, {\it Gaia} EDR3 has systematic proper motion uncertainties at least a factor of two smaller than {\it Gaia} DR2 \citep[e.g.,][]{lindegren:21}, which makes it worthwhile revisiting this analysis. We first downloaded from the {\it Gaia} science archive a list of all sources in EDR3 sitting within $0.3$ degrees of the centre of NGC 6822. We then used a series of cuts to isolate young members of NGC 6822, and keep only stars with high quality astrometry. We selected young stars as our tracers because we showed above (Section \ref{ss:maps}) that members of this population are affected by contamination at a much lower level than RGB and red clump stars. In addition, young supergiants reach brighter magnitudes than stars on the upper RGB, which increases the chance of selecting a few members with excellent astrometry.

To isolate the young population in NGC 6822, we selected stars with colour $-0.4 < G_{\rm BP} - G_{\rm RP} < 0.7$ and magnitude $G<20$. We further demanded that the parallax $|\varpi| < 2 \sigma_\varpi$, where $\sigma_\varpi$ is the error in the parallax. We limited $\sigma_\varpi<0.5$\ mas to ensure we only kept stars with fairly reliable measurements.

Because the centre of NGC 6822 is very crowded, many stars possess poor astrometry. To try and exclude such sources we made a few additional cuts. First, as recommended by \citet{gaia:21}, we limited the parameter RUWE (Renormalised Unit Weight Error) to be {\tt RUWE} $< 1.4$, as sources with RUWE greater than this are typically unresolved binaries or blends. Second, we limited {\tt astrometric\_excess\_noise} $< 1.5$, which, again, is useful for eliminating stars that are blends or binaries \citep[e.g.,][]{iorio:19}. Finally, following \citet{vasiliev:19}, we selected only stars with {\tt phot\_bp\_rp\_excess\_factor} $< 1.3 + 0.06 (G_{\rm BP} - G_{\rm RP})^2$, which helps further reduce the number of blended sources.

\begin{figure}
\centering
\includegraphics[width=0.85\columnwidth]{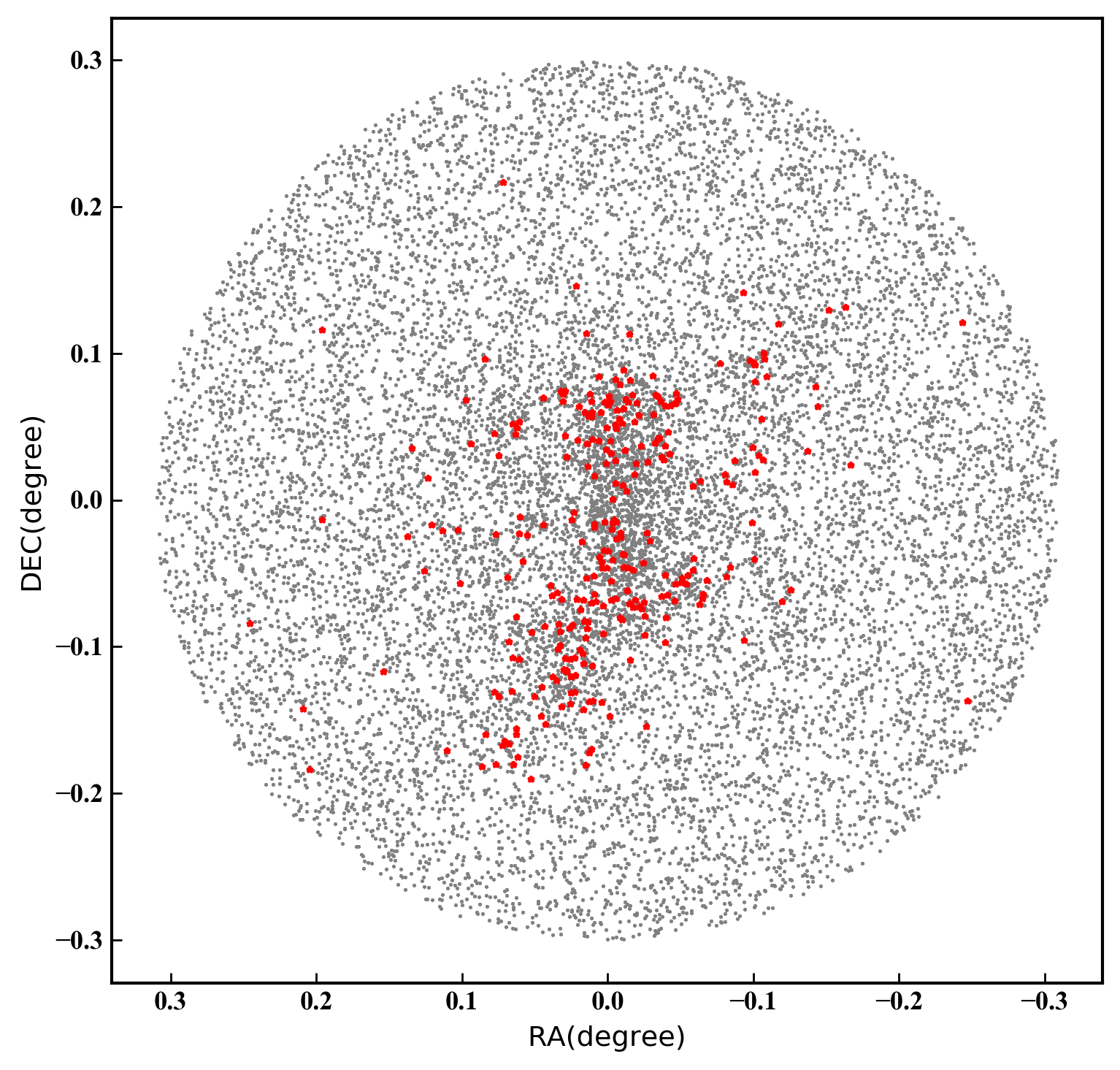}\\
\vspace{2mm}
\includegraphics[width=0.47\columnwidth]{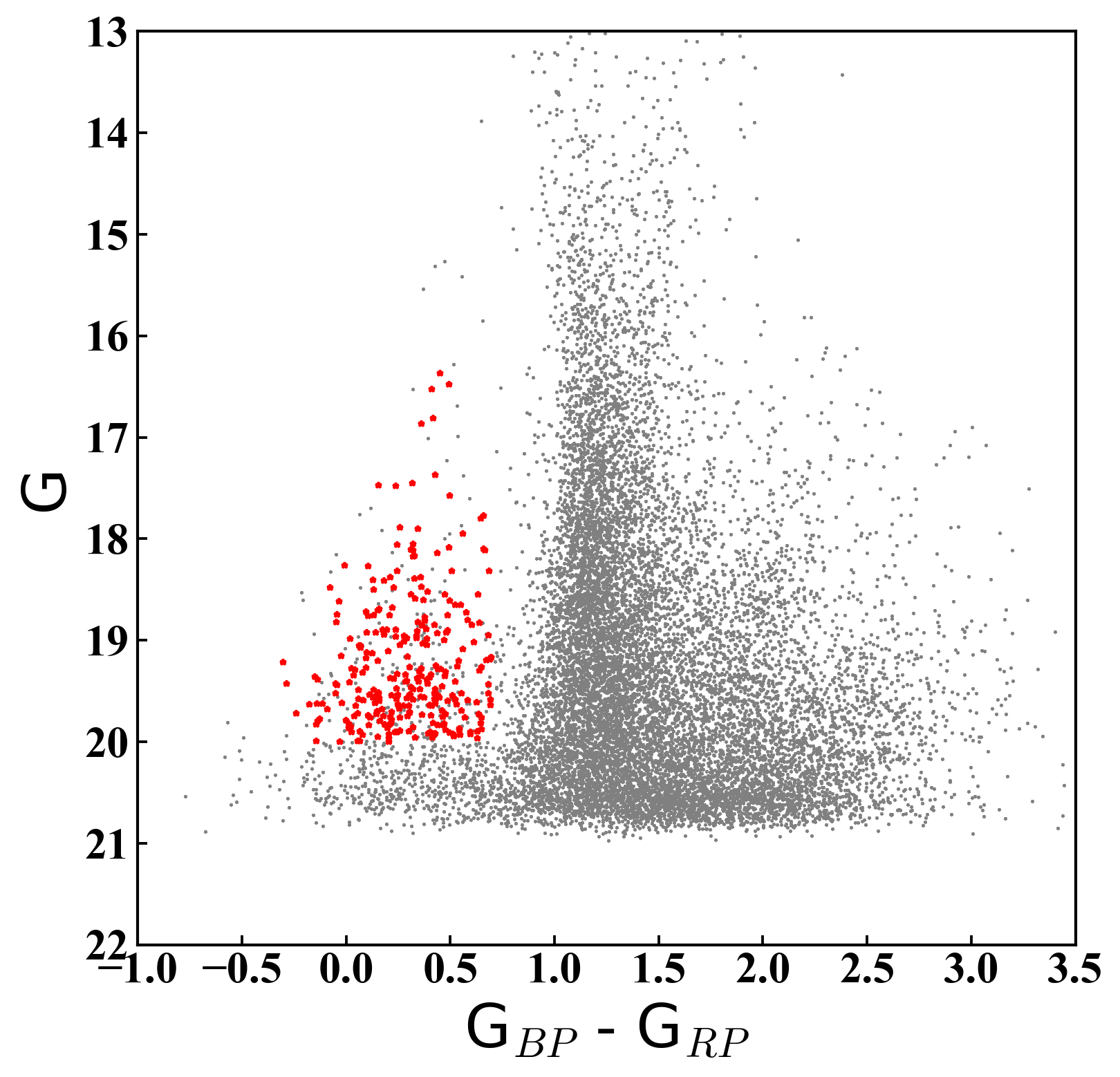}\hspace{1mm}
\includegraphics[width=0.51\columnwidth]{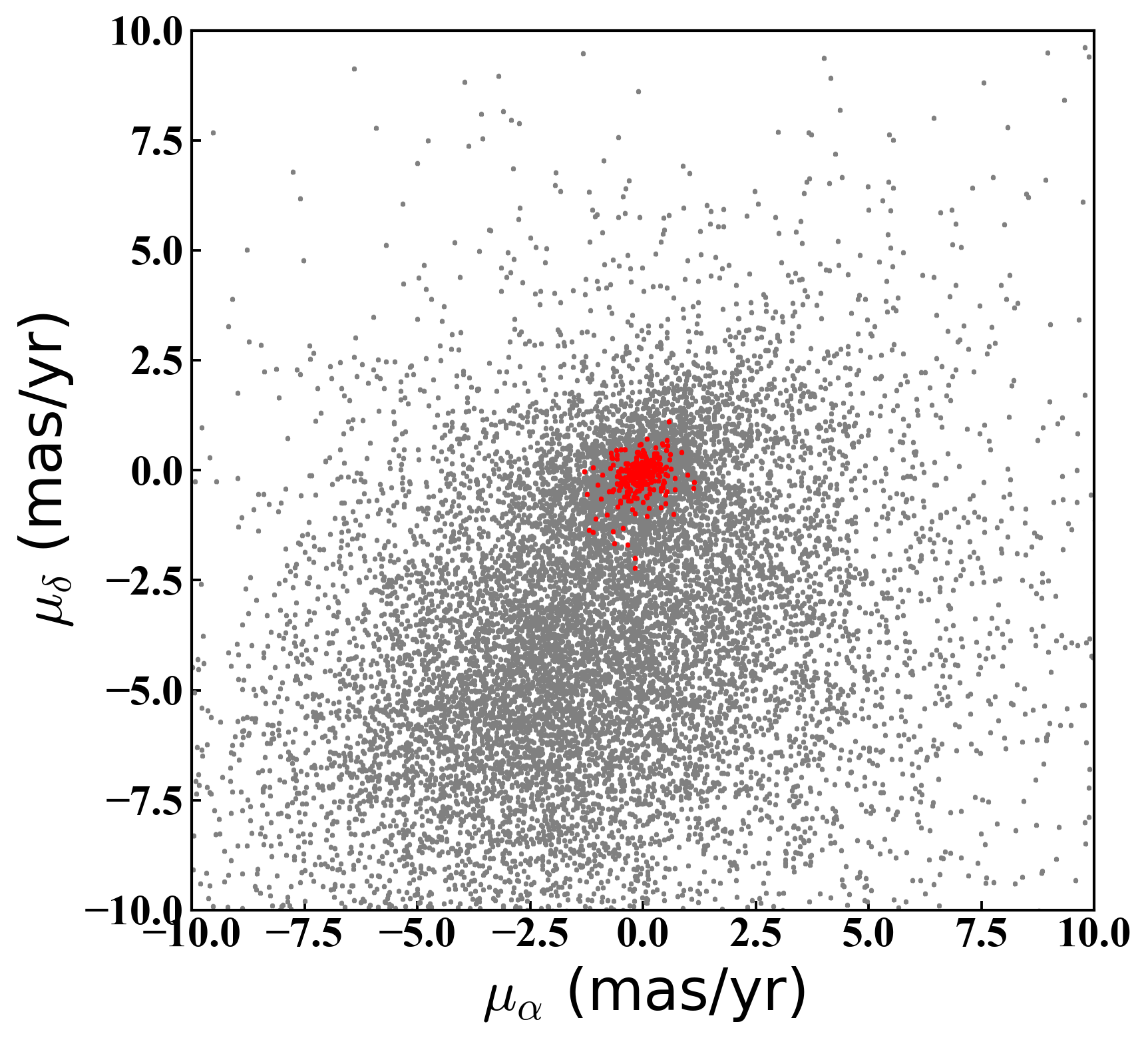}\\
\caption{Selection of high-quality NGC 6822 members from the {\it Gaia} EDR3 catalogue.  The top panel shows the position of all sources on the sky (grey points), and those passing all our selection and quality cuts as outlined in the text (red points). The lower-left panel shows the {\it Gaia} CMD, and the lower-right panel shows the proper motion plane.}
\label{f:gaiacuts}
\end{figure}
    
Figure \ref{f:gaiacuts} shows how this process improves the selection of NGC 6822 members. After our quality cuts, the remaining sources clearly trace the expected on-sky distribution of young stars in NGC 6822, and produce a very concentrated distribution on the proper motion plane. Altogether our final sample comprises $266$ stars.

To measure the proper motion of NGC 6822, we fit a 2D Gaussian to the stellar distribution on the proper motion plane. Based on our earlier examination of the young stellar populations in NGC 6822 we assume that the level of contamination by non-members is low, and elect not to fit a background contribution together with the Gaussian.  However, to ensure that proper motion outliers do not affect the fit, we calculate the median and standard deviation for each component across the sample, and retain only objects within $\pm 3\sigma$ of the median for the final fit. 

We use a maximum likelihood method to determine the most likely parameters for the 2D Gaussian given the observed stellar positions on the proper motion plane. The Gaussian has a peak at $(\mu_{\alpha,0},\, \mu_{\delta,0})$, and a width defined by $\sigma_{\mu_\alpha}$ and $\sigma_{\mu_\delta}$ plus a covariance $\rho$. The latter three parameters help account for the internal kinematics of young stars in NGC 6822, and could be non-zero in the case that the EDR3 astrometry is sufficiently good to resolve such motions. For reference, \citet{kirby:14} find that the line-of-sight velocity dispersion for (old) RGB stars in the spheroid is $23.2$\ km/s, which would equate to a proper motion of $\sim 0.01$\ mas/yr. However, the dispersion of the young populations is unknown, and plausibly substantially smaller given the likelihood of some degree of rotational support (which is seen for the H{\sc i}).

We sample the posterior probability distributions for the parameters in our model using the Markov Chain Monte Carlo sampler {\sc emcee} \citep{foreman:13}. We assume uninformative priors for all parameters. When performing the fit, we took into account the individual proper motion uncertainties per star from {\it Gaia}, as well as the covariance between the {\it Gaia} measurements of the two proper motion components for each star.

\begin{figure}
\centering
\includegraphics[width=0.99\columnwidth]{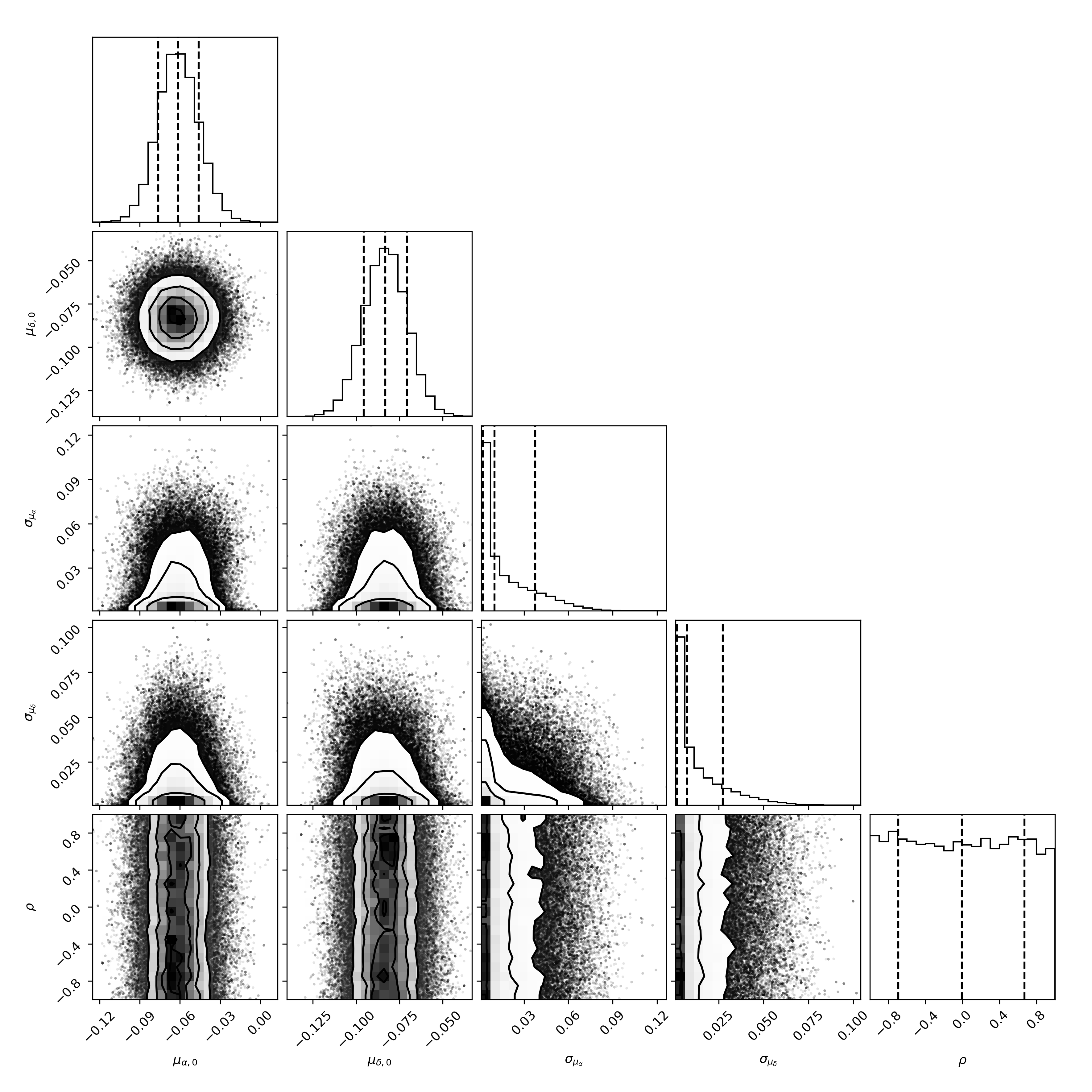}
\caption{Corner plot showing marginalized one- and two-dimensional posterior probability distributions for the five parameters in our proper motion fit for NGC 6822. The contours show the $1\sigma$, $2\sigma$, and $3\sigma$ confidence levels. Created with {\it corner.py} \citep{foreman:16}.}
\label{f:pmfit}
\end{figure}

Figure \ref{f:pmfit} shows the one- and two-dimensional marginalized posterior probability distribution functions for the five parameters in the model. From this plot it is evident that the proper motion for NGC 6822 is clearly measured and non-zero: we find $\mu_{\alpha,0} = -0.061 \pm 0.014$\ mas/yr and $\mu_{\delta,0} = -0.083 \pm 0.012$\ mas/yr. 

We also find that the dispersion parameters are essentially unresolved, although both exhibit small tails to non-zero values that suggest the data are on the cusp of being able to provide information about the internal motions of stars in NGC 6822. It would be worthwhile revisiting this possibility in future {\it Gaia} data releases.

To check the proper motion zero-point in this region of the sky, we follow \citet{vandermarel:19} and download astrometry for quasars using the {\tt agn\_cross\_id} table available in the {\it Gaia} science archive. Because these sources are relatively sparsely distributed, we needed to expand our search radius to $\sim 6\degr$\ from the centre of NGC 6822 in order to locate a sufficient number. We then applied the same cuts described above (excluding the colour-magnitude selection), and again fit a 2D Gaussian model on the proper motion plane.  This has a peak at $\mu_{\alpha,{\rm ZP}} = 0.008 \pm 0.012$\ mas/yr and $\mu_{\delta,{\rm ZP}} = 0.006 \pm 0.010$\ mas/yr.  We subtract this from our best-fit value for NGC 6822 to obtain a final systemic motion of $\mu_{\alpha,0} = -0.069 \pm 0.019$\ mas/yr and $\mu_{\delta,0} = -0.089 \pm 0.016$\ mas/yr (where the uncertainties have been combined in quadrature).

Our proper motion measurement is mildly inconsistent with the {\it Gaia} DR2 value obtained by \citet{mcconnachie:21}, even considering the substantial systematics that may affect their measurement. This is especially notable in the $\mu_\delta$ component. However, while we were writing this paper, \citet{battaglia:21} released a set of uniform {\it Gaia} EDR3 proper motion measurements for galaxies across the Local Group.  They find $\mu_{\alpha,0} = -0.06 \pm 0.01$\ mas/yr and $\mu_{\delta,0} = -0.07 \pm 0.01$\ mas/yr, derived using both young and old stellar populations and including a background contribution in their fit. They also measured a quasar zero-point of $\mu_{\alpha,{\rm ZP}} = -0.008 \pm 0.020$\ mas/yr and $\mu_{\delta,{\rm ZP}} = -0.016 \pm 0.020$\ mas/yr. Both these results are much more compatible with the results presented here.

\begin{figure}
\centering
\includegraphics[width=0.95\columnwidth]{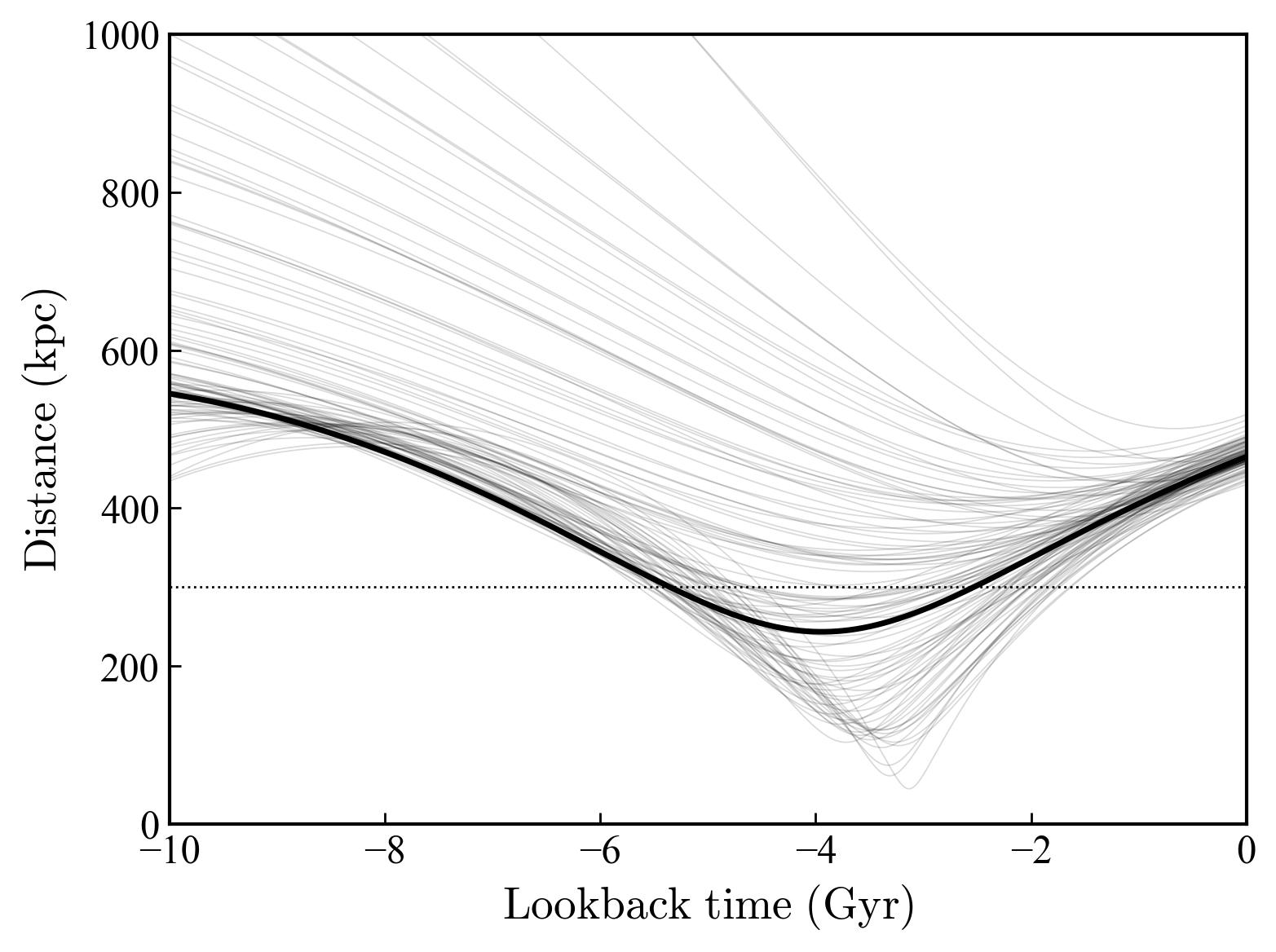}
\includegraphics[width=0.95\columnwidth]{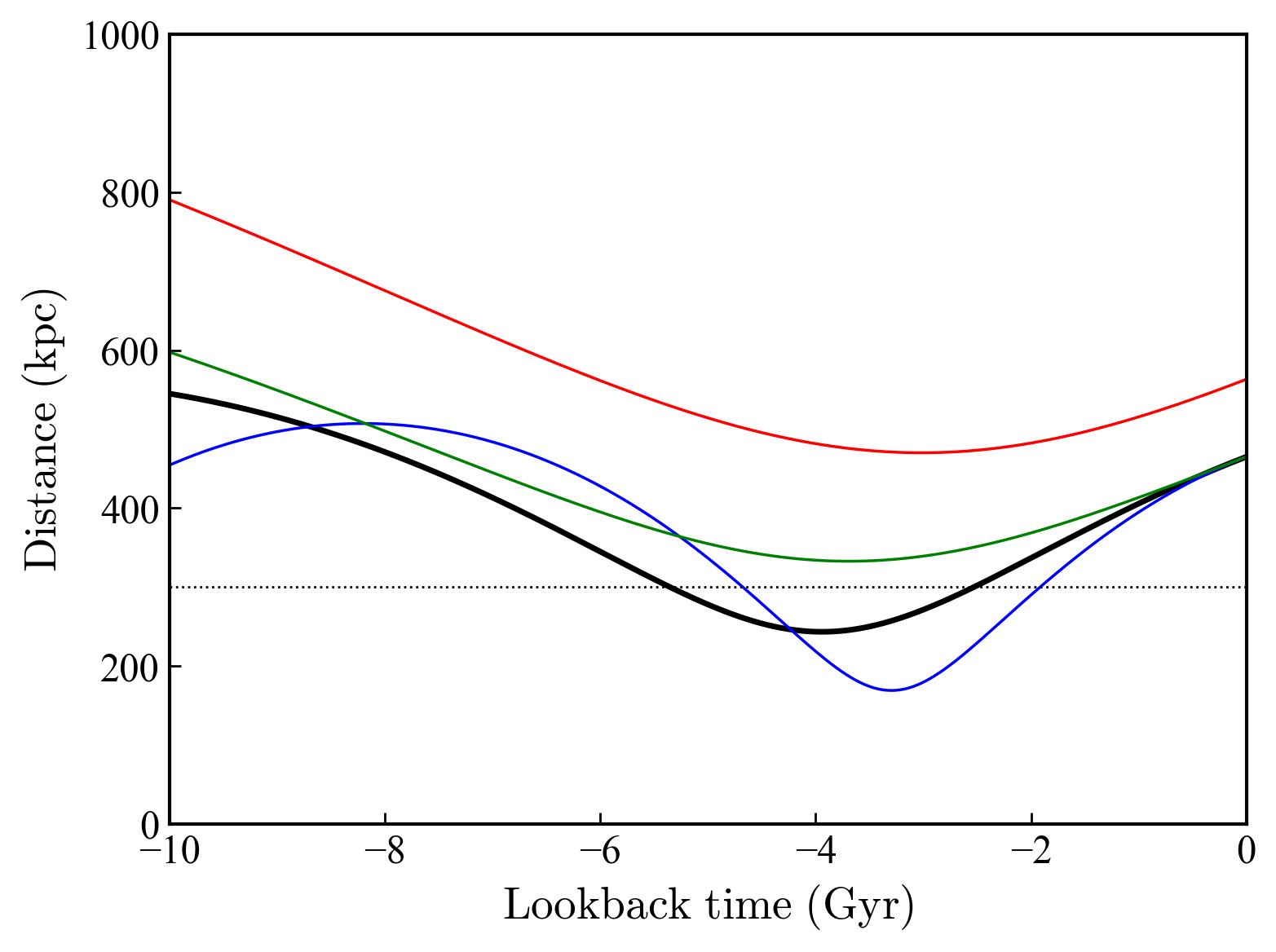}
\caption{Separation between NGC 6822 and the Milky Way over the past $10$\ Gyr assuming the Milky Way potential of \citet{mcmillan:17}. The upper panel shows the orbital solution (solid black line) for our measured proper motions and adopted position, distance, and line-of-sight velocity (see text). The dotted horizontal line indicates the approximate position of the Milky Way's virial radius at a distance of $300$\ kpc. The grey lines show $100$ different realisations of the orbit, where the initial conditions have been drawn randomly according to the measurement uncertainties on the proper motion, distance, and line-of-sight velocity. The lower panel shows the same preferred orbit as in the upper panel (solid black line), which was calculated for a Milky Way mass of $1.3\times 10^{12}\,M_\odot$, together with orbits derived assuming (i) the larger distance measured by \citet{higgs:21} (red line); (ii) a higher Milky Way mass of $2.0\times 10^{12}\,M_\odot$ (blue line); and (iii) a smaller Milky Way mass of $0.8\times 10^{12}\,M_\odot$ (green line).} 
\label{f:orbits}
\end{figure}

Having determined a proper motion for NGC 6822, we aim to briefly investigate its orbit relative to the Milky Way. To do this we use the {\sc galpy} galactic dynamics package \citep{bovy:15} to integrate the motion of NGC 6822 backwards in time for $10$\ Gyr. We assume the usual position for the centre of the galaxy at $\alpha = 19^{\rm h}\,44^{\rm m}\,56.6^{\rm s}, \delta = -14\degr\,47\arcmin\,21\arcsec$ \citep{mcconnachie:12}, our previously-adopted distance of $472$\ kpc, and the heliocentric line-of-sight velocity of $v_{\rm los} = -54.5 \pm 1.7$\ km/s measured by \citet{kirby:14}. The orbit is integrated in the Milky Way potential from \citet{mcmillan:17}, which has a total mass of $1.3\times 10^{12}\,M_\odot$. 

The top panel of Figure \ref{f:orbits} shows the distance of NGC 6822 from the centre of the Milky Way as a function of lookback time. NGC 6822 has fallen inwards from a distance of nearly $600$\ kpc, and appears to have had a weak interaction with the Milky Way, passing just inside its virial radius (assumed to be $\sim 300$\ kpc) approximately $4$\ Gyr ago. The orbit is bound -- integration further back in time reveals an apocentre of $\approx 600$\ kpc roughly $12$\ Gyr ago.

These conclusions are sensitive to measurement uncertainties as well as several of our assumptions. To check the effect of measurement uncertainties we randomly generated a set of $10^5$ new distances, proper motions, and line-of-sight velocities and then re-integrated the orbit for each combination. For the proper motions and line-of-sight velocity, we drew from normal distributions centred on the values outlined above and with widths defined by the reported uncertainty intervals. For the distance, we assumed a range of $\pm 0.08$ mag on the distance modulus, which is commensurate with typical uncertainties reported in the literature \citep[cf.][]{gorski:11,mcconnachie:12,higgs:21}. 

The upper panel of Figure \ref{f:orbits} shows $100$ example orbits from the randomly generated set. A wide variety of different trajectories are permissible given the current measurement uncertainties, ranging from orbits where NGC 6822 has fallen in from a very large distance and is essentially only now at its closest approach to the Milky Way, to orbits where NGC 6822 plunges deep within the virial radius of the Milky Way and has had a relatively strong encounter. The latter type of orbit also tends to be bound to the Milky Way, with apocentres around $\sim 500$\ kpc; however, we note that long-term integration of this type of orbit may not be reliable as our simple model does not account for dynamical friction, or changes in the mass of the Milky Way with time. Across the set of $10^5$ realisations, $\approx 57\%$ pass within $300$\ kpc of the Milky Way; the median time of pericentre for these encounters is $3.1$\ Gyr ago. Around $27\%$ of orbits are bound to the Milky Way.

Of the four sampled parameters, uncertainties on the distance are most important. Indeed, our assumed error of $\pm 0.08$ mag may be an under-estimate -- this range includes the shorter distance of $459\pm17$\ kpc listed by \citet{mcconnachie:12} \citep[originally from][]{gieren:06} as well as the larger distance of $490$\ kpc from \citet{mateo:98} and the majority of measurements compiled in \citet{gorski:11}; however, recently \citet{higgs:21} obtained a much higher estimate of $570\pm12$\ kpc using the RGB tip. The lower panel of Figure \ref{f:orbits} shows the orbit derived using this distance together with the velocities outlined above (red line). In this case, NGC 6822 does not approach the Milky Way any closer than $\approx 500$\ kpc.

The assumed gravitational potential of the Milky Way is also important.  Because NGC 6822 is quite distant, it is not generally the shape of the potential that matters, but rather the overall assumed mass. As noted above, the \citet{mcmillan:17} potential has a Milky Way mass of $1.3\times 10^{12}\,M_\odot$; the lower panel of Figure \ref{f:orbits} shows orbits integrated in the same potential but with smaller ($0.8\times 10^{12}\,M_\odot$; green line) and larger ($2.0\times 10^{12}\,M_\odot$; blue line) halo masses. Perhaps unsurprisingly, for a higher-mass Milky Way the orbit reaches more deeply within the virial radius, while for a lower-mass Milky Way the orbit does not cross the virial radius.

It is worth explicitly pointing out that our simple models do not include the effect of other galaxies such as M31 or the LMC. \citet{mcconnachie:12} showed that in general M31 remains very distant from NGC 6822 regardless of the assumed initial conditions, and we find a similar outcome here \citep[see also][]{battaglia:21}. On the sky, NGC 6822 does not pass within an angular distance of $\sim 50\degr$ of the current position of M31; because M31 is on a rather radial orbit \citep{vandermarel:12,vandermarel:19} this position does not change much with time. At the current distance of M31 an angular separation of $>50\degr$ corresponds to a projected physical distance of at least $\sim 680$\ kpc. Because M31 is currently moving towards the Milky Way (and was hence further away in the past) this is a lower limit. The LMC is more difficult to account for; indeed \citet{battaglia:21} show that it might affect the orbit of NGC 6822 during the past $\sim3-4$\ Gyr. More detailed investigation is clearly warranted.

Our finding that NGC 6822 may have passed within the virial radius of the Milky Way is consistent with recent results from both \citet{mcconnachie:21} and \citet{battaglia:21}. The former find that NGC 6822 can pass within the Milky Way's virial radius, but observe such an encounter in fewer than $10\%$ of their sampled orbits. However, as noted above their measured proper motions are somewhat different than ours. On the other hand \citet{battaglia:21}, whose measurements are entirely consistent with ours, find that NGC 6822 could certainly have passed within the virial radius of the Milky Way; indeed for their "heavy Milky Way" model the pericentre is as close as $100$\ kpc.

In summary, we measure a clear non-zero proper motion for NGC 6822 using {\it Gaia} EDR3. Integrating the orbit of this galaxy, we find that it has most likely fallen in towards the Milky Way, from a direction largely away from that of M31, and a distance of at least $500$\ kpc. There is a good chance that NGC 6822 passed within the virial radius of the Milky Way $\sim 3-4$ Gyr ago; if so, it is likely the best local example of a "backsplash" galaxy. It is also possible that NGC 6822 is on a very long-period bound orbit about the Milky Way. A previous weak gravitational interaction with our Galaxy might help explain the twisting and flattening of the outer spheroid that we have observed in our imaging data. Moreover, the star-formation history derived by \citet{weisz:14} \citep[see also][]{cannon:12} suggests that NGC 6822 formed only about $\sim30\%$ of its stars earlier than $6$\ Gyr ago, and the star-formation rate shows a substantial acceleration since $\sim3-5$ Gyr ago (somewhat location-dependent). This may also be consistent with a scenario where NGC 6822 interacted with the Milky Way around this time.

\section{Discussion and Conclusions}
We have used Blanco/DECam and Magellan/Megacam imaging to conduct a panoramic survey of the isolated Local Group dwarf galaxy NGC 6822. Our photometry reaches $\sim 2-3$ magnitudes deeper than most previous studies and we cover the widest area around the dwarf compared to any prior work.

We used young main sequence stars, and old stars from the red giant branch and red clump, to trace the distribution of stellar populations in NGC 6822 with the following main results:

\begin{enumerate}
\item{NGC 6822 has substantial young stellar populations, which closely trace the H{\sc i} gas in the dwarf. This is consistent with the findings from many previous studies \citep[e.g.,][]{komiyama:03,deblok:06}. The H{\sc i} is known to be arranged in a rotating disk-like structure \citep[e.g.,][]{deblok:00}. We do not observe any substantial young populations away from the gas.}
\item{NGC 6822 possesses a large extended spheroidal component made up of older stellar populations (age $\geq 2$\ Gyr). The major axis of the spheroid is angled by $\approx 60\degr$ to the direction of the H{\sc i} gas. Again, this is known from many previous studies \cite[e.g.,][]{komiyama:03,deblok:06,battinelli:06,higgs:21}. Our data clearly show that there is no correlation between dense clumps of young stars, and the density of older populations.}
\item{We were able to trace the extent of the spheroid to a larger radius than any previous work, $\approx 60-80\arcmin$ ($8.2-11.0$\ kpc). This is commensurate with the radial extent of the globular cluster system in NGC 6822. The spheroid appears smooth, and its density profile outside $\sim 10\arcmin$\ is well fit by a single exponential function with scale radius $9.6 \pm 0.3\arcmin$\ (random) $\pm 0.5\arcmin$ (systematic) ($1.3 \pm 0.1 \pm 0.1$\ kpc). We do not observe any break in the profile that would indicate the presence of a more extended stellar halo-like component surrounding NGC 6822.}
\item{Both the ellipticity and orientation of the spheroid change with radius. The system becomes more flattened at larger radii, and the position angle of the major axis twists by up to $40$\ degrees.}
\item{We do not observe any stellar streams or over-densities to the edge of our DECam mosaic (radius $\approx 2\degr \sim 16.5$\ kpc) and to a faint limit of $V\sim 30$\ mag$\,$arcsec$^{-2}$. There are some low-level fluctuations visible in the density maps, but these cannot be robustly distinguished from background noise.}
\end{enumerate}

The structure of NGC 6822 is a long-standing puzzle, in particular the very different orientation between the spheroid and the H{\sc i}. Studies such as those of \citet{deblok:00,deblok:03} and \citet{komiyama:03} suggested that the presence of the extended H{\sc i} structure and associated young populations could be due to a recent merger event, with the large cloud of gas to the north-west potentially indicating the location of the interacting system. \citet{demers:06} and \citet{battinelli:06} even suggested NGC 6822 could be a polar ring galaxy. However, the {\it Hubble Space Telescope} study of \citet{cannon:12} showed that the stellar populations at the location of the north-west H{\sc i} cloud were indistinguishable from the stellar populations at other locations in the dwarf. They therefore suggested that the H{\sc i} structure in NGC 6822 is simply a warped disk inclined to the line-of-sight. The observed kinematic connection between the rotation of the spheroid and the rotation of the gas is consistent with this interpretation \citep[e.g.,][]{thompson:16,belland:20}.

Our observation that there is no over-density of old stellar populations at the location of the north-west gas cloud supports the conclusion of \citet{cannon:12} that the north-west gas cloud is not an interacting companion galaxy. Moreover, we find that the spheroid appears mostly undisturbed (smooth) and observe no obvious stellar streams or over-densities of either young or old stars near to NGC 6822. This all indicates that NGC 6822 has probably not experienced any significant merger events in the past few Gyr. This may be consistent with recent hydrodynamical models which indicate that lower-mass dwarfs may largely form in place without requiring merger activity to shape their properties \citep[e.g.,][]{wheeler:19}; however, we cannot exclude the possibility that NGC 6822 underwent one or more mergers early in its history.  The origin of the substantial mis-alignment between the gas disk and spheroid major axes, reflected in the apparent prolate rotation of the spheroid \citep{belland:20}, remains a mystery. Detailed modelling is required to assess whether such a configuration could arise from an early merger event; moreover, future astrometry from the {\it Gaia} satellite might resolve internal motions to help address this question. Data from {\it Gaia} EDR3 are on the edge of being able to achieve this.

We did not observe any compact over-densities that would indicate the presence of dwarf satellites of NGC 6822, out to the edge of our DECam mosaic at $\sim 16.5$\ kpc. By inserting artificial dwarf galaxies into our smoothed density maps, we show that we could have detected any faint satellite of NGC 6822 down to a luminosity of $M_V\approx -6.5$ in the case of diffuse dwarfs ($r_h \approx 500$\ pc) or $M_V\approx -5$ in the case of more compact satellites ($r_h \approx 50-100$\ pc). This is consistent with the fact we were able to resolve a few of the better-populated globular clusters, at least with Megacam. If NGC 6822 has any low-luminosity satellites, they must be fainter than our detection limit or sit outside the surveyed area. Our data reach only to approximately half the expected virial radius of the system, so it would be worthwhile revisiting this question with future wide-field surveys such as LSST.

Finally, we compute the proper motion of NGC 6822 using astrometry from {\it Gaia} EDR3 and show that this galaxy has indeed likely spent most of its life in isolation, falling inwards from a large distance ($> 500$\ kpc) at the edge of the Local Group. This is in agreement with \citet{mcconnachie:21}. Our orbit calculations also strongly suggest that NGC 6822 passed within the virial radius of Milky Way $\sim 3-4$ Gyr ago, in agreement with recent results from \citet{battaglia:21}. A weak gravitational encounter of this nature might help explain the flattening and twisting we observed in the outer spheroid, as well as the vigorous increase in star formation seen in the last $\sim 3-5$\ Gyr \citep[e.g.,][]{weisz:14}.

\section*{Acknowledgements}
We are grateful to Erwin de Blok for supplying his zero-moment H{\sc i} map for NGC 6822. DM acknowledges support from an Australian Research Council Future Fellowship (FT160100206). 

This paper includes data gathered with the 6.5m Magellan Telescopes located at Las Campanas Observatory, Chile. Australian access to the Magellan Telescopes was supported through the National Collaborative Research Infrastructure Strategy of the Australian Federal Government. This paper uses data products produced by the OIR Telescope Data Center, supported by the Smithsonian Astrophysical Observatory.

This project used data obtained with the Dark Energy Camera (DECam), which was constructed by the Dark Energy Survey (DES) collaboration. Funding for the DES Projects has been provided by the US Department of Energy, the US National Science Foundation, the Ministry of Science and Education of Spain, the Science and Technology Facilities Council of the United Kingdom, the Higher Education Funding Council for England, the National Center for Supercomputing Applications at the University of Illinois at Urbana-Champaign, the Kavli Institute for Cosmological Physics at the University of Chicago, Center for Cosmology and Astro-Particle Physics at the Ohio State University, the Mitchell Institute for Fundamental Physics and Astronomy at Texas A\&M University, Financiadora de Estudos e Projetos, Funda\c{c}\~{a}o Carlos Chagas Filho de Amparo \`{a} Pesquisa do Estado do Rio de Janeiro, Conselho Nacional de Desenvolvimento Científico e Tecnol\'{o}gico and the Minist\'{e}rio da Ci\^{e}ncia, Tecnologia e Inova\c{c}\~{a}o, the Deutsche Forschungsgemeinschaft and the Collaborating Institutions in the Dark Energy Survey.

The Collaborating Institutions are Argonne National Laboratory, the University of California at Santa Cruz, the University of Cambridge, Centro de Investigaciones En\'{e}rgeticas, Medioambientales y Tecnol\'{o}gicas–Madrid, the University of Chicago, University College London, the DES-Brazil Consortium, the University of Edinburgh, the Eidgen\"{o}ssische Technische Hochschule (ETH) Z\"{u}rich, Fermi National Accelerator Laboratory, the University of Illinois at Urbana-Champaign, the Institut de Ci\`{e}ncies de l’Espai (IEEC/CSIC), the Institut de F\'{i}sica d’Altes Energies, Lawrence Berkeley National Laboratory, the Ludwig-Maximilians Universit\"{a}t M\"{u}nchen and the associated Excellence Cluster Universe, the University of Michigan, NSF’s NOIRLab, the University of Nottingham, the Ohio State University, the OzDES Membership Consortium, the University of Pennsylvania, the University of Portsmouth, SLAC National Accelerator Laboratory, Stanford University, the University of Sussex, and Texas A\&M University.

Based on observations at Cerro Tololo Inter-American Observatory, NSF’s NOIRLab (NOIRLab Prop. ID 2013A-0617 and 2013B-0611; PI: D. Mackey), which is managed by the Association of Universities for Research in Astronomy (AURA) under a cooperative agreement with the National Science Foundation.

The Pan-STARRS1 Surveys (PS1) and the PS1 public science archive have been made possible through contributions by the Institute for Astronomy, the University of Hawaii, the Pan-STARRS Project Office, the Max-Planck Society and its participating institutes, the Max Planck Institute for Astronomy, Heidelberg and the Max Planck Institute for Extraterrestrial Physics, Garching, The Johns Hopkins University, Durham University, the University of Edinburgh, the Queen's University Belfast, the Harvard-Smithsonian Center for Astrophysics, the Las Cumbres Observatory Global Telescope Network Incorporated, the National Central University of Taiwan, the Space Telescope Science Institute, the National Aeronautics and Space Administration under Grant No. NNX08AR22G issued through the Planetary Science Division of the NASA Science Mission Directorate, the National Science Foundation Grant No. AST-1238877, the University of Maryland, Eotvos Lorand University (ELTE), the Los Alamos National Laboratory, and the Gordon and Betty Moore Foundation.

\section*{Data Availability}
The data underlying this article will be shared on reasonable request to the corresponding author.



\bibliographystyle{mnras}
\bibliography{ms_rv} 

\begin{thebibliography}{}
\makeatletter
\relax
\def\mn@urlcharsother{\let\do\@makeother \do\$\do\&\do\#\do\^\do\_\do\%\do\~}
\def\mn@doi{\begingroup\mn@urlcharsother \@ifnextchar [ {\mn@doi@}
  {\mn@doi@[]}}
\def\mn@doi@[#1]#2{\def\@tempa{#1}\ifx\@tempa\@empty \href
  {http://dx.doi.org/#2} {doi:#2}\else \href {http://dx.doi.org/#2} {#1}\fi
  \endgroup}
\def\mn@eprint#1#2{\mn@eprint@#1:#2::\@nil}
\def\mn@eprint@arXiv#1{\href {http://arxiv.org/abs/#1} {{\tt arXiv:#1}}}
\def\mn@eprint@dblp#1{\href {http://dblp.uni-trier.de/rec/bibtex/#1.xml}
  {dblp:#1}}
\def\mn@eprint@#1:#2:#3:#4\@nil{\def\@tempa {#1}\def\@tempb {#2}\def\@tempc
  {#3}\ifx \@tempc \@empty \let \@tempc \@tempb \let \@tempb \@tempa \fi \ifx
  \@tempb \@empty \def\@tempb {arXiv}\fi \@ifundefined
  {mn@eprint@\@tempb}{\@tempb:\@tempc}{\expandafter \expandafter \csname
  mn@eprint@\@tempb\endcsname \expandafter{\@tempc}}}

\bibitem[\protect\citeauthoryear{{Amorisco}, {Evans}  \& {van de
  Ven}}{{Amorisco} et~al.}{2014}]{amorisco:14}
{Amorisco} N.~C.,  {Evans} N.~W.,   {van de Ven} G.,  2014, \mn@doi [\nat]
  {10.1038/nature12995}, \href
  {https://ui.adsabs.harvard.edu/abs/2014Natur.507..335A} {507, 335}

\bibitem[\protect\citeauthoryear{{Balbinot} et~al.,}{{Balbinot}
  et~al.}{2015}]{balbinot:2015}
{Balbinot} E.,  et~al., 2015, \mn@doi [\mnras] {10.1093/mnras/stv356}, \href
  {https://ui.adsabs.harvard.edu/abs/2015MNRAS.449.1129B} {449, 1129}

\bibitem[\protect\citeauthoryear{{Barnard}}{{Barnard}}{1884}]{barnard:84}
{Barnard} E.~E.,  1884, \mn@doi [Astronomische Nachrichten]
  {10.1002/asna.18841100805}, \href
  {https://ui.adsabs.harvard.edu/abs/1884AN....110..125B} {110, 125}

\bibitem[\protect\citeauthoryear{{Battaglia}, {Taibi}, {Thomas}  \&
  {Fritz}}{{Battaglia} et~al.}{2021}]{battaglia:21}
{Battaglia} G.,  {Taibi} S.,  {Thomas} G.~F.,   {Fritz} T.~K.,  2021, arXiv
  e-prints, \href {https://ui.adsabs.harvard.edu/abs/2021arXiv210608819B} {p.
  arXiv:2106.08819}

\bibitem[\protect\citeauthoryear{{Battinelli} \& {Demers}}{{Battinelli} \&
  {Demers}}{2011}]{battinelli:11}
{Battinelli} P.,  {Demers} S.,  2011, \mn@doi [\aap]
  {10.1051/0004-6361/201015470}, \href
  {https://ui.adsabs.harvard.edu/abs/2011A&A...525A..69B} {525, A69}

\bibitem[\protect\citeauthoryear{{Battinelli}, {Demers}  \&
  {Kunkel}}{{Battinelli} et~al.}{2006}]{battinelli:06}
{Battinelli} P.,  {Demers} S.,   {Kunkel} W.~E.,  2006, \mn@doi [\aap]
  {10.1051/0004-6361:20054718}, \href
  {https://ui.adsabs.harvard.edu/abs/2006A&A...451...99B} {451, 99}

\bibitem[\protect\citeauthoryear{{Bechtol} et~al.,}{{Bechtol}
  et~al.}{2015}]{bechtol:15}
{Bechtol} K.,  et~al., 2015, \mn@doi [\apj] {10.1088/0004-637X/807/1/50}, \href
  {https://ui.adsabs.harvard.edu/abs/2015ApJ...807...50B} {807, 50}

\bibitem[\protect\citeauthoryear{{Belland}, {Kirby}, {Boylan-Kolchin}  \&
  {Wheeler}}{{Belland} et~al.}{2020}]{belland:20}
{Belland} B.,  {Kirby} E.,  {Boylan-Kolchin} M.,   {Wheeler} C.,  2020, \mn@doi
  [\apj] {10.3847/1538-4357/abb5f4}, \href
  {https://ui.adsabs.harvard.edu/abs/2020ApJ...903...10B} {903, 10}

\bibitem[\protect\citeauthoryear{{Belokurov} \& {Erkal}}{{Belokurov} \&
  {Erkal}}{2019}]{belokurov:19}
{Belokurov} V.~A.,  {Erkal} D.,  2019, \mn@doi [\mnras]
  {10.1093/mnrasl/sly178}, \href
  {https://ui.adsabs.harvard.edu/abs/2019MNRAS.482L...9B} {482, L9}

\bibitem[\protect\citeauthoryear{{Belokurov} \& {Koposov}}{{Belokurov} \&
  {Koposov}}{2016}]{belokurov:16}
{Belokurov} V.,  {Koposov} S.~E.,  2016, \mn@doi [\mnras]
  {10.1093/mnras/stv2688}, \href
  {https://ui.adsabs.harvard.edu/abs/2016MNRAS.456..602B} {456, 602}

\bibitem[\protect\citeauthoryear{{Bertin}}{{Bertin}}{2010}]{bertin:10}
{Bertin} E.,  2010, in JENAM 2010, Joint European and National Astronomy
  Meeting. p.~227

\bibitem[\protect\citeauthoryear{{Bertin}}{{Bertin}}{2011}]{bertin:11}
{Bertin} E.,  2011, in {Evans} I.~N.,  {Accomazzi} A.,  {Mink} D.~J.,   {Rots}
  A.~H.,  eds,  Astronomical Society of the Pacific Conference Series Vol. 442,
  Astronomical Data Analysis Software and Systems XX. p.~435

\bibitem[\protect\citeauthoryear{{Bertin} \& {Arnouts}}{{Bertin} \&
  {Arnouts}}{1996}]{bertin:96}
{Bertin} E.,  {Arnouts} S.,  1996, \mn@doi [\aaps] {10.1051/aas:1996164}, \href
  {https://ui.adsabs.harvard.edu/abs/1996A&AS..117..393B} {117, 393}

\bibitem[\protect\citeauthoryear{{Bovy}}{{Bovy}}{2015}]{bovy:15}
{Bovy} J.,  2015, \mn@doi [\apjs] {10.1088/0067-0049/216/2/29}, \href
  {https://ui.adsabs.harvard.edu/abs/2015ApJS..216...29B} {216, 29}

\bibitem[\protect\citeauthoryear{{Buck}, {Macci{\`o}}, {Dutton}, {Obreja}  \&
  {Frings}}{{Buck} et~al.}{2019}]{buck:19}
{Buck} T.,  {Macci{\`o}} A.~V.,  {Dutton} A.~A.,  {Obreja} A.,   {Frings} J.,
  2019, \mn@doi [\mnras] {10.1093/mnras/sty2913}, \href
  {https://ui.adsabs.harvard.edu/abs/2019MNRAS.483.1314B} {483, 1314}

\bibitem[\protect\citeauthoryear{{Bullock} \& {Boylan-Kolchin}}{{Bullock} \&
  {Boylan-Kolchin}}{2017}]{bullock:17}
{Bullock} J.~S.,  {Boylan-Kolchin} M.,  2017, \mn@doi [\araa]
  {10.1146/annurev-astro-091916-055313}, \href
  {https://ui.adsabs.harvard.edu/abs/2017ARA&A..55..343B} {55, 343}

\bibitem[\protect\citeauthoryear{{Bullock} \& {Johnston}}{{Bullock} \&
  {Johnston}}{2005}]{bullock:05}
{Bullock} J.~S.,  {Johnston} K.~V.,  2005, \mn@doi [\apj] {10.1086/497422},
  \href {https://ui.adsabs.harvard.edu/abs/2005ApJ...635..931B} {635, 931}

\bibitem[\protect\citeauthoryear{{Cannon} et~al.,}{{Cannon}
  et~al.}{2012}]{cannon:12}
{Cannon} J.~M.,  et~al., 2012, \mn@doi [\apj] {10.1088/0004-637X/747/2/122},
  \href {https://ui.adsabs.harvard.edu/abs/2012ApJ...747..122C} {747, 122}

\bibitem[\protect\citeauthoryear{{Carlin} et~al.,}{{Carlin}
  et~al.}{2016}]{carlin:16}
{Carlin} J.~L.,  et~al., 2016, \mn@doi [\apjl] {10.3847/2041-8205/828/1/L5},
  \href {https://ui.adsabs.harvard.edu/abs/2016ApJ...828L...5C} {828, L5}

\bibitem[\protect\citeauthoryear{{Carlin} et~al.,}{{Carlin}
  et~al.}{2021}]{carlin:21}
{Carlin} J.~L.,  et~al., 2021, \mn@doi [\apj] {10.3847/1538-4357/abe040}, \href
  {https://ui.adsabs.harvard.edu/abs/2021ApJ...909..211C} {909, 211}

\bibitem[\protect\citeauthoryear{{Choi}, {Dotter}, {Conroy}, {Cantiello},
  {Paxton}  \& {Johnson}}{{Choi} et~al.}{2016}]{choi:16}
{Choi} J.,  {Dotter} A.,  {Conroy} C.,  {Cantiello} M.,  {Paxton} B.,
  {Johnson} B.~D.,  2016, \mn@doi [\apj] {10.3847/0004-637X/823/2/102}, \href
  {https://ui.adsabs.harvard.edu/abs/2016ApJ...823..102C} {823, 102}

\bibitem[\protect\citeauthoryear{{Cooper} et~al.,}{{Cooper}
  et~al.}{2010}]{cooper:10}
{Cooper} A.~P.,  et~al., 2010, \mn@doi [\mnras]
  {10.1111/j.1365-2966.2010.16740.x}, \href
  {https://ui.adsabs.harvard.edu/abs/2010MNRAS.406..744C} {406, 744}

\bibitem[\protect\citeauthoryear{{Cullinane}, {Mackey}, {Da Costa}, {Erkal},
  {Koposov}  \& {Belokurov}}{{Cullinane} et~al.}{2021}]{cullinane:21}
{Cullinane} L.~R.,  {Mackey} A.~D.,  {Da Costa} G.~S.,  {Erkal} D.,  {Koposov}
  S.~E.,   {Belokurov} V.,  2021, arXiv e-prints, \href
  {https://ui.adsabs.harvard.edu/abs/2021arXiv210603274C} {p. arXiv:2106.03274}

\bibitem[\protect\citeauthoryear{{Davidge}}{{Davidge}}{2003}]{davidge:03}
{Davidge} T.~J.,  2003, \mn@doi [\pasp] {10.1086/375389}, \href
  {https://ui.adsabs.harvard.edu/abs/2003PASP..115..635D} {115, 635}

\bibitem[\protect\citeauthoryear{{Demers}, {Battinelli}  \& {Artigau}}{{Demers}
  et~al.}{2006}]{demers:06}
{Demers} S.,  {Battinelli} P.,   {Artigau} E.,  2006, VizieR Online Data
  Catalog, \href {https://ui.adsabs.harvard.edu/abs/2006yCat..34560905D} {pp
  J/A+A/456/905}

\bibitem[\protect\citeauthoryear{{Desai} et~al.,}{{Desai}
  et~al.}{2012}]{desai:12}
{Desai} S.,  et~al., 2012, \mn@doi [\apj] {10.1088/0004-637X/757/1/83}, \href
  {https://ui.adsabs.harvard.edu/abs/2012ApJ...757...83D} {757, 83}

\bibitem[\protect\citeauthoryear{{Diemand}, {Kuhlen}  \& {Madau}}{{Diemand}
  et~al.}{2007}]{diemand:07}
{Diemand} J.,  {Kuhlen} M.,   {Madau} P.,  2007, \mn@doi [\apj]
  {10.1086/520573}, \href
  {https://ui.adsabs.harvard.edu/abs/2007ApJ...667..859D} {667, 859}

\bibitem[\protect\citeauthoryear{{Diemand}, {Kuhlen}, {Madau}, {Zemp}, {Moore},
  {Potter}  \& {Stadel}}{{Diemand} et~al.}{2008}]{diemand:08}
{Diemand} J.,  {Kuhlen} M.,  {Madau} P.,  {Zemp} M.,  {Moore} B.,  {Potter} D.,
    {Stadel} J.,  2008, \mn@doi [\nat] {10.1038/nature07153}, \href
  {https://ui.adsabs.harvard.edu/abs/2008Natur.454..735D} {454, 735}

\bibitem[\protect\citeauthoryear{{Dooley}, {Peter}, {Yang}, {Willman},
  {Griffen}  \& {Frebel}}{{Dooley} et~al.}{2017a}]{dooley:17b}
{Dooley} G.~A.,  {Peter} A. H.~G.,  {Yang} T.,  {Willman} B.,  {Griffen} B.~F.,
    {Frebel} A.,  2017a, \mn@doi [\mnras] {10.1093/mnras/stx1900}, \href
  {https://ui.adsabs.harvard.edu/abs/2017MNRAS.471.4894D} {471, 4894}

\bibitem[\protect\citeauthoryear{{Dooley}, {Peter}, {Carlin}, {Frebel},
  {Bechtol}  \& {Willman}}{{Dooley} et~al.}{2017b}]{dooley:17}
{Dooley} G.~A.,  {Peter} A. H.~G.,  {Carlin} J.~L.,  {Frebel} A.,  {Bechtol}
  K.,   {Willman} B.,  2017b, \mn@doi [\mnras] {10.1093/mnras/stx2001}, \href
  {https://ui.adsabs.harvard.edu/abs/2017MNRAS.472.1060D} {472, 1060}

\bibitem[\protect\citeauthoryear{{Dotter}, {Chaboyer}, {Jevremovi{\'c}},
  {Kostov}, {Baron}  \& {Ferguson}}{{Dotter} et~al.}{2008}]{dotter:08}
{Dotter} A.,  {Chaboyer} B.,  {Jevremovi{\'c}} D.,  {Kostov} V.,  {Baron} E.,
  {Ferguson} J.~W.,  2008, \mn@doi [\apjs] {10.1086/589654}, \href
  {https://ui.adsabs.harvard.edu/abs/2008ApJS..178...89D} {178, 89}

\bibitem[\protect\citeauthoryear{{Drlica-Wagner} et~al.,}{{Drlica-Wagner}
  et~al.}{2015}]{drlica:15}
{Drlica-Wagner} A.,  et~al., 2015, \mn@doi [\apj]
  {10.1088/0004-637X/813/2/109}, \href
  {https://ui.adsabs.harvard.edu/abs/2015ApJ...813..109D} {813, 109}

\bibitem[\protect\citeauthoryear{{Erkal}, {Belokurov}  \& {Parkin}}{{Erkal}
  et~al.}{2020}]{erkal:20}
{Erkal} D.,  {Belokurov} V.~A.,   {Parkin} D.~L.,  2020, \mn@doi [\mnras]
  {10.1093/mnras/staa2840}, \href
  {https://ui.adsabs.harvard.edu/abs/2020MNRAS.498.5574E} {498, 5574}

\bibitem[\protect\citeauthoryear{{Flaugher} et~al.,}{{Flaugher}
  et~al.}{2015}]{flaugher:15}
{Flaugher} B.,  et~al., 2015, \mn@doi [\aj] {10.1088/0004-6256/150/5/150},
  \href {https://ui.adsabs.harvard.edu/abs/2015AJ....150..150F} {150, 150}

\bibitem[\protect\citeauthoryear{{Flewelling} et~al.,}{{Flewelling}
  et~al.}{2020}]{flewelling:20}
{Flewelling} H.~A.,  et~al., 2020, \mn@doi [\apjs] {10.3847/1538-4365/abb82d},
  \href {https://ui.adsabs.harvard.edu/abs/2020ApJS..251....7F} {251, 7}

\bibitem[\protect\citeauthoryear{Foreman-Mackey}{Foreman-Mackey}{2016}]{foreman:16}
Foreman-Mackey D.,  2016, \mn@doi [The Journal of Open Source Software]
  {10.21105/joss.00024}, 1, 24

\bibitem[\protect\citeauthoryear{{Foreman-Mackey}, {Hogg}, {Lang}  \&
  {Goodman}}{{Foreman-Mackey} et~al.}{2013}]{foreman:13}
{Foreman-Mackey} D.,  {Hogg} D.~W.,  {Lang} D.,   {Goodman} J.,  2013, \mn@doi
  [\pasp] {10.1086/670067}, \href
  {https://ui.adsabs.harvard.edu/abs/2013PASP..125..306F} {125, 306}

\bibitem[\protect\citeauthoryear{{Fusco}, {Buonanno}, {Hidalgo}, {Aparicio},
  {Pietrinferni}, {Bono}, {Monelli}  \& {Cassisi}}{{Fusco}
  et~al.}{2014}]{fusco:14}
{Fusco} F.,  {Buonanno} R.,  {Hidalgo} S.~L.,  {Aparicio} A.,  {Pietrinferni}
  A.,  {Bono} G.,  {Monelli} M.,   {Cassisi} S.,  2014, \mn@doi [\aap]
  {10.1051/0004-6361/201323075}, \href
  {https://ui.adsabs.harvard.edu/abs/2014A&A...572A..26F} {572, A26}

\bibitem[\protect\citeauthoryear{{Gaia Collaboration}}{{Gaia
  Collaboration}}{2021}]{gaia:21}
{Gaia Collaboration} 2021, \mn@doi [\aap] {10.1051/0004-6361/202039657}, \href
  {https://ui.adsabs.harvard.edu/abs/2021A&A...649A...1G} {649, A1}

\bibitem[\protect\citeauthoryear{{Gaia Collaboration} et~al.,}{{Gaia
  Collaboration} et~al.}{2018}]{gaia:18}
{Gaia Collaboration} et~al., 2018, \mn@doi [\aap]
  {10.1051/0004-6361/201833051}, \href
  {https://ui.adsabs.harvard.edu/abs/2018A&A...616A...1G} {616, A1}

\bibitem[\protect\citeauthoryear{{Gaia Collaboration} et~al.,}{{Gaia
  Collaboration} et~al.}{2021}]{mag:21}
{Gaia Collaboration} et~al., 2021, \mn@doi [\aap]
  {10.1051/0004-6361/202039588}, \href
  {https://ui.adsabs.harvard.edu/abs/2021A&A...649A...7G} {649, A7}

\bibitem[\protect\citeauthoryear{{Gieren}, {Pietrzy{\'n}ski}, {Nalewajko},
  {Soszy{\'n}ski}, {Bresolin}, {Kudritzki}, {Minniti}  \&
  {Romanowsky}}{{Gieren} et~al.}{2006}]{gieren:06}
{Gieren} W.,  {Pietrzy{\'n}ski} G.,  {Nalewajko} K.,  {Soszy{\'n}ski} I.,
  {Bresolin} F.,  {Kudritzki} R.-P.,  {Minniti} D.,   {Romanowsky} A.,  2006,
  \mn@doi [\apj] {10.1086/505574}, \href
  {https://ui.adsabs.harvard.edu/abs/2006ApJ...647.1056G} {647, 1056}

\bibitem[\protect\citeauthoryear{{Gill}, {Knebe}  \& {Gibson}}{{Gill}
  et~al.}{2005}]{gill:05}
{Gill} S. P.~D.,  {Knebe} A.,   {Gibson} B.~K.,  2005, \mn@doi [\mnras]
  {10.1111/j.1365-2966.2004.08562.x}, \href
  {https://ui.adsabs.harvard.edu/abs/2005MNRAS.356.1327G} {356, 1327}

\bibitem[\protect\citeauthoryear{{G{\'o}rski}, {Pietrzy{\'n}ski}  \&
  {Gieren}}{{G{\'o}rski} et~al.}{2011}]{gorski:11}
{G{\'o}rski} M.,  {Pietrzy{\'n}ski} G.,   {Gieren} W.,  2011, \mn@doi [\aj]
  {10.1088/0004-6256/141/6/194}, \href
  {https://ui.adsabs.harvard.edu/abs/2011AJ....141..194G} {141, 194}

\bibitem[\protect\citeauthoryear{{Higgs}, {McConnachie}, {Annau}, {Irwin},
  {Battaglia}, {C{\^o}t{\'e}}, {Lewis}  \& {Venn}}{{Higgs}
  et~al.}{2021}]{higgs:21}
{Higgs} C.~R.,  {McConnachie} A.~W.,  {Annau} N.,  {Irwin} M.,  {Battaglia} G.,
   {C{\^o}t{\'e}} P.,  {Lewis} G.~F.,   {Venn} K.,  2021, \mn@doi [\mnras]
  {10.1093/mnras/stab002}, \href
  {https://ui.adsabs.harvard.edu/abs/2021MNRAS.503..176H} {503, 176}

\bibitem[\protect\citeauthoryear{{Hubble}}{{Hubble}}{1925}]{hubble:25}
{Hubble} E.~P.,  1925, \mn@doi [\apj] {10.1086/142943}, \href
  {https://ui.adsabs.harvard.edu/abs/1925ApJ....62..409H} {62, 409}

\bibitem[\protect\citeauthoryear{{Huxor}, {Ferguson}, {Veljanoski}, {Mackey}
  \& {Tanvir}}{{Huxor} et~al.}{2013}]{huxor:13}
{Huxor} A.~P.,  {Ferguson} A.~M.~N.,  {Veljanoski} J.,  {Mackey} A.~D.,
  {Tanvir} N.~R.,  2013, \mn@doi [\mnras] {10.1093/mnras/sts387}, \href
  {https://ui.adsabs.harvard.edu/abs/2013MNRAS.429.1039H} {429, 1039}

\bibitem[\protect\citeauthoryear{{Hwang}, {Lee}, {Lee}, {Park}, {Park}, {Kim}
  \& {Park}}{{Hwang} et~al.}{2011}]{hwang:11}
{Hwang} N.,  {Lee} M.~G.,  {Lee} J.~C.,  {Park} W.-K.,  {Park} H.~S.,  {Kim}
  S.~C.,   {Park} J.-H.,  2011, \mn@doi [\apj] {10.1088/0004-637X/738/1/58},
  \href {https://ui.adsabs.harvard.edu/abs/2011ApJ...738...58H} {738, 58}

\bibitem[\protect\citeauthoryear{{Ibata}, {Ibata}, {Lewis}, {Martin}, {Conn},
  {Elahi}, {Arias}  \& {Fernando}}{{Ibata} et~al.}{2014}]{ibata:14}
{Ibata} R.~A.,  {Ibata} N.~G.,  {Lewis} G.~F.,  {Martin} N.~F.,  {Conn} A.,
  {Elahi} P.,  {Arias} V.,   {Fernando} N.,  2014, \mn@doi [\apjl]
  {10.1088/2041-8205/784/1/L6}, \href
  {https://ui.adsabs.harvard.edu/abs/2014ApJ...784L...6I} {784, L6}

\bibitem[\protect\citeauthoryear{{Iorio} \& {Belokurov}}{{Iorio} \&
  {Belokurov}}{2019}]{iorio:19}
{Iorio} G.,  {Belokurov} V.,  2019, \mn@doi [\mnras] {10.1093/mnras/sty2806},
  \href {https://ui.adsabs.harvard.edu/abs/2019MNRAS.482.3868I} {482, 3868}

\bibitem[\protect\citeauthoryear{{Jethwa}, {Erkal}  \& {Belokurov}}{{Jethwa}
  et~al.}{2016}]{jethwa:16}
{Jethwa} P.,  {Erkal} D.,   {Belokurov} V.,  2016, \mn@doi [\mnras]
  {10.1093/mnras/stw1343}, \href
  {https://ui.adsabs.harvard.edu/abs/2016MNRAS.461.2212J} {461, 2212}

\bibitem[\protect\citeauthoryear{{Kallivayalil} et~al.,}{{Kallivayalil}
  et~al.}{2018}]{kallivayalil:18}
{Kallivayalil} N.,  et~al., 2018, \mn@doi [\apj] {10.3847/1538-4357/aadfee},
  \href {https://ui.adsabs.harvard.edu/abs/2018ApJ...867...19K} {867, 19}

\bibitem[\protect\citeauthoryear{{Kim} \& {Jerjen}}{{Kim} \&
  {Jerjen}}{2015}]{kim:15a}
{Kim} D.,  {Jerjen} H.,  2015, \mn@doi [\apj] {10.1088/0004-637X/799/1/73},
  \href {https://ui.adsabs.harvard.edu/abs/2015ApJ...799...73K} {799, 73}

\bibitem[\protect\citeauthoryear{{Kim}, {Jerjen}, {Mackey}, {Da Costa}  \&
  {Milone}}{{Kim} et~al.}{2015}]{kim:15b}
{Kim} D.,  {Jerjen} H.,  {Mackey} D.,  {Da Costa} G.~S.,   {Milone} A.~P.,
  2015, \mn@doi [\apjl] {10.1088/2041-8205/804/2/L44}, \href
  {https://ui.adsabs.harvard.edu/abs/2015ApJ...804L..44K} {804, L44}

\bibitem[\protect\citeauthoryear{{Kirby}, {Cohen}, {Guhathakurta}, {Cheng},
  {Bullock}  \& {Gallazzi}}{{Kirby} et~al.}{2013}]{kirby:13}
{Kirby} E.~N.,  {Cohen} J.~G.,  {Guhathakurta} P.,  {Cheng} L.,  {Bullock}
  J.~S.,   {Gallazzi} A.,  2013, \mn@doi [\apj] {10.1088/0004-637X/779/2/102},
  \href {https://ui.adsabs.harvard.edu/abs/2013ApJ...779..102K} {779, 102}

\bibitem[\protect\citeauthoryear{{Kirby}, {Bullock}, {Boylan-Kolchin},
  {Kaplinghat}  \& {Cohen}}{{Kirby} et~al.}{2014}]{kirby:14}
{Kirby} E.~N.,  {Bullock} J.~S.,  {Boylan-Kolchin} M.,  {Kaplinghat} M.,
  {Cohen} J.~G.,  2014, \mn@doi [\mnras] {10.1093/mnras/stu025}, \href
  {https://ui.adsabs.harvard.edu/abs/2014MNRAS.439.1015K} {439, 1015}

\bibitem[\protect\citeauthoryear{{Komiyama} et~al.,}{{Komiyama}
  et~al.}{2003}]{komiyama:03}
{Komiyama} Y.,  et~al., 2003, \mn@doi [\apjl] {10.1086/376551}, \href
  {https://ui.adsabs.harvard.edu/abs/2003ApJ...590L..17K} {590, L17}

\bibitem[\protect\citeauthoryear{{Koposov}, {Belokurov}, {Torrealba}  \&
  {Evans}}{{Koposov} et~al.}{2015}]{koposov:15}
{Koposov} S.~E.,  {Belokurov} V.,  {Torrealba} G.,   {Evans} N.~W.,  2015,
  \mn@doi [\apj] {10.1088/0004-637X/805/2/130}, \href
  {https://ui.adsabs.harvard.edu/abs/2015ApJ...805..130K} {805, 130}

\bibitem[\protect\citeauthoryear{{Koposov} et~al.,}{{Koposov}
  et~al.}{2018}]{koposov:18}
{Koposov} S.~E.,  et~al., 2018, \mn@doi [\mnras] {10.1093/mnras/sty1772}, \href
  {https://ui.adsabs.harvard.edu/abs/2018MNRAS.479.5343K} {479, 5343}

\bibitem[\protect\citeauthoryear{{Kuzma}, {Da Costa}, {Mackey}  \&
  {Roderick}}{{Kuzma} et~al.}{2016}]{kuzma:16}
{Kuzma} P.~B.,  {Da Costa} G.~S.,  {Mackey} A.~D.,   {Roderick} T.~A.,  2016,
  \mn@doi [\mnras] {10.1093/mnras/stw1561}, \href
  {https://ui.adsabs.harvard.edu/abs/2016MNRAS.461.3639K} {461, 3639}

\bibitem[\protect\citeauthoryear{{Kuzma}, {Da Costa}  \& {Mackey}}{{Kuzma}
  et~al.}{2018}]{kuzma:18}
{Kuzma} P.~B.,  {Da Costa} G.~S.,   {Mackey} A.~D.,  2018, \mn@doi [\mnras]
  {10.1093/mnras/stx2353}, \href
  {https://ui.adsabs.harvard.edu/abs/2018MNRAS.473.2881K} {473, 2881}

\bibitem[\protect\citeauthoryear{{Li} \& {Helmi}}{{Li} \&
  {Helmi}}{2008}]{li:08}
{Li} Y.-S.,  {Helmi} A.,  2008, \mn@doi [\mnras]
  {10.1111/j.1365-2966.2008.12854.x}, \href
  {https://ui.adsabs.harvard.edu/abs/2008MNRAS.385.1365L} {385, 1365}

\bibitem[\protect\citeauthoryear{{Lindegren} et~al.,}{{Lindegren}
  et~al.}{2021}]{lindegren:21}
{Lindegren} L.,  et~al., 2021, \mn@doi [\aap] {10.1051/0004-6361/202039709},
  \href {https://ui.adsabs.harvard.edu/abs/2021A&A...649A...2L} {649, A2}

\bibitem[\protect\citeauthoryear{{Mackey}, {Koposov}, {Erkal}, {Belokurov}, {Da
  Costa}  \& {G{\'o}mez}}{{Mackey} et~al.}{2016}]{mackey:16}
{Mackey} A.~D.,  {Koposov} S.~E.,  {Erkal} D.,  {Belokurov} V.,  {Da Costa}
  G.~S.,   {G{\'o}mez} F.~A.,  2016, \mn@doi [\mnras] {10.1093/mnras/stw497},
  \href {https://ui.adsabs.harvard.edu/abs/2016MNRAS.459..239M} {459, 239}

\bibitem[\protect\citeauthoryear{{Mackey}, {Koposov}, {Da Costa}, {Belokurov},
  {Erkal}  \& {Kuzma}}{{Mackey} et~al.}{2018}]{mackey:2018}
{Mackey} D.,  {Koposov} S.,  {Da Costa} G.,  {Belokurov} V.,  {Erkal} D.,
  {Kuzma} P.,  2018, \mn@doi [\apjl] {10.3847/2041-8213/aac175}, \href
  {https://ui.adsabs.harvard.edu/abs/2018ApJ...858L..21M} {858, L21}

\bibitem[\protect\citeauthoryear{{Martin} et~al.,}{{Martin}
  et~al.}{2016}]{martin:16}
{Martin} N.~F.,  et~al., 2016, \mn@doi [\apj] {10.3847/1538-4357/833/2/167},
  \href {https://ui.adsabs.harvard.edu/abs/2016ApJ...833..167M} {833, 167}

\bibitem[\protect\citeauthoryear{{Mart{\'\i}nez-Delgado}
  et~al.,}{{Mart{\'\i}nez-Delgado} et~al.}{2012}]{md:12}
{Mart{\'\i}nez-Delgado} D.,  et~al., 2012, \mn@doi [\apjl]
  {10.1088/2041-8205/748/2/L24}, \href
  {https://ui.adsabs.harvard.edu/abs/2012ApJ...748L..24M} {748, L24}

\bibitem[\protect\citeauthoryear{{Mateo}}{{Mateo}}{1998}]{mateo:98}
{Mateo} M.~L.,  1998, \mn@doi [\araa] {10.1146/annurev.astro.36.1.435}, \href
  {https://ui.adsabs.harvard.edu/abs/1998ARA&A..36..435M} {36, 435}

\bibitem[\protect\citeauthoryear{{McConnachie}}{{McConnachie}}{2012}]{mcconnachie:12}
{McConnachie} A.~W.,  2012, \mn@doi [\aj] {10.1088/0004-6256/144/1/4}, \href
  {https://ui.adsabs.harvard.edu/abs/2012AJ....144....4M} {144, 4}

\bibitem[\protect\citeauthoryear{{McConnachie}, {Higgs}, {Thomas}, {Venn},
  {C{\^o}t{\'e}}, {Battaglia}  \& {Lewis}}{{McConnachie}
  et~al.}{2021}]{mcconnachie:21}
{McConnachie} A.~W.,  {Higgs} C.~R.,  {Thomas} G.~F.,  {Venn} K.~A.,
  {C{\^o}t{\'e}} P.,  {Battaglia} G.,   {Lewis} G.~F.,  2021, \mn@doi [\mnras]
  {10.1093/mnras/staa3740}, \href
  {https://ui.adsabs.harvard.edu/abs/2021MNRAS.501.2363M} {501, 2363}

\bibitem[\protect\citeauthoryear{{McLeod} et~al.,}{{McLeod}
  et~al.}{2015}]{mcleod:15}
{McLeod} B.,  et~al., 2015, \mn@doi [\pasp] {10.1086/680687}, \href
  {https://ui.adsabs.harvard.edu/abs/2015PASP..127..366M} {127, 366}

\bibitem[\protect\citeauthoryear{{McMillan}}{{McMillan}}{2017}]{mcmillan:17}
{McMillan} P.~J.,  2017, \mn@doi [\mnras] {10.1093/mnras/stw2759}, \href
  {https://ui.adsabs.harvard.edu/abs/2017MNRAS.465...76M} {465, 76}

\bibitem[\protect\citeauthoryear{{McMonigal} et~al.,}{{McMonigal}
  et~al.}{2016}]{mcmonigal:16}
{McMonigal} B.,  et~al., 2016, \mn@doi [\mnras] {10.1093/mnras/stv2690}, \href
  {https://ui.adsabs.harvard.edu/abs/2016MNRAS.456..405M} {456, 405}

\bibitem[\protect\citeauthoryear{{Minniti} \& {Zijlstra}}{{Minniti} \&
  {Zijlstra}}{1996}]{minniti:96}
{Minniti} D.,  {Zijlstra} A.~A.,  1996, \mn@doi [\apjl] {10.1086/310189}, \href
  {https://ui.adsabs.harvard.edu/abs/1996ApJ...467L..13M} {467, L13}

\bibitem[\protect\citeauthoryear{{Minniti}, {Zijlstra}  \& {Alonso}}{{Minniti}
  et~al.}{1999}]{minniti:99}
{Minniti} D.,  {Zijlstra} A.~A.,   {Alonso} M.~V.,  1999, \mn@doi [\aj]
  {10.1086/300735}, \href
  {https://ui.adsabs.harvard.edu/abs/1999AJ....117..881M} {117, 881}

\bibitem[\protect\citeauthoryear{{Minniti}, {Borissova}, {Rejkuba}, {Alves},
  {Cook}  \& {Freeman}}{{Minniti} et~al.}{2003}]{minniti:03}
{Minniti} D.,  {Borissova} J.,  {Rejkuba} M.,  {Alves} D.~R.,  {Cook} K.~H.,
  {Freeman} K.~C.,  2003, \mn@doi [Science] {10.1126/science.1088529}, \href
  {https://ui.adsabs.harvard.edu/abs/2003Sci...301.1508M} {301, 1508}

\bibitem[\protect\citeauthoryear{{Moore}, {Ghigna}, {Governato}, {Lake},
  {Quinn}, {Stadel}  \& {Tozzi}}{{Moore} et~al.}{1999}]{moore:99}
{Moore} B.,  {Ghigna} S.,  {Governato} F.,  {Lake} G.,  {Quinn} T.,  {Stadel}
  J.,   {Tozzi} P.,  1999, \mn@doi [\apjl] {10.1086/312287}, \href
  {https://ui.adsabs.harvard.edu/abs/1999ApJ...524L..19M} {524, L19}

\bibitem[\protect\citeauthoryear{{Myeong}, {Jerjen}, {Mackey}  \& {Da
  Costa}}{{Myeong} et~al.}{2017}]{myeong:17}
{Myeong} G.~C.,  {Jerjen} H.,  {Mackey} D.,   {Da Costa} G.~S.,  2017, \mn@doi
  [\apjl] {10.3847/2041-8213/aa6fb4}, \href
  {https://ui.adsabs.harvard.edu/abs/2017ApJ...840L..25M} {840, L25}

\bibitem[\protect\citeauthoryear{{Navarrete} et~al.,}{{Navarrete}
  et~al.}{2019}]{navarrete:19}
{Navarrete} C.,  et~al., 2019, \mn@doi [\mnras] {10.1093/mnras/sty3347}, \href
  {https://ui.adsabs.harvard.edu/abs/2019MNRAS.483.4160N} {483, 4160}

\bibitem[\protect\citeauthoryear{{Roderick}, {Jerjen}, {Mackey}  \& {Da
  Costa}}{{Roderick} et~al.}{2015}]{roderick:15}
{Roderick} T.~A.,  {Jerjen} H.,  {Mackey} A.~D.,   {Da Costa} G.~S.,  2015,
  \mn@doi [\apj] {10.1088/0004-637X/804/2/134}, \href
  {https://ui.adsabs.harvard.edu/abs/2015ApJ...804..134R} {804, 134}

\bibitem[\protect\citeauthoryear{{Roderick}, {Jerjen}, {Da Costa}  \&
  {Mackey}}{{Roderick} et~al.}{2016a}]{roderick:16a}
{Roderick} T.~A.,  {Jerjen} H.,  {Da Costa} G.~S.,   {Mackey} A.~D.,  2016a,
  \mn@doi [\mnras] {10.1093/mnras/stw949}, \href
  {https://ui.adsabs.harvard.edu/abs/2016MNRAS.460...30R} {460, 30}

\bibitem[\protect\citeauthoryear{{Roderick}, {Mackey}, {Jerjen}  \& {Da
  Costa}}{{Roderick} et~al.}{2016b}]{roderick:16b}
{Roderick} T.~A.,  {Mackey} A.~D.,  {Jerjen} H.,   {Da Costa} G.~S.,  2016b,
  \mn@doi [\mnras] {10.1093/mnras/stw1541}, \href
  {https://ui.adsabs.harvard.edu/abs/2016MNRAS.461.3702R} {461, 3702}

\bibitem[\protect\citeauthoryear{{Saha} et~al.,}{{Saha} et~al.}{2010}]{saha:10}
{Saha} A.,  et~al., 2010, \mn@doi [\aj] {10.1088/0004-6256/140/6/1719}, \href
  {https://ui.adsabs.harvard.edu/abs/2010AJ....140.1719S} {140, 1719}

\bibitem[\protect\citeauthoryear{{Sales}, {Navarro}, {Kallivayalil}  \&
  {Frenk}}{{Sales} et~al.}{2017}]{sales:17}
{Sales} L.~V.,  {Navarro} J.~F.,  {Kallivayalil} N.,   {Frenk} C.~S.,  2017,
  \mn@doi [\mnras] {10.1093/mnras/stw2816}, \href
  {https://ui.adsabs.harvard.edu/abs/2017MNRAS.465.1879S} {465, 1879}

\bibitem[\protect\citeauthoryear{{Salpeter}}{{Salpeter}}{1955}]{salpeter:55}
{Salpeter} E.~E.,  1955, \mn@doi [\apj] {10.1086/145971}, \href
  {https://ui.adsabs.harvard.edu/abs/1955ApJ...121..161S} {121, 161}

\bibitem[\protect\citeauthoryear{{Schlafly} \& {Finkbeiner}}{{Schlafly} \&
  {Finkbeiner}}{2011}]{schlafly:11}
{Schlafly} E.~F.,  {Finkbeiner} D.~P.,  2011, \mn@doi [\apj]
  {10.1088/0004-637X/737/2/103}, \href
  {https://ui.adsabs.harvard.edu/abs/2011ApJ...737..103S} {737, 103}

\bibitem[\protect\citeauthoryear{{Schlegel}, {Finkbeiner}  \&
  {Davis}}{{Schlegel} et~al.}{1998}]{schlegel:98}
{Schlegel} D.~J.,  {Finkbeiner} D.~P.,   {Davis} M.,  1998, \mn@doi [\apj]
  {10.1086/305772}, \href
  {https://ui.adsabs.harvard.edu/abs/1998ApJ...500..525S} {500, 525}

\bibitem[\protect\citeauthoryear{{Simon}}{{Simon}}{2019}]{simon:19}
{Simon} J.~D.,  2019, \mn@doi [\araa] {10.1146/annurev-astro-091918-104453},
  \href {https://ui.adsabs.harvard.edu/abs/2019ARA&A..57..375S} {57, 375}

\bibitem[\protect\citeauthoryear{{Springel} et~al.,}{{Springel}
  et~al.}{2008}]{springel:08}
{Springel} V.,  et~al., 2008, \mn@doi [\mnras]
  {10.1111/j.1365-2966.2008.14066.x}, \href
  {https://ui.adsabs.harvard.edu/abs/2008MNRAS.391.1685S} {391, 1685}

\bibitem[\protect\citeauthoryear{{Swan}, {Cole}, {Tolstoy}  \& {Irwin}}{{Swan}
  et~al.}{2016}]{swan:16}
{Swan} J.,  {Cole} A.~A.,  {Tolstoy} E.,   {Irwin} M.~J.,  2016, \mn@doi
  [\mnras] {10.1093/mnras/stv2774}, \href
  {https://ui.adsabs.harvard.edu/abs/2016MNRAS.456.4315S} {456, 4315}

\bibitem[\protect\citeauthoryear{{Teyssier}, {Johnston}  \&
  {Kuhlen}}{{Teyssier} et~al.}{2012}]{teyssier:12}
{Teyssier} M.,  {Johnston} K.~V.,   {Kuhlen} M.,  2012, \mn@doi [\mnras]
  {10.1111/j.1365-2966.2012.21793.x}, \href
  {https://ui.adsabs.harvard.edu/abs/2012MNRAS.426.1808T} {426, 1808}

\bibitem[\protect\citeauthoryear{{Thompson}, {Ryan}  \& {Sibbons}}{{Thompson}
  et~al.}{2016}]{thompson:16}
{Thompson} G.~P.,  {Ryan} S.~G.,   {Sibbons} L.~F.,  2016, \mn@doi [\mnras]
  {10.1093/mnras/stw1193}, \href
  {https://ui.adsabs.harvard.edu/abs/2016MNRAS.462.3376T} {462, 3376}

\bibitem[\protect\citeauthoryear{{Valdes}, {Gruendl}  \& {DES
  Project}}{{Valdes} et~al.}{2014}]{valdes:14}
{Valdes} F.,  {Gruendl} R.,   {DES Project} 2014, in {Manset} N.,  {Forshay}
  P.,  eds,  Astronomical Society of the Pacific Conference Series Vol. 485,
  Astronomical Data Analysis Software and Systems XXIII. p.~379

\bibitem[\protect\citeauthoryear{{Vasiliev}}{{Vasiliev}}{2019}]{vasiliev:19}
{Vasiliev} E.,  2019, \mn@doi [\mnras] {10.1093/mnras/stz171}, \href
  {https://ui.adsabs.harvard.edu/abs/2019MNRAS.484.2832V} {484, 2832}

\bibitem[\protect\citeauthoryear{{Veljanoski} et~al.,}{{Veljanoski}
  et~al.}{2015}]{veljanoski:15}
{Veljanoski} J.,  et~al., 2015, \mn@doi [\mnras] {10.1093/mnras/stv1259}, \href
  {https://ui.adsabs.harvard.edu/abs/2015MNRAS.452..320V} {452, 320}

\bibitem[\protect\citeauthoryear{{Weisz}, {Dolphin}, {Skillman}, {Holtzman},
  {Gilbert}, {Dalcanton}  \& {Williams}}{{Weisz} et~al.}{2014}]{weisz:14}
{Weisz} D.~R.,  {Dolphin} A.~E.,  {Skillman} E.~D.,  {Holtzman} J.,  {Gilbert}
  K.~M.,  {Dalcanton} J.~J.,   {Williams} B.~F.,  2014, \mn@doi [\apj]
  {10.1088/0004-637X/789/2/147}, \href
  {https://ui.adsabs.harvard.edu/abs/2014ApJ...789..147W} {789, 147}

\bibitem[\protect\citeauthoryear{{Weldrake}, {de Blok}  \& {Walter}}{{Weldrake}
  et~al.}{2003}]{weldrake:03}
{Weldrake} D.~T.~F.,  {de Blok} W.~J.~G.,   {Walter} F.,  2003, \mn@doi
  [\mnras] {10.1046/j.1365-8711.2003.06170.x}, \href
  {https://ui.adsabs.harvard.edu/abs/2003MNRAS.340...12W} {340, 12}

\bibitem[\protect\citeauthoryear{{Wheeler}, {O{\~n}orbe}, {Bullock},
  {Boylan-Kolchin}, {Elbert}, {Garrison-Kimmel}, {Hopkins}  \&
  {Kere{\v{s}}}}{{Wheeler} et~al.}{2015}]{wheeler:15}
{Wheeler} C.,  {O{\~n}orbe} J.,  {Bullock} J.~S.,  {Boylan-Kolchin} M.,
  {Elbert} O.~D.,  {Garrison-Kimmel} S.,  {Hopkins} P.~F.,   {Kere{\v{s}}} D.,
  2015, \mn@doi [\mnras] {10.1093/mnras/stv1691}, \href
  {https://ui.adsabs.harvard.edu/abs/2015MNRAS.453.1305W} {453, 1305}

\bibitem[\protect\citeauthoryear{{Wheeler} et~al.,}{{Wheeler}
  et~al.}{2019}]{wheeler:19}
{Wheeler} C.,  et~al., 2019, \mn@doi [\mnras] {10.1093/mnras/stz2887}, \href
  {https://ui.adsabs.harvard.edu/abs/2019MNRAS.490.4447W} {490, 4447}

\bibitem[\protect\citeauthoryear{{de Blok} \& {Walter}}{{de Blok} \&
  {Walter}}{2000}]{deblok:00}
{de Blok} W.~J.~G.,  {Walter} F.,  2000, \mn@doi [\apjl] {10.1086/312777},
  \href {https://ui.adsabs.harvard.edu/abs/2000ApJ...537L..95D} {537, L95}

\bibitem[\protect\citeauthoryear{{de Blok} \& {Walter}}{{de Blok} \&
  {Walter}}{2003}]{deblok:03}
{de Blok} W.~J.~G.,  {Walter} F.,  2003, \mn@doi [\mnras]
  {10.1046/j.1365-8711.2003.06669.x}, \href
  {https://ui.adsabs.harvard.edu/abs/2003MNRAS.341L..39D} {341, L39}

\bibitem[\protect\citeauthoryear{{de Blok} \& {Walter}}{{de Blok} \&
  {Walter}}{2006}]{deblok:06}
{de Blok} W.~J.~G.,  {Walter} F.,  2006, \mn@doi [\aj] {10.1086/497829}, \href
  {https://ui.adsabs.harvard.edu/abs/2006AJ....131..343D} {131, 343}

\bibitem[\protect\citeauthoryear{{van der Marel}, {Fardal}, {Besla}, {Beaton},
  {Sohn}, {Anderson}, {Brown}  \& {Guhathakurta}}{{van der Marel}
  et~al.}{2012}]{vandermarel:12}
{van der Marel} R.~P.,  {Fardal} M.,  {Besla} G.,  {Beaton} R.~L.,  {Sohn}
  S.~T.,  {Anderson} J.,  {Brown} T.,   {Guhathakurta} P.,  2012, \mn@doi
  [\apj] {10.1088/0004-637X/753/1/8}, \href
  {https://ui.adsabs.harvard.edu/abs/2012ApJ...753....8V} {753, 8}

\bibitem[\protect\citeauthoryear{{van der Marel}, {Fardal}, {Sohn}, {Patel},
  {Besla}, {del Pino}, {Sahlmann}  \& {Watkins}}{{van der Marel}
  et~al.}{2019}]{vandermarel:19}
{van der Marel} R.~P.,  {Fardal} M.~A.,  {Sohn} S.~T.,  {Patel} E.,  {Besla}
  G.,  {del Pino} A.,  {Sahlmann} J.,   {Watkins} L.~L.,  2019, \mn@doi [\apj]
  {10.3847/1538-4357/ab001b}, \href
  {https://ui.adsabs.harvard.edu/abs/2019ApJ...872...24V} {872, 24}

\makeatother
\end{thebibliography}


\bsp	
\label{lastpage}
\end{document}